\documentclass[a4paper,11pt]{article}
\pdfoutput=1 % if your are submitting a pdflatex (i.e. if you have
\usepackage{jcappub} % for details on the use of the package, please
\usepackage[T1]{fontenc} % if needed

\usepackage{hyperref}
\hypersetup{
colorlinks=true,
linkcolor=blue,
citecolor=blue}
\usepackage{url}
\usepackage{caption}
\usepackage{graphicx}
\usepackage{float}
\usepackage{subcaption}
\usepackage{multirow}
\usepackage{makecell}
\usepackage{cases}
\usepackage{pifont}% http://ctan.org/pkg/pifont
\newcommand{\xmark}{\ding{55}}%

\usepackage{natbib}

\allowdisplaybreaks[3]
\def \bal#1\eal  {\begin{align} #1 \end{align}}
\def\({\left(}
\def\){\right)}
\def\[{\left[}
\def\]{\right]}
\def\<{\langle}
\def\>{\rangle}
\def\d{\mathrm{d}}

\newcommand{\eref}[1]{Eq.(\ref{#1})}
\newcommand{\f}[2]{\frac{#1}{#2}}
\newcommand{\bim} {\begin{itemize}[noitemsep]}

\newcommand{\eim}{\end{itemize}}
\newcommand{\be} {\begin{equation}}
\newcommand{\ee} {\end{equation}}
\newcommand{\bc}{\begin{center}}
\newcommand{\ec}{\end{center}}

\newcommand{\nn} {\nonumber\\}

\newcommand{\marrow}{~~\Longrightarrow~~}

\newcommand{\nd} {\nabla}
\newcommand{\pd} {\partial}
\newcommand{\mc} {\mathcal}

\newcommand{\li}{{\lambda}}

\newcommand{\thi}{\theta}

\newcommand{\Li}{\Lambda}
\newcommand{\tis}{\tilde{s}}
\newcommand{\tit}{\tilde{t}}
\newcommand{\tiu}{\tilde{u}}
\newcommand{\bA}{\bar{A}}
\newcommand{\bG}{\bar{G}}

\newcommand{\notcheckmark}{\checkmark\makebox[0pt][r]{\normalfont\symbol{'26}}}

\title{\huge \boldmath Triple crossing positivity bounds, mass dependence and cosmological scalars: Horndeski theory and DHOST}
\author[a]{Hao Xu}
\author[a,b]{and Shuang-Yong Zhou}
\affiliation[a]{Interdisciplinary Center for Theoretical Study, University of Science and Technology of China,
Hefei, Anhui 230026, China}
\affiliation[b]{Peng Huanwu Center for Fundamental Theory, Hefei, Anhui 230026, China}

\emailAdd{haoxu@mail.ustc.edu.cn}
\emailAdd{zhoushy@ustc.edu.cn}

\abstract{
Scalars are widely used in cosmology to model novel phenomena such as the late-time cosmic acceleration. These are effective field theories with highly nonlinear interactions, including Horndeski theory/generalized galileon and beyond. We use the latest fully crossing symmetric positivity bounds to constrain these cosmological EFTs. These positivity bounds, based on fundamental principles of quantum field theory such as causality and unitarity, are able to constrain the EFT coefficients both from above and below. We first map the mass dependence of the fully crossing symmetric bounds, and find that a nonzero mass generically enlarges the positivity regions. We show that fine-tunings in the EFT construction can significantly reduce the viable regions and sometimes can be precarious. Then, we apply the positivity bounds to several models in the Horndeski class and beyond, explicitly listing the ready-to-use bounds with the model parameters, and discuss the implications for these models. The new positivity bounds are found to severely constrain some of these models, in which positivity requires the mass to be parametrically close to the cutoff of the EFT, effectively ruling them out. The examples include massive galileon, the original beyond Horndeski model, and DHOST theory with unity speed of gravity and nearly constant Newton's coupling.
}

\begin{document}

\hfill{ {\small USTC-ICTS/PCFT-23-17} }

\maketitle
\flushbottom

\section{Introduction and summary}

Effective field theory (EFT) is powerful modeling technique for understanding physics at low energy scales. By isolating the relevant physical degrees of freedom in the problem, EFTs can greatly simplify calculations, and provide systematic parametrizations of the underlying physical processes, which allows for model-independent constraints to be placed in comparison with experimental data. Nonetheless, the vast parameter spaces associated with EFTs can present a significant obstacle to identifying and characterizing new physics in experiments as well as making useful practical predictions. Fortunately, positivity bounds \cite{Adams:2006sv} offer a powerful means of constraining the parameter spaces of low energy EFTs. Bootstrapped from fundamental principles of quantum field theories, these bounds can be used to place rigorous constraints on the allowed values of the Wilson coefficients, significantly narrowing down our aim at the search of viable theories. In recent years, this topic has attracted theorists from diverse backgrounds and seen rapid developments both in the improvements of the strength of the bounds and in their applications in particle physics and cosmology \cite{deRham:2022hpx}.
Among the ingredients used to derive the positivity bounds, causality plays a central role, as it implies analyticity of the scattering amplitudes, which is used to bring UV unitarity down to the low energy scales. Thus, these bounds are also known as causality bounds.

Scalar fields are the simplest kind of fields that are invariant, rather than covariant, under the symmetries of nature. The Higgs field is an established example of them, which in the Standard Model provides a mechanism to generate masses for the other particles. They are widely used in cosmology, as their bosonic nature allows them to form classical condensates, and are  inherently compatible with the cosmological principle. Indeed, inflation in the early universe is typically modeled with a scalar field \cite{Achucarro:2022qrl, Baumann:2009ds}, and a plethora of attempts have been made to model the other accelerated cosmic expansion in the late universe with scalar fields \cite{Clifton:2011jh, Joyce:2014kja}. Recently, especially after the discovery of gravitational waves \cite{LIGOScientific:2016lio}, significant efforts have been devoted to search for possible sign of extra scalars in strong gravity regimes such as near black holes or neutron stars \cite{Berti:2015itd, Barack:2018yly}.
Since general relativity (GR) has been accurately tested in weak gravity regimes, the effects of the extra scalars need to be suppressed in those environments, and yet they need to be sizable to achieve accelerated late-time cosmic expansion or significant deviations from GR near compact stars. This often means that the scalar field is adequately nonlinear, and the essential effects beyond GR are to a large extent accounted by only the scalar field.

Horndeski theory \cite{Horndeski:1974wa, Nicolis:2008in, Deffayet:2009mn, Kobayashi:2011nu} is a large, popular class of models with a highly nontrivial scalar mode that have been extensively explored phenomenologically \cite{Kobayashi:2019hrl, Deffayet:2010qz, Koyama:2013paa,  Sotiriou:2013qea, Sotiriou:2014pfa, Barausse:2015wia, BeltranJimenez:2015sgd, Pogosian:2016pwr, Babichev:2016rlq, Alonso:2016suf, Copeland:2018yuh, Sakstein:2017xjx, Creminelli:2017sry, Ezquiaga:2017ekz, Peirone:2017ywi, Noller:2018wyv, Raveri:2019mxg, Frusciante:2019puu, Lee:2022cyh}.  It was first constructed by Horndeski in 1970s \cite{Horndeski:1974wa} and rediscovered in the wake of constructing generalized (covariant) galileon theory \cite{Nicolis:2008in, Deffayet:2009mn}. The theory is obtained by requiring the equations of motion to only have second derivatives, so it is a special class of EFTs that eliminates the Ostragradski ghosts to all scales. Beyond Horndeski models have also been widely investigated \cite{Zumalacarregui:2013pma, Gleyzes:2014dya, Gleyzes:2014qga, Langlois:2015cwa} where the equations of motion have apparent higher order derivatives but they can still avoid the Ostragradski ghosts. The reason why they can still evade the Ostradgradski theorem is that the beyond Horndeski models are in fact singular field theories where the kinetic matrix is degenerate. The degeneracy can reside only in the scalar part of the kinetic matrix, which is the case for part of Horndeksi theory or early beyond Horndeski models \cite{Zumalacarregui:2013pma, Gleyzes:2014dya, Gleyzes:2014qga}, or in general exists between the scalar and tensor mode, which is often referred to as degenerate higher order scalar tensor (DHOST) theory \cite{Langlois:2015cwa, Langlois:2015skt}. Horndeski theory and beyond contain free functions of the scalar and its kinetic term, so they are large classes of EFTs. In this paper, we investigate how the recent developments in positivity bounds can help reduce the parameter spaces of these theories.

Positivity bounds have been previously used to constrain cosmological EFTs \cite{Cheung:2016yqr, Bonifacio:2016wcb, deRham:2017imi, Bellazzini:2017fep, deRham:2018qqo,  Ye:2019oxx, Melville:2019wyy, deRham:2019ctd, Alberte:2019xfh, Herrero-Valea:2019hde, Wang:2020xlt, deRham:2021fpu, Traykova:2021hbr, Melville:2022ykg, Freytsis:2022aho, deRham:2022sdl, Creminelli:2022onn, Hong:2023zgm, Bellazzini:2023nqj, Tokuda:2020mlf, Serra:2022pzl,Baumann:2015nta,Kim:2019wjo,Kennedy:2020ehn,Grall:2021xxm,Davis:2021oce,Aoki:2021ffc}.
For example, the lowest order forward limit bound requires positivity of the coefficient of $(\pd\phi)^4$ \cite{Adams:2006sv}, the leading term in the $P(X)$ models that are used to model inflation and dark energy. Ref \cite{deRham:2017imi} uses the Y-type positivity bounds \cite{deRham:2017avq, deRham:2017zjm} to constrain the massive galileon models, which are motivated by quite a few IR modifications of GR such as the Dvali-Gabadadze-Porrati (DGP) braneworld model \cite{Dvali:2000hr} and dRGT massive gravity \cite{deRham:2010kj}. The Y-type bounds are a set of easy-to-use analytic positivity bounds that utilize the positivity of Legendre polynomials and analyticity of the amplitude in the positive $t$ direction to extend the positivity away from the forward limit. These positivity bounds have been used to constrain Horndeski theory and related models and confront with cosmological data \cite{Melville:2019wyy, deRham:2021fpu, Tokuda:2020mlf, Traykova:2021hbr, Melville:2022ykg,Baumann:2015nta,Kim:2019wjo,Kennedy:2020ehn,Grall:2021xxm,Davis:2021oce,Aoki:2021ffc}. Other fruitful phenomenological applications of positivity bounds include constraining the enormous parameter space of the Standard Model EFT (SMEFT) (see, {\it e.g.,} \cite{Zhang:2018shp, Zhang:2020jyn, Li:2021lpe, Bellazzini:2018paj, Bi:2019phv, Remmen:2019cyz,  Fuks:2020ujk, Yamashita:2020gtt, Gu:2020ldn, Remmen:2020vts, Bonnefoy:2020yee, Bellazzini:2014waa, Zhang:2021eeo, Li:2022tcz,  Li:2022rag}).
The SMEFT systematically parameterizes new physics beyond the Standard Model so as to facilitate experimental searches. It has been found that reduction of the parameter space by a couple of orders of magnitude is attainable even for small subsets of the theory \cite{Zhang:2018shp, Bi:2019phv, Remmen:2019cyz, Li:2021lpe, Yamashita:2020gtt}. Conversely, positivity bounds also allow us to reserve bootstrap the UV Regge behavior from the Standard Model data such as the electron's mass and charge \cite{Alberte:2021dnj, deRham:2022gfe}.

Recently, it has been realized that the strength of positivity bounds can be significantly enhanced by utilizing the full crossing symmetry of the amplitudes \cite{Tolley:2020gtv, Caron-Huot:2020cmc}. The Y-type bounds or other earlier positivity bounds only use the $su$ crossing symmetry inherent in the fixed-$t$ dispersion relations. A simple way to achieve the full triple crossing symmetry is to impose the $st$ crossing symmetry on the $su$-symmetric dispersion relations, which leads to an infinity set of null constraints. From the EFT point of view, these null constraints are already contained in the Wilson coefficient structure of the crossing symmetric EFT amplitude. From the UV point of view, they imply that the UV amplitudes need to be delicately balanced among different spins at different energy scales so as to give rise to the desired low energy EFT model. Making use of these null constraints, one can bound the Wilson coefficients both from above and below, constraining the scalar EFT parameter space to be within a finite region, parametrically $\mc{O}(1)$ in a suitable normalization. The optimal bounds can generally be obtained numerically by linear programing \cite{Caron-Huot:2020cmc} (more generally semi-definite programming \cite{Li:2021lpe, Du:2021byy} for multiple degrees of freedom), or semi-analytically by formulating it as a bi-variate moment problem \cite{Chiang:2021ziz, Arkani-Hamed:2020blm, Bellazzini:2020cot}. (Alternatively, at least for the case of an identical scalar, one may directly use triple crossing symmetric dispersion relations \cite{Sinha:2020win, Song:2023quv}.)
We would like to emphasize that the positivity bounds are robust and model-independent since they are solely based on the fundamental principles of quantum field theory. Therefore, these results remain valid irrespective of any potential intricate design of the UV completion.

In this paper, we use the latest fully crossing symmetric positivity bounds to constrain the parameter spaces of the EFTs of cosmological scalars. We shall utilize the simple two-channel, $su$-symmetric dispersion relation, and add on the null constraints to account for the $st$ crossing symmetry. While the two-sided positivity bounds have been explicitly computed in the massless limit, how the scalar mass affects the bounds has yet to be mapped. We first extend the formalism of obtaining the triple crossing bounds to the case of a massive scalar. This can be done by simply expanding the sum rules/dispersion relation in terms  of $m^2$, truncating at sufficient high orders of $m^2/\Li^2$, where $m$ is the mass of the scalar and $\Li$ is the EFT cutoff. As mentioned, to obtain optimal bounds numerically, one can formulate the problem as a linear program with a continuous decision variable and employ the efficient SDPB package. The continuous decision variable essentially parametrizes all possible scales of UV states above the EFT cutoff, and the SDPB cleverly handles it by transforming it to a standard linear program with a finite set of constraints. This method constrains the ratios of higher order coefficients against the lowest order coefficient, {\it i.e.,} the $s^2$ coefficient. In the decoupling limit of the graviton, the lower bound of the $s^2$ coefficient is zero, while its upper bound can be obtained by using the upper bound of partial wave unitarity with the help of some null constraints.

These fully crossing bounds are quite restrictive. We find that, for a generic scalar EFT, a nonzero mass generally enlarges the viable region of positivity bounds, allowing more parameter space for phenomenological applications. However, the mass effects only affect the positivity bounds within an order of magnitude, even when the threshold scale almost reaches the cutoff ($2m\sim \Lambda$). We will see that some of these features can be explained qualitatively by inspecting the dispersion relation. On the other hand, fine-tuned EFTs are also widely used in cosmology, in which some coefficients are set to zero or tuned to be unnaturally small. We will see that, from the positivity point of view, setting some coefficients to strictly zero is often inconsistent. If we relax the restriction and allow those coefficients to be small but nonzero, positivity bounds will then require some other coefficients to be extremely small, possibly rendering the phenomenologically desired features unrealizable. We also explore in detail different regions and boundaries of the positivity bounds for the lowest order coefficients, which are phenomenologically more relevant, and discuss their relations with possible UV models.

We then apply these fully crossing symmetric bounds to a few popular cosmological EFTs involving only one scalar. As mentioned, in these models, the scalar mode is responsible for many of the salient features of the beyond GR effects. We compute the scattering amplitudes in these models, obtain the bounds on the Wilson coefficients, and illustrate them with 2D and 3D plots, charting the effects of a nonzero mass. These plots vividly demonstrate how the mass can increase the availability for the relevant models.

A caveat is in order. The prerequisite of applying these flat space positivity bounds is that the corresponding model has a (or an approximate) Minkowski background around which the scalar plays the dominant role in the relevant dynamics, as these bounds are based on the flat space S-matrix principles. Particularly, these bounds would not necessarily impose any constraint for observables on a time-dependent, cosmological background, on which the S-matrix theory is less understood and additional assumptions may be needed to derive useful bounds. See Ref \cite{Grall:2021xxm} and the references within for a recent discussion on the positivity bounds on Lorentz breaking backgrounds.

For Horndeski theory, we first derive the bounds for the generic theory, and then discuss the implications of these bounds for a few popular models with a shift symmetry that is softly broken by the mass. In this restricted class of Horndeski theory, there are three constrainable coefficients $\bG_{2,XX}$, $\bG_{3,X}$ and $\bG_{4,XX}$ in the Lagrangian around a generic background $\phi_0$. The positivity bounds depict a closed region in the space spanned by the three parameters. In fact, $\bG_{3,X}$ always appears as $\bG^2_{3,X}$ in the positivity bounds, and in the space of $\bG_{2,XX}$, $\bG^2_{3,X}$ and $\bG_{4,XX}$ the positivity bounds carve out a convex polyhedral region, as the bounds (\ref{sshtcon}) are only linear inequalities in terms of $\bG_{2,XX}$, $\bG^2_{3,X}$ and $\bG_{4,XX}$. In the more direct parameter space with $\bG_{3,X}$, the positivity region then becomes curved. We show how some specific models such as k-essence, kinetic braiding and modified galileon are located in the 3D positivity region.

The galileon model is the one that motivated the original study of positivity bounds \cite{Adams:2006sv}. With the strict galileon symmetry $\phi\to \phi+b_\mu x^\mu+c$ \cite{Nicolis:2004qq}, the model is marginally ruled out the lowest order forward limit bound. A viable positivity region can be obtained by adding a mass for the galileon. Massive galileon is motivated by the fact that galileon's parent theories such as massive gravity models are endowed with massive degrees of freedom \cite{deRham:2016nuf}. Alternatively, one may supply the galileon with a $(\pd\phi)^4$ term, in order to be consistent with positivity bounds. Both a small mass and $(\pd\phi)^4$ term restrict the positivity region to be a narrow strip in the $\bG_{3,X}$ and $\bG_{4,XX}$ plane \cite{Tolley:2020gtv}.  A larger mass and $(\pd\phi)^4$ term widen the positivity region, and interestingly adding a mass and adding a $(\pd\phi)^4$ term widen the region in orthogonal directions.

For beyond Horndeski theories, we focus on the (quadratic) DHOST models \cite{Langlois:2015cwa}. We first derive the positivity bounds for the generic theory, and then discuss the implications of the bounds for a few specific classes of models. It has been pointed out that the current gravitational wave observations, particularly the constraints on the speed of gravitational waves, impose strong restrictions of the forms of the DHOST models \cite{Langlois:2017dyl, Crisostomi:2017lbg, Crisostomi:2019yfo}, although these constraints have been cautioned based on the validity of the EFT in interpreting the gravitational wave data \cite{deRham:2018red}. If the speed of gravity constraints are valid, the surviving DHOST models belong to the N-I class. We find that these surviving models need to have a mass $m/\Li\simeq 0.4$ to be consistent with positivity bounds. That is, positivity bounds essentially rule out these models, as a healthy hierarchy between $m$ and $\Li$ is prerequisite for a  useful EFT.

We also discuss the implications for other DHOST models. In particular, the beyond Horndeski models of \cite{Gleyzes:2014dya, Gleyzes:2014qga} belong to the M-I class. The fully crossing symmetric positivity bounds require that a mass $m/\Li \gtrsim 0.4$ for original beyond Horndeski models, which again rules out these models for the lack of a healthy hierarchy between $m$ and $\Li$. For the M-II class, the positivity bounds require the model to lie on a narrow strip in the $\bar{A}_{1,X}$ and $\bar{A}_4$ plane for a reasonably small mass.

\begin{table}[]
\centering
  \renewcommand\arraystretch{1.5}
\scalebox{0.80}{
\begin{tabular}{|c|ccc|cc|cc|}
\hline
\multirow{2}{*}{} & \multicolumn{3}{c|}{\multirow{2}{*}{\textbf{Class}}}                                             & \multicolumn{2}{c|}{\multirow{2}{*}{\textbf{Positivity bounds}}}    & \multicolumn{2}{c|}{\textbf{Consistency}}    \\ \cline{7-8}
                  & \multicolumn{3}{c|}{}                                                              & \multicolumn{2}{c|}{}                     & \multicolumn{1}{c|}{\textbf{massless}} &\textbf{massive}  \\ \hline
\multirow{4}{*}{\makecell[c]{~\\Horndeski\\(\ref{hornlag1}-\ref{hornlag5})}} & \multicolumn{3}{c|}{Generic}                                                              & \multicolumn{2}{c|}{(\ref{app10}\text{-}\ref{genb})}                     & \multicolumn{1}{c|}{\checkmark} &\checkmark  \\ \cline{2-8}
                  & \multicolumn{1}{c|}{\multirow{3}{*}{\makecell[c]{$G_i=G_i(X)$}}} & \multicolumn{2}{c|}{\makecell[c]{K-essence\\ ($G_3=G_4=G_5=0$)}}                     & \multicolumn{1}{c|}{\multirow{3}{*}{\makecell[c]{~\\(\ref{sshtcon})\\ Fig \ref{fig:sshtm0}\text{-}\ref{fig:sX}}}} &(\ref{keb})  & \multicolumn{1}{c|}{\checkmark} &\checkmark  \\ \cline{3-4} \cline{6-8}
                  & \multicolumn{1}{c|}{}                  & \multicolumn{2}{c|}{\makecell[c]{Kinetic braiding\\($G_4=G_5=0$)}}                     & \multicolumn{1}{c|}{}                  &(\ref{kbcons})  & \multicolumn{1}{c|}{\checkmark} &\checkmark  \\ \cline{3-4} \cline{6-8}
                  & \multicolumn{1}{c|}{}                  & \multicolumn{2}{c|}{\makecell[c]{Galileon (\ref{gallag})}}                     & \multicolumn{1}{c|}{}                  &(\ref{galcons})  & \multicolumn{1}{c|}{$\times$} &\notcheckmark  \\ \hline
\multirow{6}{*}{\makecell[c]{~\\~\\Beyond\\ Horndeski}} & \multicolumn{1}{c|}{\multirow{5}{*}{\makecell[c]{~\\~\\ DHOST\\(\ref{dhost})}}} & \multicolumn{1}{c|}{\multirow{2}{*}{\makecell[c]{N-I\\(\ref{dhost})+(\ref{NI})}}} &Generic  & \multicolumn{1}{c|}{\multirow{5}{*}{\makecell[c]{~\\ ~\\(\ref{gendh})}}} &  & \multicolumn{1}{c|}{\checkmark} &\checkmark  \\ \cline{4-4} \cline{6-8}
                  & \multicolumn{1}{c|}{}                  & \multicolumn{1}{c|}{}                  &\makecell[c]{GW constraints (\ref{NIGW})}   & \multicolumn{1}{c|}{}                  &\makecell[c]{(\ref{dGWcons})\\Fig \ref{fig:NI}}  & \multicolumn{1}{c|}{$\times$} &\xmark  \\ \cline{3-4} \cline{6-8}
                  & \multicolumn{1}{c|}{}                  & \multicolumn{1}{c|}{\multirow{2}{*}{\makecell[c]{M-I\\(\ref{dhost})+(\ref{MIde})}}} &Generic  & \multicolumn{1}{c|}{}                  &\makecell[c]{(\ref{MIcons})\\Fig \ref{fig:MI}}  & \multicolumn{1}{c|}{$\times$} &\notcheckmark  \\ \cline{4-4} \cline{6-8}
                  & \multicolumn{1}{c|}{}                  & \multicolumn{1}{c|}{}                  &\makecell[c]{beyond Horndeski \cite{Gleyzes:2014dya}\\+(\ref{bHde})}  & \multicolumn{1}{c|}{}                  &(\ref{MIbH})  & \multicolumn{1}{c|}{$\times$} &\xmark  \\ \cline{3-4} \cline{6-8}
                  & \multicolumn{1}{c|}{}                  & \multicolumn{2}{c|}{\makecell[c]{M-II (\ref{dhost})+(\ref{MII})}}                     & \multicolumn{1}{c|}{}                  &\makecell[c]{(\ref{MIIcons})\\Fig \ref{fig:MII}}  & \multicolumn{1}{c|}{$\times$} &\notcheckmark  \\ \cline{2-8}
                  & \multicolumn{3}{c|}{\makecell[c]{Combining Horndeski \& DHOST (\ref{cDGW})}}                                                              & \multicolumn{2}{c|}{(\ref{cDGWcons}), Fig \ref{fig:cDGW}}                     & \multicolumn{1}{c|}{\checkmark} &\checkmark  \\ \hline
\end{tabular}
}
\caption{Summary of triple crossing positivity bounds on various popular models. The ``Consistency'' denotes whether a model is consistent with these bounds, with ``massless'' and ``massive'' referring to the mass of the scalar.  "$\times$" means that the model is marginally ruled out because it lives on the boundary of the bounds, "\xmark" means that the model is ruled out in the sense that the positivity bounds have excluded the possibility of a healthy hierarchy between the scalar mass $m$ and the EFT cut-off $\Lambda$, and ``\notcheckmark" means that the positivity bounds require some fine-tuning of the model parameters to achieve a healthy hierarchy between $m$ and $\Li$. We adopt the \cite{Crisostomi:2016czh} classification for DHOST theory \cite{Langlois:2015cwa}. }\label{sumtab}
\end{table}

The main results regarding the triple crossing symmetric positivity bounds for a generic massive scalar can be found in the figures in Section \ref{sec:sec3}.
For a check reference to the main results regarding applying these bounds to the popular cosmological models we have surveyed in this paper, readers are referred to Table \ref{sumtab}.

The rest of the paper is organized as follows. In Section \ref{sec:sec2}, we derive the $su$-symmetric dispersion relation for a massive theory, and extract the sum rules for the Wilson coefficients and the null constraints. The optimization framework as a linear program is then presented, with some details of the numerical implementation and convergence tests deferred to Appendix \ref{app:appA}.  In Section \ref{sec:sec3}, we first compute the upper bound on the leading forward-limit coefficient for various masses, and then compute the agnostic 1D and 2D bounds on the leading few coefficients; the bounds on fine-tuned EFTs are also discussed. In Section \ref{sec:sec4} and \ref{sec:sec5}, we apply the fully crossing symmetric bounds to Horndeski theory and beyond, explicitly listing ready-to-use positivity bounds for respective models; We illustrate where several popular models are located in the whole positivity region and discuss the implications of the bounds; We point out that some of these models are essentially ruled out by triple crossing positivity at a theoretical level for lack of a healthy hierarchy between the mass and the EFT cutoff.

\section{Dispersion relation and triple crossing bounds}
\label{sec:sec2}

Analyticity along with locality allows us to derive dispersion relations to bridge IR physics to unknown UV completions, passing down the UV unitarity constraints to the low energy EFT coefficients. Systematic procedures have been recently developed to extract optimal unitarity constraints (particularly the positivity of unitarity constraints) using the dispersion relations, taking into account of full crossing symmetry. The use of full crossing symmetry is crucial, as it permits derivation of two-sided bounds, as opposed to previous bounds that often restrict the Wilson coefficients from one side. In this section, we shall review this framework to derive these triple crossing bounds, with added consideration that the mass of the particles might be sizable. As these bounds are to be used to constrain cosmological EFTs, we shall assume that the EFT is weakly coupled in the IR and make use of tree-level amplitudes.

\subsection{Dispersion relation}

Let us first introduce the dispersion relation that will be used to derive positivity bounds. We shall use the fixed-$t$ and $su$-symmetric dispersion relation. The leftover $st$ crossing symmetry can be later imposed on the $su$-symmetric dispersion relation, which leads to a series of null constraints \cite{Tolley:2020gtv}. We will focus on the $\phi\phi\to \phi\phi$ scattering amplitude $A(s,t)$, for which analyticity is well established, and define the usual Mandelstam variables as follows
\begin{equation}
\label{man1}
  s=-(p_1+p_2)^2,\qquad t=-(p_1+p_3)^2,\qquad u=-(p_1+p_4)^2 ,
\end{equation}
where $p_i$ denotes the four-momentum of the $i$-th particle. The three variables are not independent, subject to the constraint of energy-momentum conservation $s+t+u=4m^2$, where $m$ is the scalar mass. Later, we will more often make use of the tilded Mandelstam variables
\be
\tilde{z}=z-\f{4}3 m^2, ~~(z=s,\,t,\,u),~~ {\rm and}~~~v \equiv \tis + \f{\tit}2 =s+\f{t}2-2m^2,
\ee
so that we have $\tis+\tit+\tiu=0$. The $v$ variable is convenient to use as a proxy variable, simplifying some expressions, thanks to the property that $v \rightarrow -v$ under an $su$ crossing. The scattering angle $\thi$ between particle 1 and 3 can be expressed using Mandelstam variables as $\cos \theta = 1+2t/(s-4m^2)$.

Utilizing Cauchy's integral formula on the complex $s$ plane (with $t$ fixed) along with the Froissart-Martin bound and the $su$ crossing symmetry, the twice subtracted dispersion relation can be written as (see, {\it e.g.,} \cite{deRham:2017avq}):
\begin{align}\label{2}
    B(s,t)&\equiv A(s,t)-\left(\frac{\lambda}{m^2-s}
    +\frac{\lambda}{m^2-t}+\frac{\lambda}{m^2-u}\right) \notag \\
    &=a(t) +\int_{\Lambda^2}^{\infty}\frac{\mathrm{d}\mu}{\pi (\mu-\mu_p)^2}\left[ \frac{(s-\mu_p)^2 \mathrm{Im} A(\mu,t)}{\mu-s}+\frac{(u-\mu_p)^2 \mathrm{Im} A(\mu,t)}{\mu-u} \right],
\end{align}
where $\Lambda$ is the cut-off of the EFT and we set the subtraction point at $\mu_p = 2m^2 - t/2$. By shifting the integral variable $\mu\rightarrow\mu+4m^2$ and using the $v$ variable, the dispersion relation can be cast as
\begin{equation}\label{disp4}
  {B}(s,t)=a(t)+v^2\int_{\tilde\Lambda^2}^{\infty} \frac{ \mathrm{d}\mu}{\pi (\mu+2m^2+\frac{t}{2})^2}
   \frac{2 \mathrm{Im}\, A(\mu+4m^2,t)}{(\mu+2m^2+\frac{t}{2})^2-v^2},
\end{equation}
where we have defined
\be
\tilde{\Lambda}^2\equiv \Lambda^2-4m^2 .
\ee
This shift will simplify the formulas below, and $\tilde{\Lambda}^2$ is positive as we require the cutoff to be above the threshold. Note that there are only even powers of $v$, reflecting the $su$ crossing symmetry. The usefulness of the dispersion relation above is that, choosing $|s|,|t|\ll \Li^2$, we can use the low energy EFT amplitude to approximate $B(s,t)$ on the left hand side, while the equality links it to the right hand side where the amplitude $A(\mu,t)$ is necessarily described the UV amplitude, because its center-of-mass energy is above the EFT cutoff, {\it i.e.,} $\mu>\Lambda^2$. That is, one can extract the UV information from the integral on the right hand side and pass it down to the IR EFT. In the following, we will show that, merely assuming that $A(\mu,t)$ respects unitarity at high energies, coupled with additional $st$ crossing symmetry, the dispersion relation can impose strong constraints on the low energy Wilson coefficients.

To this end, we expand the amplitude in terms of its partial waves
\begin{equation}\label{amp5}
 A(\mu,t)=16 \pi\sqrt{\frac{\mu}{\mu-4m^2}} \sum_{l=0}^{\infty}(2l+1)P_l\left(1+\frac{2t}{\mu-4m^2}\right) a_l(\mu),
\end{equation}
where $P_l(x)$ are the Legendre polynomials and $a_l(\mu)$ is the spin-$l$ partial wave amplitude. For an identical scalar, the summation over the partial waves is only over even $l$, due to the $tu$ crossing symmetry. This is because, thanks to the property of the Legendre polynomials $P_l(-x)=(-1)^lP_l(x)$, we have
\begin{equation}
\label{eq2-3}
 A(\mu,u)=16\pi\sqrt{\frac{\mu}{\mu-4m^2}} \sum_{l=0}^{\infty}(2l+1)(-1)^lP_l\left(1+\frac{2t}{\mu-4m^2}\right) a_l(\mu)  .
\end{equation}
Comparing the above equation to \eref{amp5}, $A(\mu,t)=A(\mu,u)$ requires $a_l(\mu)=(-1)^l a_l(\mu)$, which means that $a_{l}(\mu)=0$ when $l={\rm odd}$. More importantly, conservation of the angular momentum means that each partial wave should preserve unitarity, which gives the generalized optical theorem for each partial wave:
\begin{equation}\label{amp6}
  2\mathrm{Im}\,a_{l}= \sum_X |a_l^{X}|^2 ,
\end{equation}
where $a_l^{X}$ is the partial wave amplitude from $\phi\phi$ to UV state $X$. Thus the generalized optical theorem/partial wave unitarity indicates that the imaginary part of each partial wave is positive definite $\mathrm{Im}\,a_{l}>0$. Furthermore, only keeping $X=\phi\phi$ and dropping the other terms in the summation, we see that \eref{amp6} requires $2\mathrm{Im}\,a_l\geq  |a_l|^2$, which in turn implies $\mathrm{Im}\,a_l<2$. We will make use of both of these unitarity conditions in this paper.

For later convenience, we introduce an abbreviation for the integration over all UV scales with $\mu$ and the sum over all spins with $l$:
\begin{equation}
  \bigg\langle\cdots\bigg\rangle  \equiv  \sum_{ l~\text{even}}\int_{\tilde{\Lambda}^2}^{\infty}  \mathrm{d} \mu \, 16(2l+1)\sqrt{(\mu+4m^2) / \mu}\, \mathrm{Im}\,a_l \bigg(\cdots\bigg) ,
\end{equation}
where $16(2l+1)\sqrt{(\mu+4m^2) / \mu}\, \mathrm{Im}\,a_l~ \d \mu$ is a positive measure. So, after the partial wave expansion, the dispersion relation can be written as
\begin{equation}
\label{BFinalDisp}
 {B}(s,t) - a(t) =   \bigg\langle \frac{ 2 v^2 P_l\left(1+{2t}/{\mu}\right) }{(\mu+2m^2+\frac{t}{2})^2((\mu+2m^2+\frac{t}{2})^2-v^2)} \bigg\rangle.
\end{equation}
As mentioned, the left hand side of the above equation can be approximated by the EFT amplitude, which can be expanded as
\be
\label{Bexpvt}
{B}(s,t)-{a}(t)  =  \sum_{i= 1}^{\infty}\sum_{j=0}^{\infty} c^{2i,j}v^{2i}\tilde t^j .
\ee
(As already mentioned, odd powers of $v$ vanish because of the $su$ crossing symmetry.) Also expanding the right hand side of \eref{BFinalDisp} and matching term by term with the left hand side, we get sum rules for the EFT coefficients
\be
\label{cbra}
{c}^{2i,j} = \left\langle \sum_{q=j}^{\infty} \binom{q}{j}\left(\frac{4m^2}{3}\right)^{q-j}  \sum_{p=0}^{q} \frac{2L_l^pH_{2i+1}^{q-p}}{(\mu+2m^2)^{2i+q+1-p}\mu^p} \right\rangle\equiv\left\langle C^{2i,j}_l(\mu)  \right\rangle.
\ee
where $\binom{q}{j}\equiv\Gamma(q+1)/(\Gamma(j+1)\Gamma(q-j+1))$ denotes the binomial coefficient, and
\begin{equation}\label{11}
  L_l^p\equiv\frac{2^p}{p!}\frac{\mathrm{d}^p}{\mathrm{d}x^p}P_l(x)|_{x=1}
  =\frac{\Gamma(l+p+1)}{p!\Gamma(l-p+1)\Gamma(p+1)}\geq 0,
\end{equation}
and
\begin{equation}\label{12}
  H_{2i+1}^q \equiv  \frac{\Gamma(2i+1+q)}{(-2)^q\Gamma(q+1)\Gamma(2i+1)}.
\end{equation}
The sum rules Eq.(\ref{cbra}) are one of the main ingredients that will be used to numerically obtain optimal positivity bounds later.
In the practical numerical computations, we need to truncate $C^{2i,j}_l(\mu)$ to a finite order in $m^2$, which will be referred to as $O_{\rm m}$. More precisely, it is essentially an expansion in terms of $m^2/\tilde{\Li}^2$, because the integration over $\mu$ starts from $\tilde{\Li}^2$. We will see later that numerical convergence can be achieved even for a relatively large ratio of $m^2/\tilde{\Li}^2$ by keeping the leading few orders of $m^2$.

To fully utilize the crossing symmetry of the amplitude, we can impose the $st$ crossing symmetry on the $su$-symmetric EFT coefficients $c^{2i,j}$, which gives a set of homogenous linear equalities on $c^{2i,j}$. That is, the final $stu$ symmetric EFT parameter space is a linear subspace of the space spanned by $c^{2i,j}$. Using the sum rules Eq.(\ref{cbra}), these equalities on $c^{2i,j}$ can be expressed as a set of null sum rules (both the EFT coefficient equalities and the null sum rules will be referred to as null constraints later), which will be used in our linear programs to get the optimal positivity bounds.

A convenient way to get the coefficient equalities that link different $c^{2i,j}$ is to expand $v=\tilde s+\tilde t/2$ and express $B({s},{t})$ in terms of powers of $\tilde{s}^p\tilde{t}^q$
\begin{equation}\label{Bst17}
  B({s},{t}) = \sum_{i,j}c^{2i,j}(\tis+\tit/2)^{2i}\tit^j \equiv \sum_{p,q}a^{p,q}\tilde{s}^p\tilde{t}^q ,
\end{equation}
and then the $st$ crossing symmetry simply requires that $a^{p,q}=a^{q,p}$. Therefore, the extra null constraints one need to impose on the $su$-symmetric sum rules Eq.(\ref{cbra}) are
\begin{equation}\label{npq19}
  n^{p,q}=a^{p,q}-a^{q,p}=\sum_{a=p}^{p+q}\frac{\Gamma(a+1)c^{a,p+q-a}}{2^{a-p}\Gamma(p+1)\Gamma(a-p+1)}
  -\sum_{b=q}^{p+q}\frac{\Gamma(b+1)c^{b,p+q-b}}{2^{b-q}\Gamma( q+1)\Gamma(b-q+1)}
\end{equation}
These null constraints are the other main ingredient used to numerically obtain the optimal positivity bounds later.
Explicitly, the first three nontrivial null constraints are given by
\begin{align}\label{19m}
  n^{1,3}&=c^{2,2}-\frac{3}{2}c^{4,0}, \notag \\
  n^{1,4}=n^{2,3}&=c^{2,3}-\frac{1}{2}c^{4,1}, \notag \\
  n^{1,5}=n^{2,4}&=c^{2,4}+\frac{1}{2}c^{4,2}-\frac{45}{16}c^{6,0}.
\end{align}
Plugging Eq.(\ref{cbra}) into \eref{npq19}, we can write the null conditions as
\begin{equation}
\label{null20}
  n^{p,q}\equiv \langle N^{p,q}_l(\mu) \rangle =0.
\end{equation}
In the numerical optimization, we only need to impose the lowest few orders of null constraints to get numerical convergence, as we shall see in the next section. The significant constraining power of the null constraints, which go like $\langle \cdots \rangle = \sum_l \int\d \mu  \mathrm{Im}\,a_l  ( \# l^i +  \# l^{i-1}+...)/\mu^j=0$, can be seen from their implications that $\mathrm{Im}\,a_l$ must be highly suppressed for high spin $l$ and that the lower and higher spin part in the summation must exactly cancel.

Since the amplitude is $stu$ symmetric, the independent coefficients in the pole-subtracted amplitude $B(s,t)$ can be counted by expanding $B(s,t)$ in terms of elementary symmetric polynomials $x=-(\tis\tit+\tit\tiu+\tis\tiu)$ and $y=-\tis \tit \tiu$ (note that $\tis + \tit+\tiu=0$):
\be
B(s,t) = \sum_{p, q} g_{p,q}x^p y^q .
\ee
Explicitly, the relations between the $g_{p,q}$ coefficients and a conveniently chosen subset of the $c^{2i,j}$ coefficients are as follows
\bal
\label{gcoefs}
g_{1,0}&= c^{2,0},&\quad        &g_{0,1}=c^{2,1},&\quad                  &g_{2,0}=c^{4,0},&  \qquad          &g_{1,1}=c^{4,1}, \notag \\
g_{3,0}&=c^{6,0},&          &g_{0,2}=c^{4,2}-\frac{9}{4}c^{6,0},&   &g_{2,1}=c^{6,1},&             &g_{4,0}=c^{8,0}, \notag \\
g_{1,2}&=c^{6,2}-3c^{8,0},&    &g_{3,1}=c^{8,1},&                    &g_{0,3}=c^{6,3} -2c^{8,1}, &        &  \dots
\eal
As the order increases,
In the next section, we shall express the positivity bounds in terms of independent $c^{2i,j}$, the coefficients directly from our sum rules, and the conversion to the bounds on $g_{p,q}$ is straightforward with the above relations.

Since we aim to obtain sharp bounds on the Wilson coefficients which are ``naturally'' $\lesssim{\cal O}(1)$, it is essential we also take necessary factors of $4\pi$ into account. This kind of dimensional analysis/power counting is often referred to as naive dimensional analysis \cite{Manohar:1983md}. Specifically, for a general scalar EFT, naive dimensional analysis suggests that should a generic Lagrangian operator take the following form
\begin{equation}\label{app1}
  \mathcal{L}\supset\sum_{I} C_I \widehat{O}_{I},~~~~  \widehat{O}_I= \f{\Lambda^4}{(4\pi)^2}\left(\frac{\partial}{
  \Lambda}\right)^{N_p}\left(\frac{4\pi \phi}{\Lambda}\right)^{N_\phi}.
\end{equation}
then the Wilson coefficient $C_I$ is typically $\lesssim{\cal O}(1)$. In the $\phi\phi\rightarrow\phi\phi$ scattering, the contributions to the amplitude come from double insertions of $4\pi\partial^{N_p}\phi^{3}/\Lambda^{N_p-1}$ (connected by a propagator) and contact interactions $(4\pi)^2\partial^{N_p}\phi^{4}/\Li^{N_p}$, both of which result in a factor of $(4\pi)^2$ in the amplitude. Thefore, generically, the coefficients $\hat c^{2i,j}$ in \eref{Bexpvt} are typically order $\mc{O}((4\pi)^2)$ rather than $\mc{O}(1)$:
\be
|c^{2i,j}|\lesssim \f{\mc{O}((4\pi)^2)}{\Li^{4i+2j}}
\ee

\subsection{Optimal bounds by linear programming}
\label{sec:optbounds}

We shall first compute the bounds on the $c^{2i,j}$ coefficients, as projected onto the lowest order coefficient $c^{2,0}$,
\be
\bar{c}^{2i,j}\equiv {\Lambda}^{2(2i+j-2)} \f{c^{2i,j}}{c^{2,0}},
\ee
where
\be
c^{2,0} >0,
\ee
which is the forward limit bound of  \cite{Adams:2006sv} and can be easily seen from the $c^{2,0}$ sum rule of \eref{cbra}, as all the terms in $\langle~\rangle$ is positive and the average $\langle~\rangle$ has a positive measure. Note that the $(4\pi)^2$ factors cancel, so the dimensionless coefficients $\bar{c}^{2i,j}$ are expected to $\lesssim\mc{O}(1)$. To convert to the bounds on the $stu$ symmetric coefficients $g_{p,q}$, one may simply use the relations in Eq.(\ref{gcoefs}). Let us now outline how to formulate a linear program to compute the optimal positivity bounds on the projected coefficients $\bar{c}^{2i,j}$, using the two ingredients obtained in the last subsection \eref{cbra} and \eref{npq19}. More detailed numerical setup with the SDPB package \cite{Simmons-Duffin:2015qma} is explained in Appendix \ref{app:appA}. We shall outline the strategy to obtain the upper bound on $c^{2,0}$ in Section \ref{sec:s2upper}.

Let us first see how we can extract the constraint on $\bar{c}^{2,1} = {\Lambda}^{2} {c^{2,1}}/{c^{2,0}}$ from the $c^{2,0}$ and $c^{2,1}$ sum rules
\bal
\label{c20explicit}
c^{2,0}&=\Big\langle C^{2,0}_l(\mu) \Big\rangle \\
&=\Big\langle\frac{1}{9(\mu+2m^2)^5}\left( 9\mu^2+6m^2(3+2l+2l^2)\mu+4m^4(6+4l+5l^2+2l^3+l^4)+\dots \right)\Big\rangle ,
\notag \\
\label{c21explicit}
 c^{2,1}&=\Big \langle C^{2,1}_l(\mu)\Big\rangle \notag \\
 &=\Big\langle\frac{1}{54(\mu+2m^2)^6}\big( 27(-3+2l+2l^2)\mu^2+36m^2(-3+l+2l^2+2l^3+l^4)\mu \notag \\
   &\hspace{2em} +4m^4(-63+24l+35l^2+24l^3+17l^4+6l^5+2l^6)+\dots \big)\Big\rangle ,
\eal
and the null constraints
\bal
n^{1,3} &=\Big\langle N^{1,3}_l(\mu) \Big\rangle \notag \\
    &=\Big\langle\frac{l(l+1)}{108\mu(\mu+2m^2)^6}\big( 27(-8+l+l^2)\mu^2+12m^2 (-6-8l-7l^2+2l^3+l^4) \mu \notag \\
    &\hspace{2em} +2m^4(-288+30l+22l^2-15l^3-5l^4+3l^5+l^6)+\dots \big)\Big\rangle ,
\\
\label{n14explicit}
n^{1,4}&=\Big\langle N^{1,4}_l(\mu) \Big\rangle  \notag \\
&=\Big\langle \frac{l(l+1)}{3240\mu(\mu+2m^2)^7}\big( 45(150-43l-41l^2+4l^3+2l^4)\mu^2 \notag \\
&\hspace{2em} +30m^2(54+45l+25l^2-39l^3-17l^4+3l^5+l^6)\mu \notag \\
&~~ +2m^4(11340-2538l-2422l^2+197l^3+13l^4 -97l^5-23l^6+8l^7+2l^8) +...\big)  \Big\rangle
\\
 &~\; \vdots \notag
\eal
where we have only explicitly shown the expansion of $m^2$ up to $\mc{O}(m^4)$. In the numerical results when $m^2$ is close to $\tilde\Lambda^2$, we need to include higher orders of $m^2$. Note that, we do not need to expand the $(\mu+2m^2)^n$ factor in the denominators in terms of $m^2$, for the reason that will be explained in Appendix \ref{app:appA}.

With these explicit forms, we observe that if there exists constant $\alpha_{2,0}$ and $\lambda_{p,q}$ such that the following inequalities
\begin{equation}\label{c21}
    -\alpha_{2,0}\,C^{2,0}_l(\mu)+{\Lambda}^2 C^{2,1}_l(\mu) +\sum_{p,q} \lambda_{p,q} {\Lambda}^{2(p+q-2)} N_l^{p,q}(\mu) \geq 0 ,
\end{equation}
are satisfied for all $\mu\geq\tilde{\Lambda}^2$ and all even $l$, upon acting $\langle ~ \rangle$ on both sides of \eref{c21}, we will obtain the following bound
\be
-\alpha_{2,0}\,c^{2,0}+{\Lambda}^2 c^{2,1}\geq 0 \marrow    \bar{c}^{2,1} \geq \alpha_{2,0} .
\ee
Running through all possible values of $\alpha_{2,0}$ and $\lambda_{p,q}$, we get the best lower bound for $\bar{c}^{2,1}$, given by the maximal value of $\alpha_{2,0}$ that solves the linear inequalities Eq.(\ref{c21}). That is, the optimal bound can be obtained by a linear program. Obviously, the optimal lower bound will vary depending on how many null constraints are included, more null constraints leading to more possible values of $\alpha_{2,0}$, hence a larger lower bound on $\bar{c}^{2,1}$. In practice, as we shall see, the lower bound numerically converges quickly with only a few lowest orders of null constraints. Also, we only impose constraints Eq.(\ref{c21}) for a finite number of $l$, but we can impose the constraints for all $\mu\geq\tilde{\Lambda}^2$ exactly (see Appendix \ref{app:appA}).

Similarly, the upper bound of $\bar{c}^{2,1}$ can be obtained by looking for constant $\beta_{2,0}$ and $\lambda_{p,q}$ that solve the following inequalities
\begin{equation}\label{c21up}
\beta_{2,0} \,C^{2,0}_l(\mu)- {\Lambda}^2 C^{2,1}_l(\mu) + \sum_{p,q} \lambda_{p,q} {\Lambda}^{2(p+q-2)} N_l^{p,q}(\mu) \geq 0 ,
\end{equation}
for all $\mu\geq\tilde{\Lambda}^2$ and even $l$. The optimal upper bound on $\bar{c}^{2,1}$ is given by the smallest possible $\beta_{2,0}$. Therefore, with the crossing symmetry fully used, as will be shown shortly, we can get the two-sided bounds on $\bar{c}^{2,1}$:
\begin{equation}\label{26}
  \alpha_{2,0}^{\rm max}  \leq\bar{c}^{2,1}\leq\beta_{2,0}^{\rm min}   .
\end{equation}
At a technical level, the reason why full crossing symmetry can impose the bounds from both sides is that the null constraints $\sum_{p,q} \lambda_{p,q} {\Lambda}^{2(p+q-2)} N_l^{p,q}$ can contribute significantly in inequalities Eq.(\ref{c21}) and Eq.(\ref{c21up}), and yet they vanish upon the $\langle ~ \rangle$ evaluation.

It is straightforward to generalize the above procedure to constrain multiple coefficients. For instance, let us consider bounds in the $\bar{c}^{2,2}$-$\bar{c}^{2,1}$ coefficient plane. A simple way to do this is to constrain $\bar{c}^{2,2}$ for a fixed $\bar{c}^{2,1} = \bar{c}^{2,1}_{0}$. That is, for sufficiently many discrete $\bar{c}^{2,1} = \bar{c}^{2,1}_{0}$, we go through the same procedure as the above to find the lower bound $\bar{c}^{2,2} \geq \alpha_{2,1}\bar{c}^{2,1}_{0}+\alpha_{2,0}$ and the upper bound $\bar{c}^{2,2}\leq \beta_{2,1}\bar{c}^{2,1}_{0}+\beta_{2,0}$. This gives us a 2D bound in the $\bar{c}^{2,2}$-$\bar{c}^{2,1}$ plane.

More detailed discussions about how to implement the linear programs above are presented in Appendix \ref{app:appA}. In particular, whilst the constraints enumerated by the continuous variable $\mu$ can be imposed exactly, we will truncate to finite orders for the UV spin, mass expansion and null constraints. We will demonstrate how the numerical results converge with the truncation orders in Appendix \ref{app:appA}.

\section{Mass dependence of positivity bounds}
\label{sec:sec3}

In this section, we will calculate the numerical positivity bounds on the leading few Wilson coefficients, obtained by using the method described in the last section, and as well as compute the upper bound on $c^{2,0}$ with a separate method. We will carefully chart the dependence of the bounds with respect to the mass of the scalar, demonstrating that the bounds are generally relaxed in the presence of a mass term. The bounds on $c^{2,0}$ and ${c}^{2,1}$  will be used to constrain some popular cosmological EFTs in the next sections.

\subsection{Upper bound on the $s^2$ coefficient}
\label{sec:s2upper}

The method described in the last section allows us to derive two-sided bounds on the coefficient ratios $\bar{c}^{2i,j}= {\Lambda}^{2(2i+j-2)} {c}^{2i,j}/c^{2,0}$. As for $c^{2,0}$, we only deduced that it is positive.
Before computing the bounds on the coefficient ratios in the next subsections, we first would like to derive the upper bound on $c^{2,0}$, with which we can convert the projected bound on $\bar{c}^{2i,j}$ into the bound on ${c}^{2i,j}$ itself. An upper bound on $c^{2,0}$ can be obtained by directly evaluating the $c^{2,0}$ sum rule using the upper bound of partial wave unitarity, with the help of the first null constraint \cite{Caron-Huot:2020cmc}.

\begin{figure}[ht]
    \centering
    \includegraphics[width=0.5\linewidth]{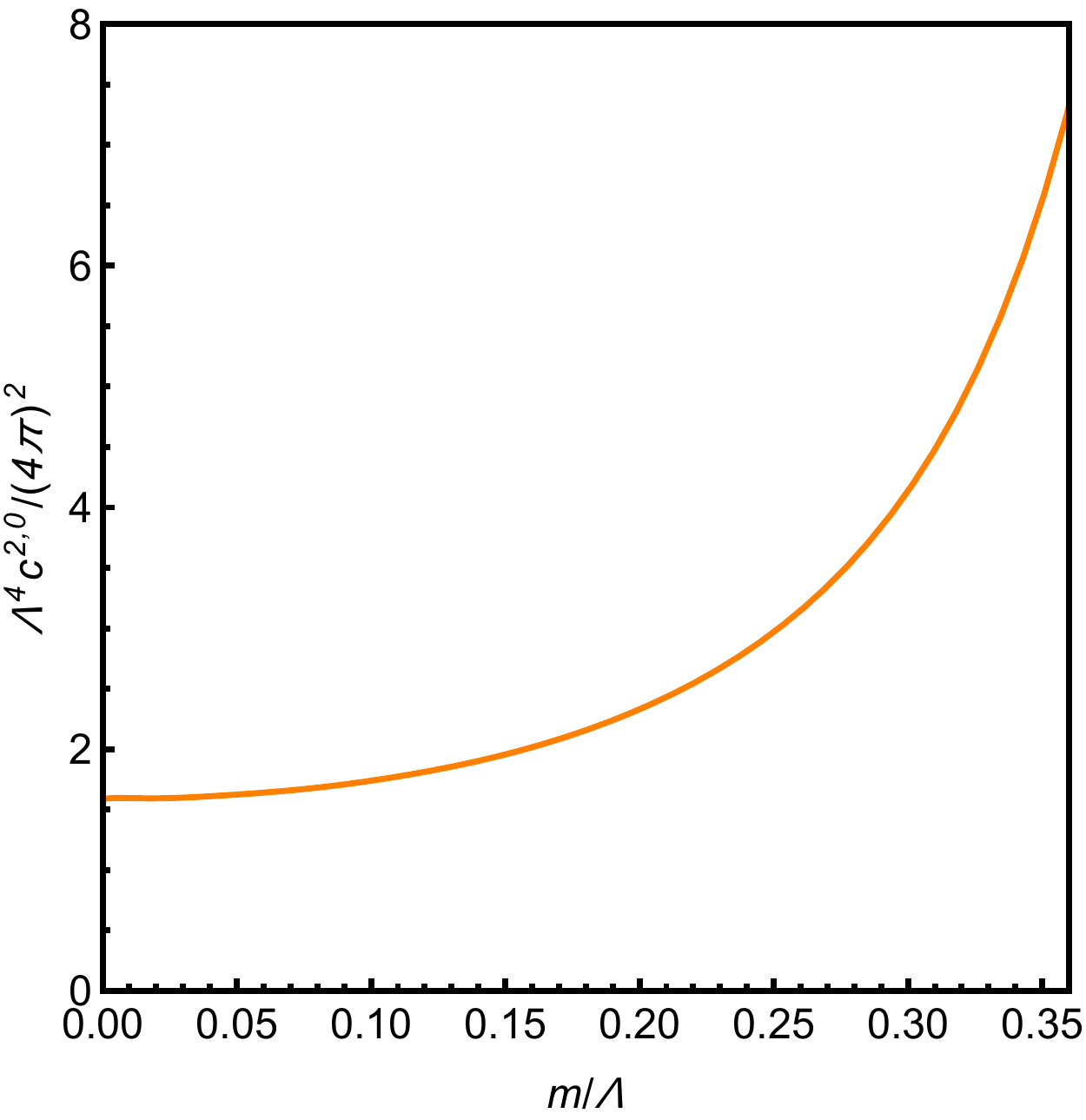}
    \caption{\label{fig:1} Upper bound on $c^{2,0}$ for various mass $m$. It is computed with the $m^2$ truncation order $O_\mathrm{m}=8$ and the maximal spin number $l_{\rm max}=60$. }
\end{figure}

First, note that, combining with the first null constraint, we can write the $c^{2,0}$ coefficient as the dispersive integral
\begin{equation}\label{c20int}
  c^{2,0}=\sum_{l} \int_{\tilde{\Lambda}^2}^{\infty} \mathrm{d}\mu 16(2l+1)\sqrt{\frac{\mu+4m^2}{\mu}}\text{Im}\,a_l
  \[ C_l^{2,0}(\mu) -\lambda N_{l}^{1,3}(\mu) \],
\end{equation}
where the constant $\li$ is taken to be positive and we recall that $\text{Im}\,a_l$ is positive definite. A crucial fact about the integrand in \eref{c20int} is that for a sufficiently large $l$ the part in the square bracket becomes negative for a range of $\mu$, $\tilde{\Lambda}^2\leq\mu< \mu_{l,\li}$, where $\mu_{l,\li}$ is determined by $C_l^{2,0}(\mu_{l,\li}) -\lambda N_{l}^{1,3}(\mu_{l,\li})=0$ and depends on $l$ and $\li$. Otherwise, the integrand of \eref{c20int} is positive. Also, recall that partial wave unitarity implies that $\mathrm{Im}\,a_l$ has an upper bound $\mathrm{Im}\,a_l<2$.  Utilizing this and dropping the negative parts in the dispersive integral, we get an inequality
\begin{equation}\label{c20up}
  c^{2,0}\leq\sum_{l=0}\int_{\mu_{l,\li}}^{\infty} \mathrm{d}\mu\, 16(2l+1)\sqrt{\frac{(\mu+4m^2)}{\mu}}
  \Big[ C_l^{2,0}(\mu) -\lambda N_{l}^{1,3}(\mu) \Big].
\end{equation}
Optimizing over $\li$, we can get an upper bound on $c^{2,0}$. In numerical evaluations, as an approximation, we need to truncate $l$ and the $m^2$ expansion to finite orders, denoted as $l_{\rm max}$ and $O_{\rm m}$ respectively. As shown in Appendix \ref{app:appA}, numerical convergence can be achieved with reasonable $l_{\rm max}$ and $O_{\rm m}$.

Numerical results confirm that the right hand side of the inequality Eq.(\ref{c20up}) is indeed finite. (It would not be finite if we directly use $\mathrm{Im}\,a_l<2$ without dropping the negative parts.) Figure \ref{fig:1} shows that the causality upper bound increases as the particle becomes more massive.  We see that the rate of the increase is actually accelerating as the mass increases, but even so, when the threshold scale $2m$ approaches the cutoff $\Lambda$, $\Lambda^4 c^{2,0}/(4\pi)^2$ is still about $\mc{O}(1)$, as suggested by native dimensional analysis.

\subsection{One dimensional bounds}

\subsubsection{Generic EFTs}

Now, we shall use the linear program method of the last section to derive the positivity bounds on the individual Wilson coefficient ratios $\bar{c}^{2i,j}= {\Lambda}^{2(2i+j-2)} {c}^{2i,j}/c^{2,0}$. We shall first focus on the bounds on one coefficient while being agnostic about the values of all the other coefficients. These agnostic bounds are conservative ones, valid regardless of the values of the other coefficients. Later, we will also show that if some of the rest coefficients are known, the bounds may be improved. All these bounds are two-sided, thanks to the implementation of full crossing symmetry. Using the upper bound on $c^{2,0}$, we can convert the bounds on $\bar{c}^{2i,j}$ to the bounds on ${c}^{2i,j}$.

\begin{figure}[ht]
    \centering\!\!\!
    \begin{subfigure}{0.47\linewidth}
    \centering
    \includegraphics[width=0.99\linewidth]{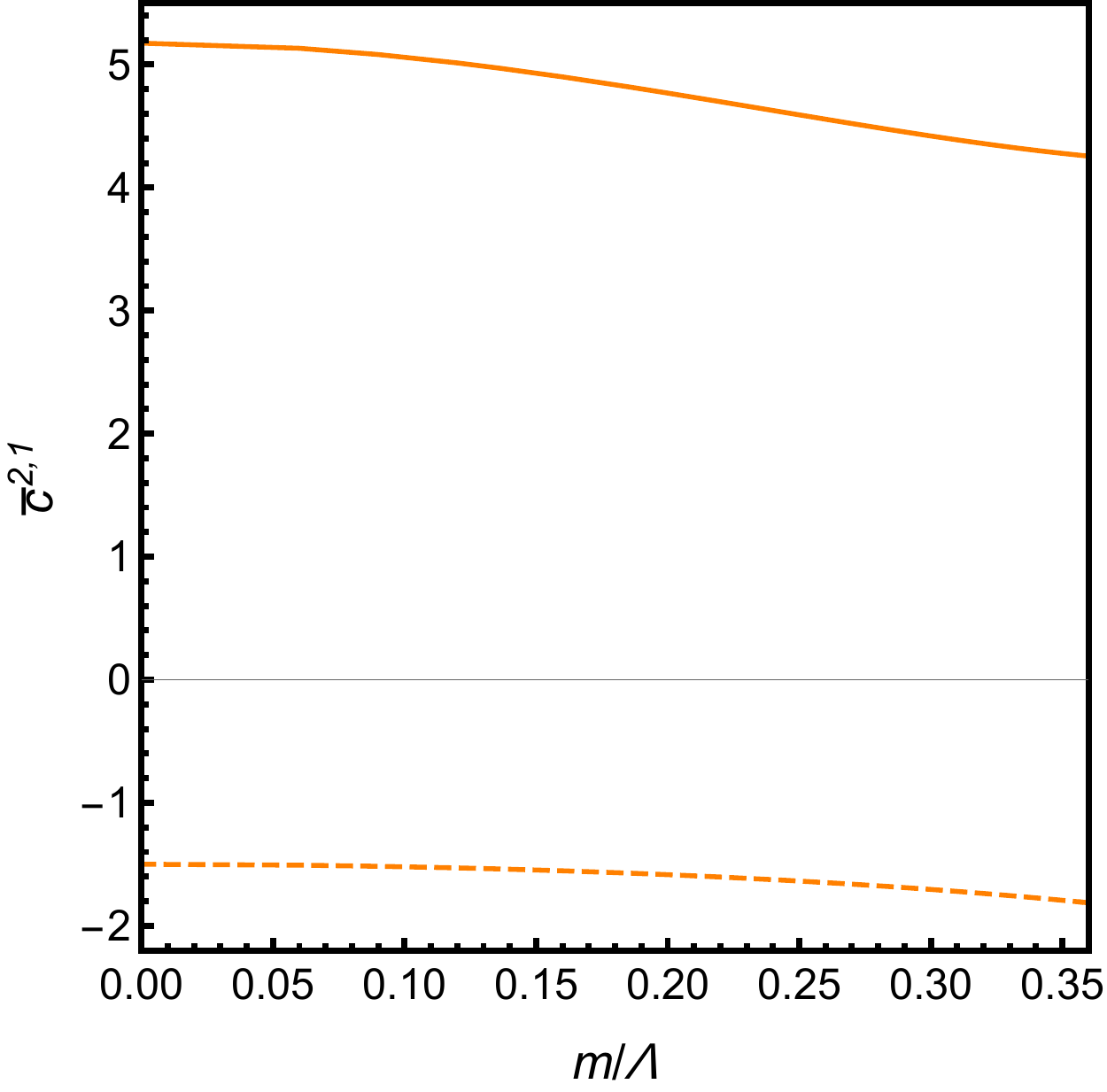}
    \end{subfigure}
    \begin{subfigure}{0.49\linewidth}
    \centering~~
    \includegraphics[width=0.975\linewidth]{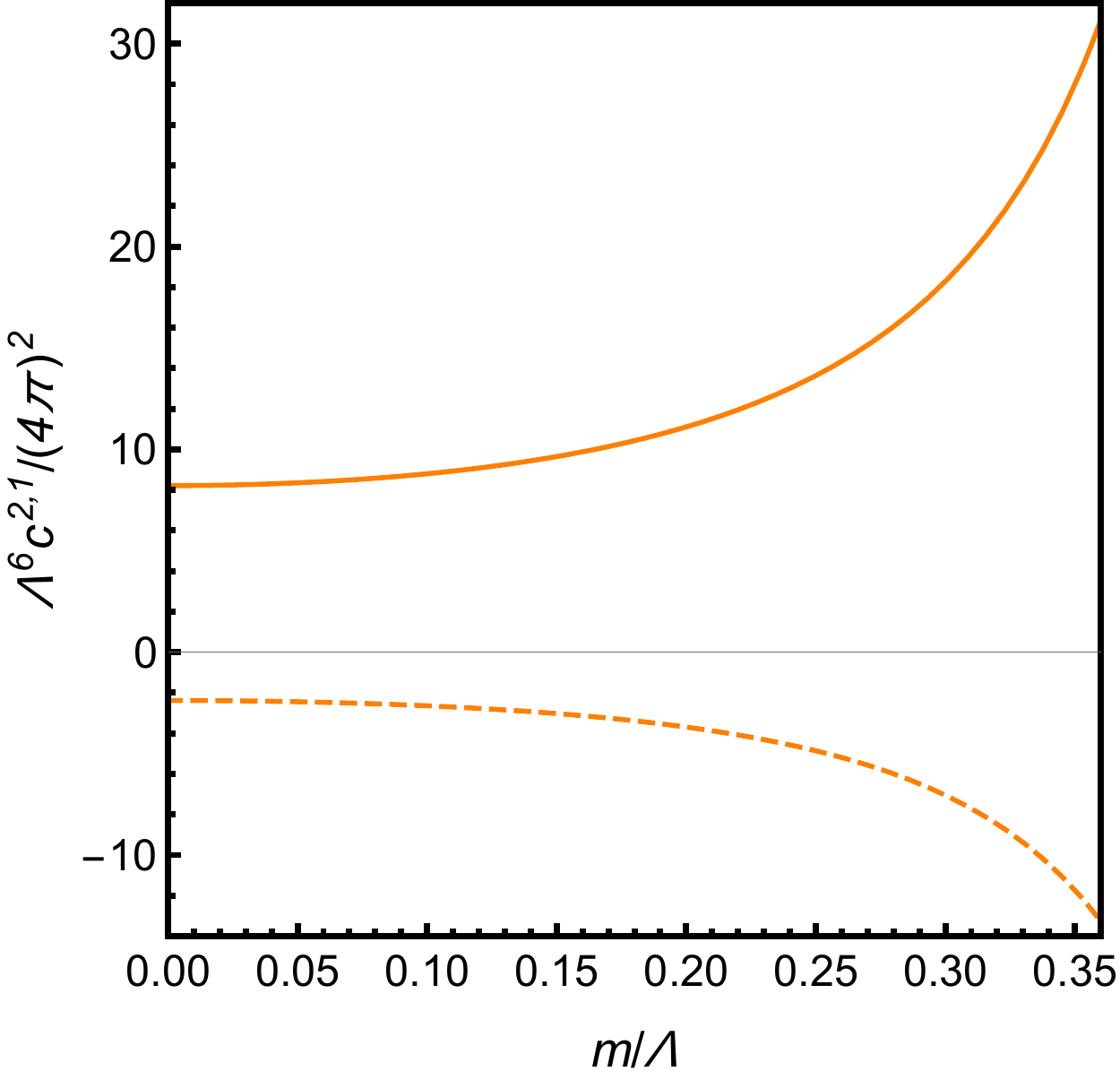}
    \end{subfigure}
\caption{({\it left}) Triple crossing positivity bounds on $\bar{c}^{2,1}= {\Lambda}^{2} {c}^{2,1}/c^{2,0}$ for various scalar mass $m$, agnostic about all the other coefficients. The solid curve is the upper bound, and the dashed curve the lower bound. ({\it right}) Triple crossing bounds on ${c}^{2,1}$ itself, incorporating the upper bound on $c^{2,0}$. The triple crossing symmetric coefficient $g_{0,1}={c}^{2,1}$ (see \eref{gcoefs}). }
\label{fig:c21}
\end{figure}

\begin{figure}[ht]
\centering
    \begin{subfigure}{0.485\linewidth}
    \centering \!\!\!
    \includegraphics[width=0.99\linewidth]{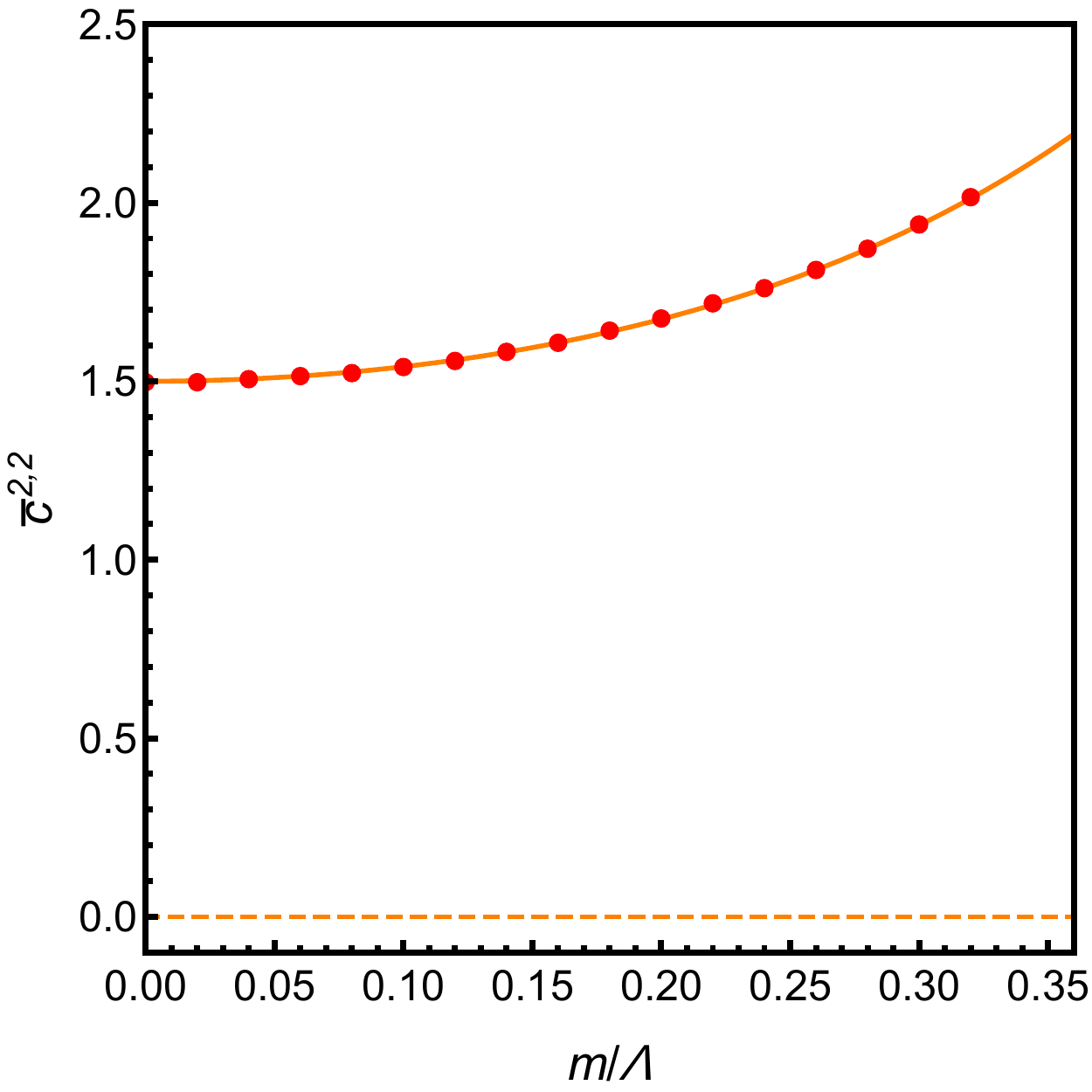}
    \end{subfigure}
    \centering
    \begin{subfigure}{0.47\linewidth}
    \centering~~
    \includegraphics[width=0.99\linewidth]{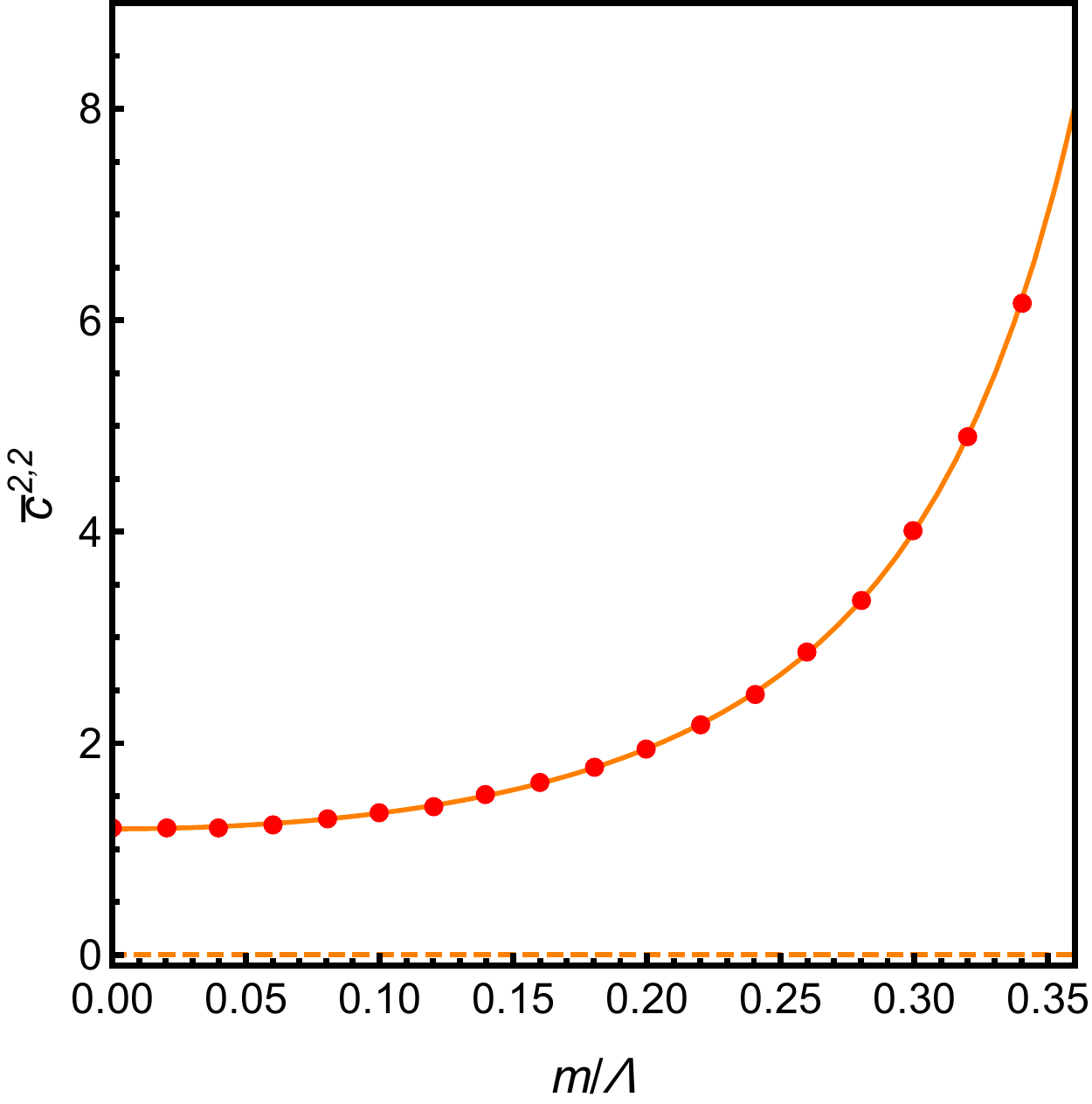}
    \end{subfigure}
    \centering
    \caption{Mass dependence of two-sided positivity bounds on $\bar{c}^{2,2}$ ({\it left}). The dots are obtained by computing $\bar{c}^{4,0}$ and then converting to $\bar{c}^{2,2} = 3\bar{c}^{4,0}/2$, which constitutes a sanity check. The solid lines are the upper bounds and the dashed lines are the lower bounds. ({\it right}) Two-sided bounds on ${c}^{2,2}$ itself, incorporating the upper bound on $c^{2,0}$. }
    \label{fig:2240}
\end{figure}

For the coefficients $\bar{c}^{2i,j}$ with $2i+j=3$, there is only one independent coefficient, which can be chosen to be $\bar{c}^{2,1}$. In Figure \ref{fig:c21}, we plot the upper and lower bounds on $\bar{c}^{2,1}$ and ${c}^{2,1}$ for various masses of the scalar. We have plotted the mass dependence up to around $m/\Lambda=0.35$, for which the cutoff is already fairly low and close to the threshold scale $2m$. For a relatively large $m$, as expected, it is essential that we include many orders of $m^2$ in the sum rules to achieve numerical convergence (See Appendix \ref{app:appA}). We see that $\bar{c}^{2,1}$ has very mild dependence on the scalar mass, the upper bound actually slightly decreasing as $m$ increases, while ${c}^{2,1}$ becomes greater for larger $m$. That is, ${c}^{2,1}$'s mass dependence is similar to that of ${c}^{2,0}$. The situation is similar at order $2i+j=4$, in which case the only independent coefficient can be chosen to be $\bar{c}^{2,2}$, the $\bar{c}^{4,0}$ coefficient being related to it by $\bar{c}^{4,0}=2\bar{c}^{2,2}/3$. In Figure \ref{fig:2240}, we present the two-sided bounds for $\bar{c}^{2,2}$ and $\bar{c}^{4,0}$ respectively, and confirm their upper bounds are indeed linearly dependent. Whilst the lower bound on $\bar{c}^{2,2}$ is always zero, the upper bound on $\bar{c}^{2,2}$ increases slowly with $m$. Again, the upper bound on $c^{2,2}$ itself increases much faster because $c^{2,0}$ is more sensitive to $m$.

\begin{figure}[ht]
\centering
    \begin{subfigure}{0.495\linewidth}
    \centering
    \includegraphics[width=0.99\linewidth]{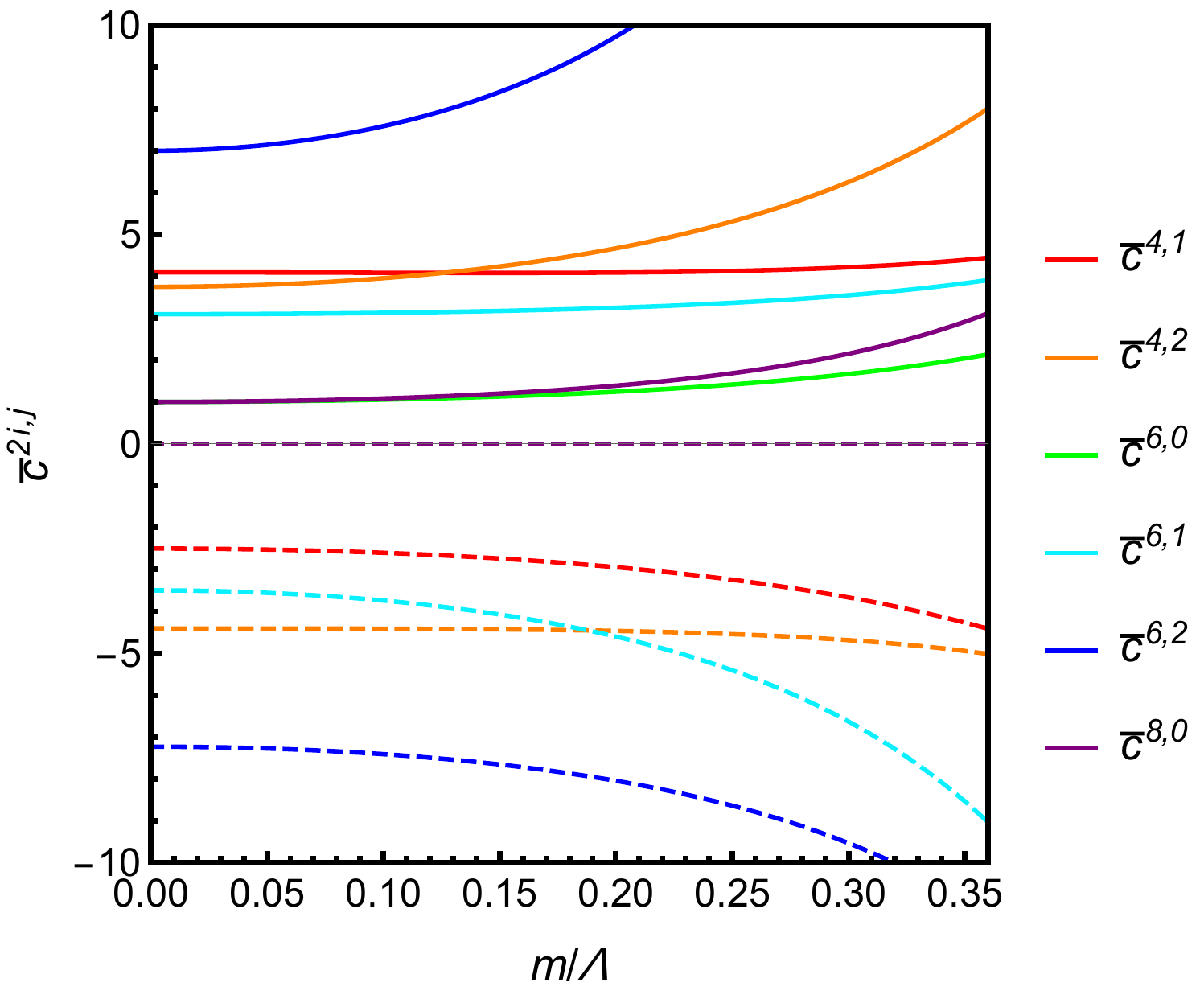}
    \end{subfigure}
    \centering
    \begin{subfigure}{0.495\linewidth}
    \centering
    \includegraphics[width=0.99\linewidth]{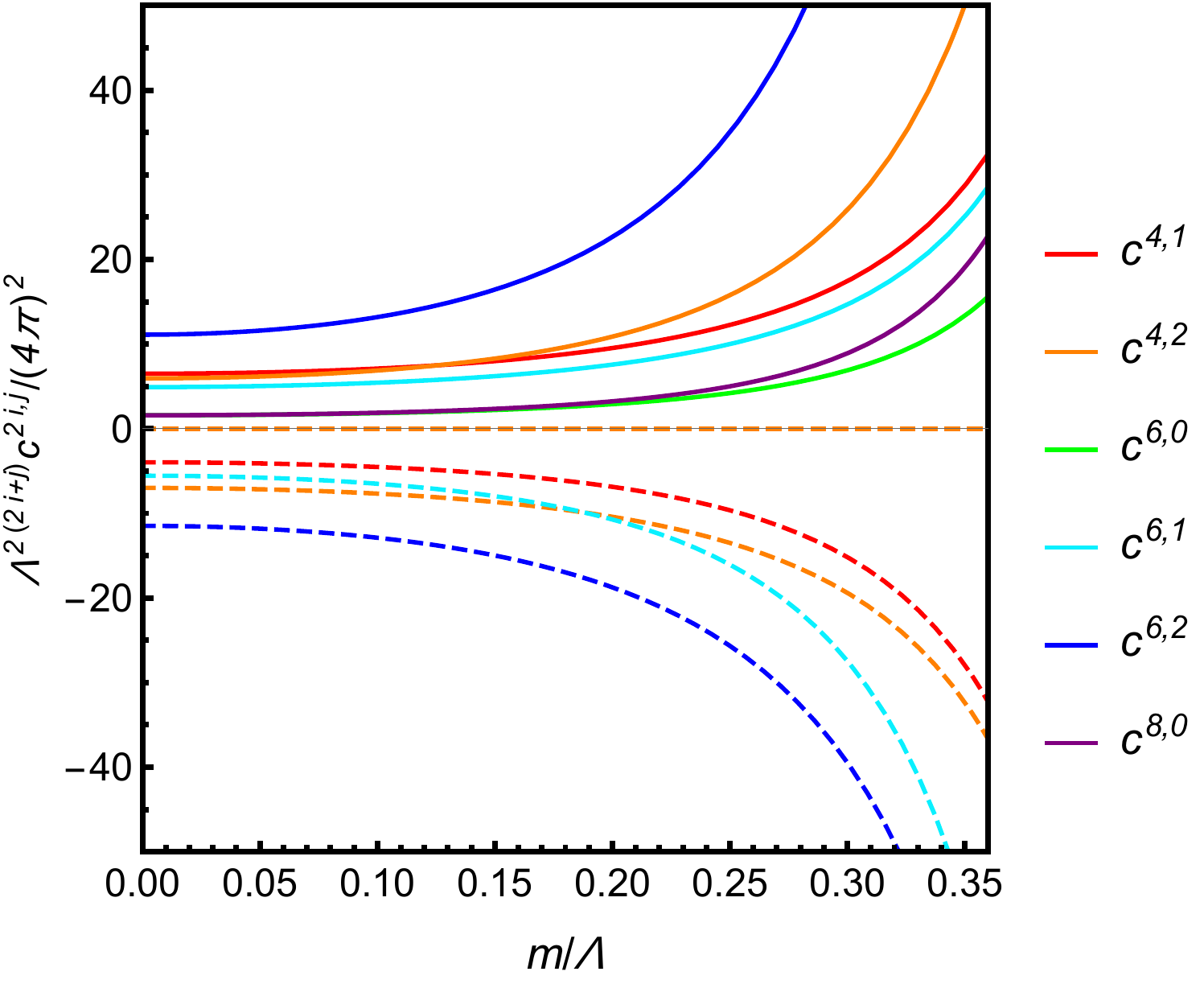}
    \end{subfigure}
    \centering
    \caption{Two-sided positivity bounds on higher order coefficients. The solid lines are the upper bounds and the dashed lines the lower bounds. We define $\bar{c}^{2i,j}= {\Lambda}^{2(2i+j-2)} {c}^{2i,j}/c^{2,0}$, and the $c^{2i,j}$ coefficients are related to the $stu$ symmetric coefficients by \eref{gcoefs}. }
    \label{fig:highcoef}
\end{figure}

For the higher order coefficients, we find that generally the bounds on $\bar{c}^{2i,j}$ are similarly relaxed for larger $m$, with some of the lower bounds being exactly zero; see Figure \ref{fig:highcoef}. The sensitivity of the coefficients to $m$ seems to depend on the intricate forms of the sum rules, not having a clear monotonic trend. The $c^{2i,j}$ coefficients are related to the $stu$ symmetric coefficients via \eref{gcoefs}.

The reason why a nonzero mass generally relaxes positivity bounds can be readily seen from sum rules Eq.(\ref{cbra}) or more explicitly \eref{c20explicit} and \eref{c21explicit} for the first two sum rules. Recall that the  average $\langle~\rangle$ notation entails an integration over $\mu\geq \tilde\Lambda^2$ and a summation for spin $l$. Schematically, the general structure of $C^{2i,j}_l(\mu)$ goes like
\be
 C^{2i,j}_l(\mu) \sim  \f{1}{\mu^a (\mu+2m^2)^b} \sum_{c,d} f^{2i,j}_{cd}(l)\mu^c (m^{2})^d .
\ee
Compared with the massless case, the $1/\mu^a (\mu+2m^2)^b$ factor becomes smaller for the massive case because $\mu$ starts from $\mu\geq \tilde\Lambda^2=\Lambda^2-4m^2$, which after acting the average $\langle~\rangle$ would increase $|c^{2i,j}|$ via the sum rule if the spin polynomials $f^{2i,j}_{cd}(l)$ are sign semi-definite. For the coefficients such as $c^{2,0}$ and $c^{2,2}$, all the $f^{2i,j}_{cd}(l)$ are indeed positive semi-definite, and the $(m^2)^{d}$ terms with $d>0$ act to further increase $|c^{2i,j}|$, compared with the massless case. Hence, as the mass increases, the upper bounds of these coefficients also increase, while the lower bounds of these coefficients remain zero.

For the coefficients such as $c^{2,1}$, on the other hand, $f^{2i,j}_{cd}(l)$ are only sign semi-definite for  $l=0$ and $l\geq 2$ separately; more precisely $f^{2i,j}_{cd}(l=0)$ are negative semi-definite and $f^{2i,j}_{cd}(l\geq 2)$ are positive semi-definite. Therefore, for the outcome of the optimization procedure, one expects that the upper and lower bounds will be of the opposite signs, with one of them dominated by the $l=0$ spin and the other dominated by the $l\geq 2$ spins. Again, for either sides of the bounds, these $|c^{2i,j}|$ will increase as the scalar mass increases. The above argument has yet taken into account the roles of null constraints, but from the $c^{2i,j}$ space point of view the null constraints just tell us to restrict to the linear subspace as depicted by \eref{npq19}, which can be done after running the above argument.

\begin{figure}[ht]

\centering
    \includegraphics[width=0.487\linewidth]{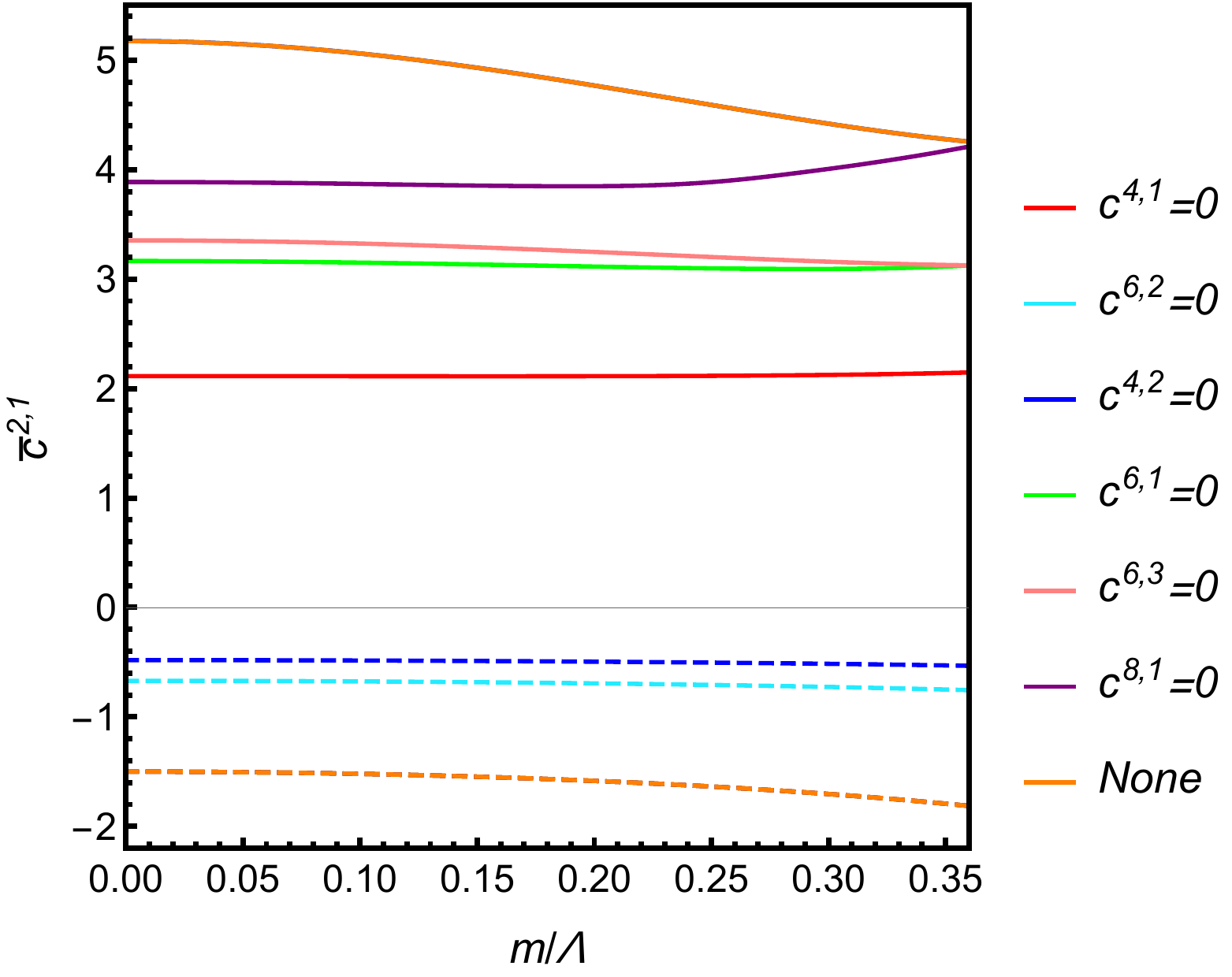}
    \includegraphics[width=0.49\linewidth]{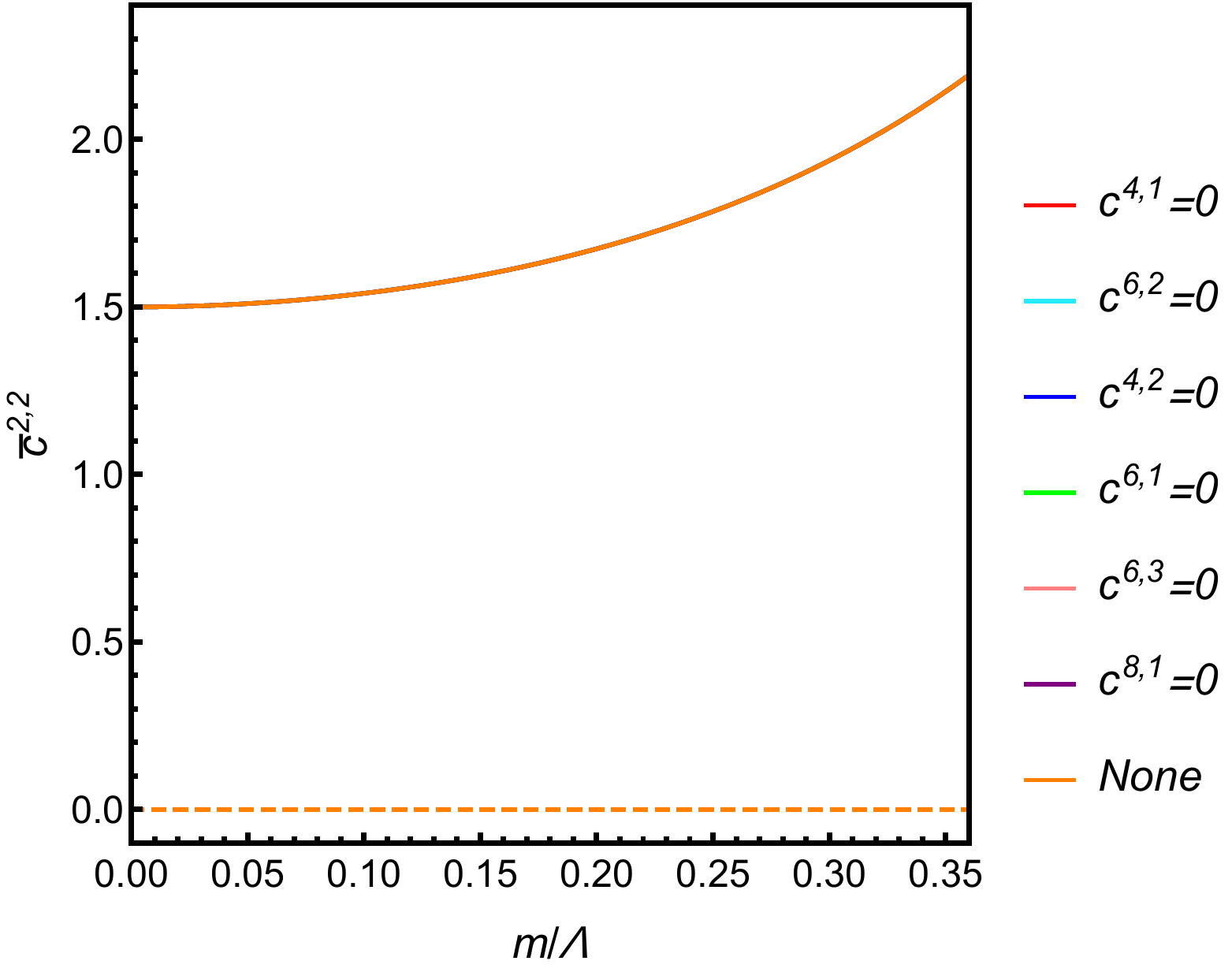}
    \caption{Triple crossing bounds when one higher order coefficient is fine-tuned to zero. The solid lines are upper bounds and the dashed lines lower bounds. Setting a higher order coeffiicient $c^{2i,1}$ to zero tightens the upper bounds on $\bar{c}^{2,1}$ while setting $c^{2i,2}=0$ improves the lower bounds on $\bar{c}^{2,1}$. On the other hand, the bounds on $\bar{c}^{2,2}$ are insensitive to suppressing the higher order coefficients. }
    \label{fig:hoc}
\end{figure}

\subsubsection{Fine-tuned EFTs}

 In model-building, it is a common practice to set certain coefficients to zero or have them suppressed. However, from a positivity perspective, such practices can sometimes be precarious. This is because positivity bounds require some coefficients such as ${c}^{2i,0}$ to be strictly positive, ${c}^{2i,0}>0$, so it is inconsistent to have a theory with exact $\bar{c}^{2i,0}=0$. By crossing symmetry, coefficients such as $c^{2,2}=3 {c}^{4,0}/2$ are also required to be strictly positive. (In the present of gravity, however, these coefficients are allowed to slightly negative, the negativity being suppressed by $M_P^2$.) Of course, phenomenologically, a zero higher order coefficient often simply means some extra suppression in that coefficient. However, suppressing a coefficient that can not be set to strict zero will often result in some other coefficients also being required to be highly suppressed, in order to be consistent with positivity bounds. That is, the positivity region in a fined EFT can be much smaller than that of a generic EFT.  For example, if we have a suppressed $c^{4,0}$, then the bounds on ${c}^{2,1}$ will be highly suppressed. This can be seen in Figure \ref{fig:2Dbounds} in the next subsection.

For the coefficients that can be both positive and negative, setting them to zero essentially adds more null constraints to the theory, which provides extra constraints on the EFT parameter space. In the left subfigure of Figure \ref{fig:hoc}, we plot how the bounds on $\bar{c}^{2,1}$ change if we individually set various higher order coefficients to zero. In some cases, such fine-tunings can significantly reduce the mass dependence of the bounds. For example, setting $c^{4,2}$ or $c^{6,2}$ to zero makes the lower bound on $\bar{c}^{2,1}$ essentially independent of the mass, and setting $c^{4,1}$ to zero makes the upper bound on $\bar{c}^{2,1}$ essentially independent of the mass. Setting a higher order coeffiicient $c^{2i,1}$ to zero improves the upper bounds on $\bar{c}^{2,1}$ while setting $c^{2i,2}=0$ improves the lower bounds on $\bar{c}^{2,1}$. On the other hand, the bounds on $\bar{c}^{2,2}$ do not change if we set these higher order coefficients to zero; see the right subfigure of Figure \ref{fig:hoc}.
The reason for this is that the bounds on $\bar{c}^{2,2}=3\bar{c}^{4,0}/2$ can be obtained in the forward limit, which is insensitive to the non-forward limit coefficients that are set to zero in Figure \ref{fig:hoc}.

\subsection{Two-dimensional bounds}

\begin{figure}[ht]
\centering
    \begin{subfigure}{0.48\linewidth}
    \centering
    \includegraphics[width=0.99\linewidth]{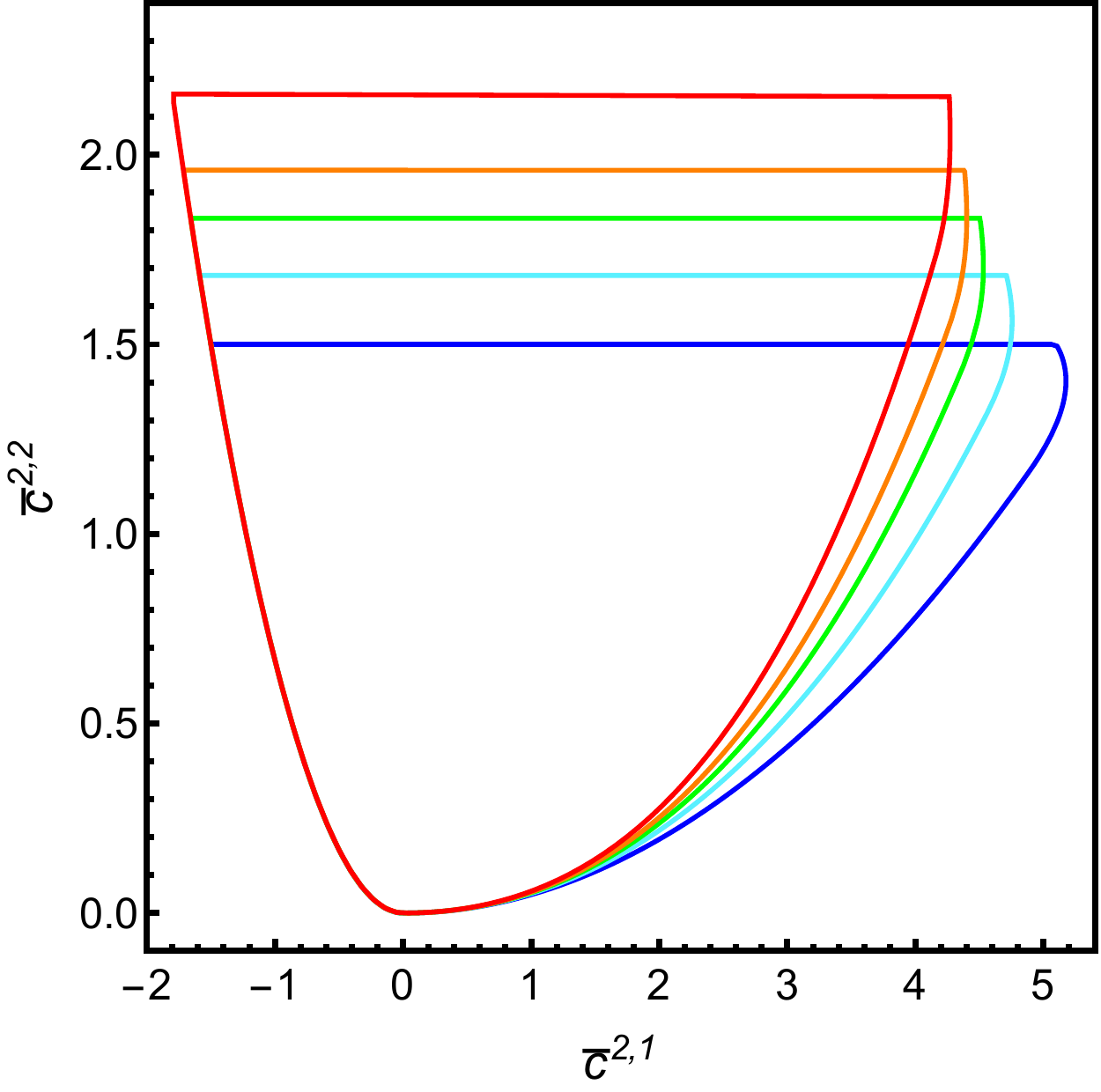}
    \end{subfigure}
    \centering~~
    \begin{subfigure}{0.485\linewidth}
    \centering
    \includegraphics[width=0.99\linewidth]{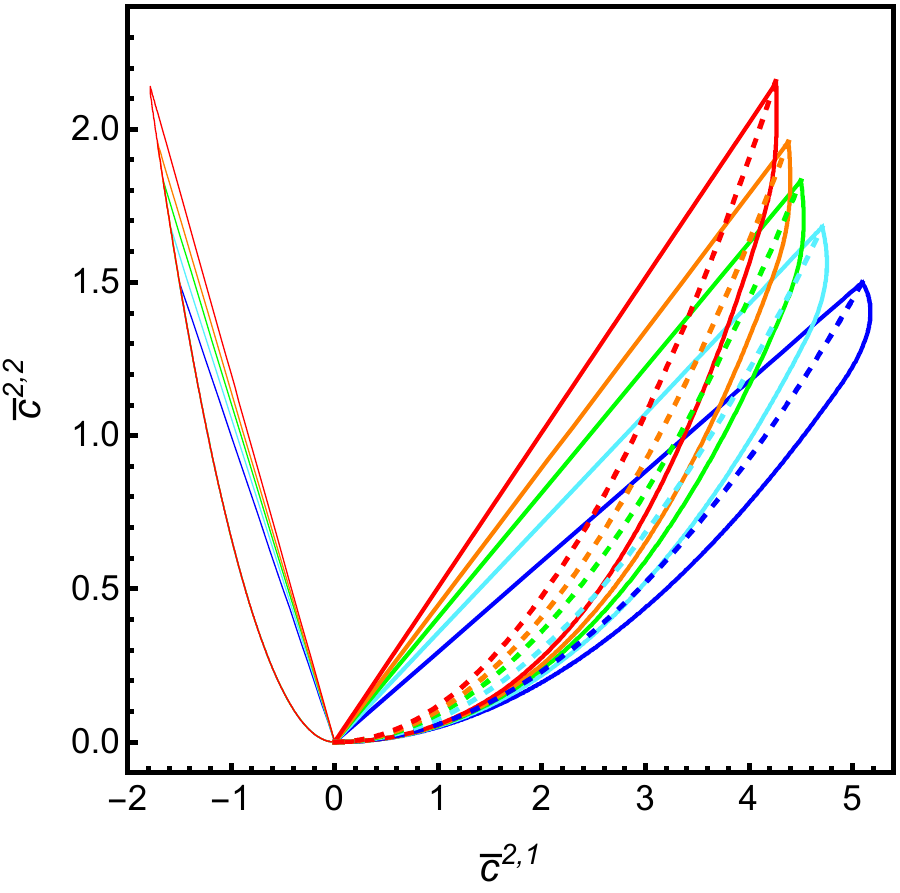}
    \end{subfigure}
    \centering

    \centering
    \begin{subfigure}{0.68\linewidth}
    \centering
        \includegraphics[width=1\linewidth]{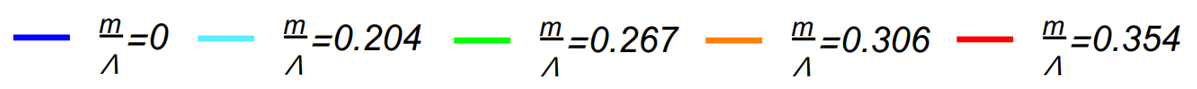}
    \end{subfigure}
    \caption{\label{fig:2Dbounds} ({\it left}) Two dimensional regions allowed by the triple crossing bounds in the $\bar{c}^{2,2}$-$\bar{c}^{2,1}$ plane for various scalar masses $m$. ({\it right}) Two dimensional positivity regions assuming two different UV scenarios. The solid lines on the left are when the UV states are only spin-0 particles, while the thick solid lines on the right correspond to UV states with spin $l\geq 2$. The dashed lines are the positivity boundaries generated by a $stu$ pole with the spin-0 contribution subtracted (see \eref{lhsumrule}). The top tips of the $l\geq 2$ numerical positivity curves match exactly those of the $stu$ pole with different $m$. The full positivity region on the left figure is the convex hull of the corresponding two regions on the right figure. }
\end{figure}

We now move on to consider the positivity constraints on a two-dimensional parameter plane. Figure \ref{fig:2Dbounds} displays the variations of the 2D bounds on $\bar{c}^{2,2}$ and $\bar{c}^{4,0}$ as the mass changes. The left subfigure is for a generic UV completion with all possible spins for the UV states. The right subfigure consists of two separate regions for a fixed $m$, which correspond to two scenarios: the left region is obtained by only allowing UV states with $l=0$; the right region is obtained by allowing UV states with $l\geq 2$. (In the case of a tree level UV completion, $l$ denotes the spin of the UV particle; in the case of a loop level UV completion, $l$ simply denotes the spin of the UV partial wave component.) These restrictions can be easily implemented in the optimization with SDPB. (As mentioned previously around \eref{eq2-3}, spin-1 states must be absent due to crossing symmetry.) The full positivity constrained region is the convex hull of the corresponding two regions for a fixed $m$. As expected, if the mass increases, the positivity region shrinks in the $\bar{c}^{2,1}$ direction and is relaxed in the $\bar{c}^{2,2}$ direction.

Note that, for the $l\geq 2$ case, numerical results indicate that it is important that we have multiple UV states. For example, if the UV only has a spin-2 state but no higher spin states, we will not have any available positivity region in the scalar EFT. The same is true if the UV has only $l\geq 4$ states but no spin-2 state.  In fact, it appears that one at least needs $l=2,4,6,8$ to get any positivity region.

Interestingly, as the low energy scalar mass varies, the left boundary of the positivity region, which comes from contributions of the spin-0 UV states, is unchanged, except for extension of the boundary upwards; see the right subfigure of Figure \ref{fig:2Dbounds}. The reason why the left boundary is unchanged can be understood as follows. In the case of a tree-level UV completion, the amplitude from the contributions of UV scalars must be of the form
\be
\label{A0UVS}
\bar A_0(s,t) = \sum_i  \( \f{\li_i}{\tilde M_i^2 -\tilde s} + \f{\li_i}{\tilde M_i^2 -\tilde t}+ \f{\li_i}{\tilde M_i^2 -\tilde u} \) ,
\ee
where $\tilde M_i^2=M_i^2 - 4m^2/3$, $M_i$ are the masses of the UV scalars and all the $\li_i$ need to be of the same sign to avoid ghost instabilities. Expanding in terms of small $\tilde s,\tilde t,\tilde u$, we have
\be
\bar A_0(s,t) \supset \( \sum_i \f{2\li_i}{\tilde M_i^6} \) x + \( \sum_i \f{3\li_i}{\tilde M_i^8} \) y  + \( \sum_i   \f{2\li_i}{\tilde M_i^{10}}  \) x^2  = g^{(0)}_{1,0} x +  g^{(0)}_{0,1} y  + g^{(0)}_{2,0} x^2
\ee
where $x=-(\tilde s \tilde t  + \tilde s \tilde u +  \tilde t\tilde u)$, $y=-\tilde s \tilde t \tilde u$ and $\supset$ means that the terms on the right hand side are included on the left hand side.
The left boundary of Figure \ref{fig:2Dbounds} can be obtained if all the $\li_i$ and $M_i$ are the same $\li_i=\li,~M_i=M$, that is, if there is only one kind of UV scalars, for which case we have
\be
A^{(0)}(s,t) =   \f{\li}{\tilde M^2 -\tilde s} + \f{\li}{\tilde M^2 -\tilde t}+ \f{\li}{\tilde M^2 -\tilde u} \supset \f{2\li}{\tilde M^6} x  - \f{3\li}{\tilde M^8}  y  +   \f{2\li}{\tilde M^{10}}  x^2 .
\ee
Thus, for the left positivity boundary, we clearly have $g^{(0)}_{1,0}= c^{2,0}_{(0)}={2\li}/{\tilde M^6}$, $g^{(0)}_{0,1} = c^{2,1}_{(0)}=-{3\li}/{\tilde M^8}$ and $g^{(0)}_{2,0}=2c^{2,2}_{(0)}/3={2\li}/{\tilde M^{10}}$. The over-barred values are $\bar{c}^{2,1}=-3\Lambda^2/(2\tilde{M}^2)$, $\bar{c}^{2,2}=3\Lambda^4/(2\tilde{M}^4)$. Therefore, we see that there is no $m$ dependence in the relation between $\bar c^{2,2}$ and $\bar c^{2,1}$ at the left boundary: $\bar c^{2,2}_{(0)} = 2(\bar c^{2,1}_{(0)})^2/3$.

On the other hand, the right boundary, which comes from contributions of the UV particles with spin for a tree-level UV completion, changes significantly as $m$ varies. However, the top tip of the right boundary changes according to a simple rule. To see this, note that this tip can be described by the following amplitude without any spin-0 contribution \cite{Caron-Huot:2020cmc}
\be
\label{lhsumrule}
A^{(\geq 2)}_{stu} (s,t) = \f{\lambda  M^4}{(\tilde M^2 -\tilde s)(\tilde M^2 -\tilde t)(\tilde M^2 -\tilde u)} - \gamma_0(m) A^{(0)}(s,t) \supset g^{(\geq 2)}_{1,0} x + g^{(\geq 2)}_{0,1} y  + g^{(\geq 2)}_{2,0} x^2 ,
\ee
where $\gamma_0(m)$ can be fixed by simply setting the spin-0 partial wave amplitude to zero. Explicitly, using the orthogonality of the Legendre polynomials  $\frac{1}{2}\int_{-1}^{1}\mathrm{d}z P_l(z)P_{l^\prime}(z)=\delta_{ll^{\prime}}/(2l+1)$, we can get the $l$ partial wave amplitude
\begin{equation}\label{spal}
  a_l(s)=\frac{1}{32}\sqrt{\frac{s-4m^2}{s}}\int_{-1}^{1}\mathrm{d}z~ P_l(z) A^{(\geq 2)}_{stu} \(s,\f12 (s-4m^2) (z-1) \).
\end{equation}
If we seek an amplitude without the spin-0 component, we must have $a_0(s)=0$, which gives
\be
\label{gam0res}
  \gamma_0(m)=\frac{2\ln(2-4m^2/M^2)}{(1-4m^2/M^2)(3-4m^2/M^2)} .
\ee
Expanding Eq.(\ref{lhsumrule}) in terms of small $\tilde s, \tilde t, \tilde u$, we can extract the Wilson coefficients
\begin{align}\label{sphg}
  g^{(\geq 2)}_{1,0}  = \frac{\li M^4}{\tilde{M}^{10}} -\f{2\li\gamma_0(m)}{\tilde{M}^6} ,   ~
  g^{(\geq 2)}_{0,1}  =   -\frac{\li M^4}{\tilde{M}^{12}}+\frac{3\lambda \gamma_0(m)}{\tilde{M}^8}  ,   ~
  g^{(\geq 2)}_{2,0}  =  \frac{\li M^4}{\tilde{M}^{14}} -\frac{2 \lambda \gamma_0(m)}{\tilde{M}^{10}} ,
\end{align}
and the corresponding expressions for the ratios of the coeffcients $\bar{c}^{2,1}_{(\geq 2)}$ and $\bar{c}^{2,2}_{(\geq 2)}$ are
\be
\label{bcsph}
  \bar{c}^{2,1}_{(\geq 2)}  = -\frac{1}{(\tilde{M}/\Lambda)^2}\frac{(M/\tilde{M})^4-3\gamma_0(m)}{(M/\tilde{M})^4-2\gamma_0(m)} ,~~~~~
  \bar{c}^{2,2}_{(\geq 2)}  =\frac{3}{2}\frac{1}{(\tilde{M}/\Lambda)^4}.
\ee
We see that $\bar{c}^{2,2}_{(\geq 2)}$ has exactly the same value as the UV scalar case (\ref{A0UVS}). We have numerically verified that, under the flow of the IR scalar mass, the top tip of the right positivity region in Figure \ref{fig:2Dbounds} (computed with the linear programing) follows very well the trajectory of
\be
\label{UVcur}
  \bar{c}^{2,2}_{(\geq 2)}  = \frac{3}{2} \(\frac{(M/\tilde{M})^4-2\gamma_0(m)}{(M/\tilde{M})^4-3\gamma_0(m)}\)^2 \( \bar{c}^{2,1}_{(\geq 2)}\)^2  ,
\ee
once the UV mass scale $M$ is taken to be the EFT cutoff $\Lambda$. However, we see that the $l\geq 2$ numerical positivity region is greater than the region bounded by the curves (\ref{UVcur}) (the dashed lines in Figure \ref{fig:2Dbounds}).

With the above UV amplitudes, one can understand why the lower bound of $\bar{c}^{2,1}$ and the upper bound of $\bar{c}^{2,2}$ are insensitive to the number of null constraints and the maximal spin partial wave $l_{\mathrm{max}}$ as shown in the convergence tests in Appendix \ref{app:appA}. As we have seen above, the lower bound of $\bar{c}^{2,1}$ is determined by the spin-0 partial wave, while the importance of the null constraints lies on the fact that they imply the higher spin partial wave amplitudes are highly suppressed, so this indeed means that higher spin partial waves must have little effects on the lower bound of $\bar{c}^{2,1}$. On the other hand, although the upper bound of $\bar{c}^{2,2}$ seems to depend on the $A^{(\geq 2)}_{stu} (s,t)$ amplitude, we accidentally have $\bar{c}^{2,2}_{(\geq 2)} = 3/(2(\tilde{M}/\Lambda)^2) =\bar{c}^{2,2}_{(0)}$ for the upper bound of $\bar{c}^{2,2}$, in which case the effects of higher spin partial waves are again unimportant.

\section{Bounds on Horndeski theories}
\label{sec:sec4}

Recent cosmological and astronomical observations have revealed several surprising phenomena that can not be accommodated by GR with known matter sources. This raises the possibility that there may be some extra fields such as a scalar field whose dynamics can account for the observational anomalies. Many salient features of these models can be captured mostly by the scalar mode alone.

Scalar-tensor theories are a diverse class of models and have a wide range of applicabilities in astrophysics. The positivity bounds derived here provide a powerful tool to reduce their vast parameter spaces. In the following subsections, we will apply these fully crossing symmetric positivity constraints to several branches of scalar-tensor theories. To use the bounds obtained in the last section, we can take the decoupling limit around flat space and focus on the scalar sector of the theories. The reduction of the parameter spaces by the positivity bounds will facilitate the comparisons of the models with observational data.

\subsection{Generic theory}
\label{sec:appsec1}

Horndeski theory \cite{Horndeski:1974wa} was derived early on as the most general scalar-tensor theory that has second order equations of motion which can be obtained from a Lagrangian. It has recently been re-discovered as generalized galileon theory \cite{Deffayet:2009mn, Nicolis:2008in}, and an explicit map between the two formulations was found in Ref \cite{Kobayashi:2011nu}. The equations of motion being second order eliminates Ostrogradski ghosts explicitly. The Lagrangian of a generic Horndeski theory consists of five terms, which are given by
\bal
\label{hornlag1}
 \mathcal{L}_2^{\rm H}&= M_P^2 \Lambda^2 G_2(\phi,X) ,\\
 \mathcal{L}_3^{\rm H}&=\frac{M_P^2}{\Lambda} G_3(\phi,X) \square\phi ,\\
   \label{hornlag4}
 \mathcal{L}_4^{\rm H}&=M_P^2 G_4(\phi,X) R +\frac{M_P^2}{\Lambda^4}G_{4,X}   \left((\square\phi)^2-(\nabla_\mu\nabla_\nu\phi)^2\right) ,
  \\
   \label{hornlag5}
 \mathcal{L}_5^{\rm H}&=\frac{M_P^2}{\Lambda^3} G_{5}(\phi,X)G_{\mu\nu}\nabla^\mu\nabla^\nu\phi -\frac{ M_P^2G_{5,X}}{6\Lambda^7} \left((\square\phi)^3-3\square\phi(\nabla_\mu\nabla_\nu\phi)^2+2(\nabla_\mu\nabla_\nu\phi)^3 \right) ,
\eal
where $G_i(\phi,X)$ are (dimensionless) generic functions of $\phi$ and $X\equiv -\pd_\mu \phi \pd^\mu \phi/(2\Lambda^4)$ and $G_{i,X}(\phi,X)$ is the $X$ derivative of $G_{i}(\phi,X)$.
%, and $\Lambda$ is a mass scale.

Of course, from the EFT point of view, Lagrangian terms that lead to higher than second order equations of motion are perfectly acceptable as long as the resulting Ostrogradski ghosts are heavier than the EFT cutoff. In fact, as mentioned in the last section, the higher order terms must be non-vanishing in order to be consistent with positivity bounds. Therefore, Horndeski theory at face value, as listed by Eqs.~(\ref{hornlag1}-\ref{hornlag5}), is inconsistent with the fundamental principles of quantum field theory and does not have a standard UV completion. In this paper, we instead take the viewpoint that the higher order terms are generally present and are suppressed compared to the Horndeski terms. If the higher order terms are assumed to be further suppressed than the usual EFT power counting, much stronger positivity bounds can be obtained.

To utilize the scalar positivity bounds derived from the dispersion relation, we take the decoupling limit of the graviton by expanding the theory around Minkowski space ($g_{\mu\nu}=\eta_{\mu\nu}+h_{\mu\nu}/M_P$) and taking $M_P\to\infty$ and keeping $\Lambda$ fixed. This freezes out all interactions with the graviton, preserving the scalar interactions that are most relevant for the targeted novel astrophysical phenomena, and $\Lambda$ can be identified as the cutoff of the scalar theory. To see this, first note that locality suggests that the generic functions $G_i(\phi,X)$ can be parameterized by expanding around the background value $\phi_0$ and $X_0=-\pd_\mu\phi_0\pd^\mu\phi_0/(2\Lambda^4)$. In this paper, we will only consider the background $\phi_0=0$ and thus $X_0=0$, which gives
\bal
\label{eq5-gi}
    \frac{M_P^2}{\Lambda^2}G_i(\phi,X)&=\( \frac{M_P^2}{\Lambda^2}\)^{\eta_i}\bG_i+\frac{\bG_{i,\phi}}{\Lambda} \phi+\frac{\bG_{i,\phi\phi}}{2\Lambda^2}\phi^2+\bG_{i,X}X+\frac{\bG_{i,\phi X}}{\Lambda}
    \phi X\nn
    &\quad~ +\frac{\bG_{i,\phi\phi X}}{\Lambda^2}\phi^2X+\frac{1}{2}\bG_{i,XX}X^2+\dots
\eal
where, for example, $\bG_{i,\phi X}\equiv \pd^2 G_i/\pd (\phi_0/\Lambda)\pd X_0$ is dimensionless. The $\bG_{i,*}$ coefficients are expected to be parametrically $\mc{O}(1)$ or smaller but ultimately should be constrained by the positivity bounds.  Indeed, the coefficients ${\bG}_{4,X}$ and ${\bG}_{5,\phi}$ generally need to be suppressed if the speed of gravity is close to the speed of light, which is consistent with the constraints from the gravitational wave observations \cite{Copeland:2018yuh, Sakstein:2017xjx, Creminelli:2017sry, Ezquiaga:2017ekz}. $\mathcal{L}_4^{\rm H}$ and $\mathcal{L}_5^{\rm H}$ explicitly contain the curvature tensors, so, to accommodate the gravitational nature of the relevant terms (and recover GR in suitable limits), we require $\eta_i=0$ for $i=2,3$ and  $\eta_i=1$ for $i=4,5$. In this way, the scalar degree of freedom plays the dominant role near the cutoff $\Lambda$, which largely motivated the recent proposals of these models. With these considerations, the $\phi\phi h$ vertices are suppressed in the decoupling limit. Concretely, from $\mathcal{L}_2^{\rm H}$ to $\mathcal{L}_5^{\rm H}$, we find that the relevant vertices are given by $\(\frac{1}{2}\bG_{2,X}+\bG_{3,\phi}\)(\phi^\mu\phi^\nu h_{\mu\nu}-\frac{1}{2}\phi^{\mu}\phi_\mu h^\nu_\nu)/M_P$,  $-{\bG}_{4,\phi\phi}\phi^2\delta^{\mu\nu}_{\alpha\beta}h^{\alpha,\beta}_{\mu,\nu}/(2M_P)$ and $\(-{\bG}_{4,X}+{\bG}_{5,\phi}  \)\delta^{\mu\nu\rho}_{\alpha\beta\gamma}\phi\phi^\alpha_\mu h^{\beta,\gamma}_{\nu,\rho}/(2M_P\Lambda^2)$, which all vanish in the decoupling limit. (Note that $\bG_{2,X}+2\bG_{3,\phi}$ is the coefficient in front of the scalar kinetic term, which will be set to 1 exactly. If $\bG_{4,5,*}$ coefficients were very large, to be $\mc{O}(M_P/\Lambda)$, the second and third $\phi\phi h$ vertex would contribute a non-negligible diagram mediated by a graviton, but it can be field-redefined to be equivalent to adding $\phi\phi\phi\phi$ contact interactions \cite{Melville:2019wyy}. Thus, in that scenario, our formalism still works, and to get the corresponding positivity bounds one only needs to append $-2\Lambda^2(\bG_{4,X}-\bG_{5,\phi})\bG_{4,\phi\phi}/(\bG_4 M_P^2)-m^2(\bG_{4,X}-\bG_{5,\phi})^2/(2\bG_4 M_P^2)$ and $-3\Lambda^2(\bG_{4,X}-\bG_{5,\phi})^2/(2\bG_4 M_P^2)$ to ${\Lambda^4} c^{2,0}(m)$ and ${\Lambda^4} c^{2,1}(m)$ in Eq.~(\ref{app10}) and Eq.~(\ref{app11}) respectively.)

\begin{figure}[ht]
    \centering
    \begin{subfigure}{0.48\linewidth}
    \centering
        \includegraphics[width=0.9\linewidth]{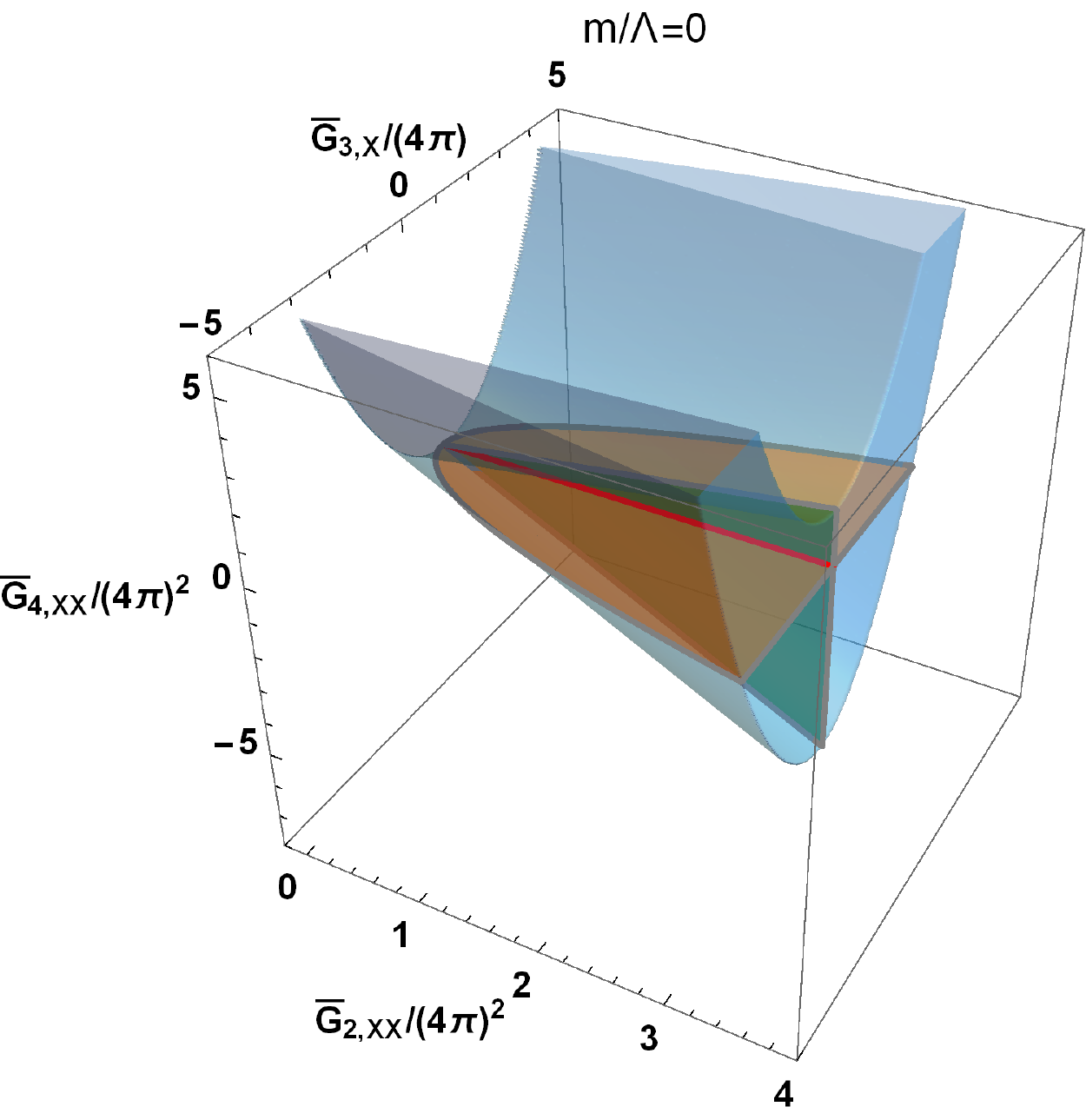}
    \end{subfigure}
     \centering
    \begin{subfigure}{0.48\linewidth}
    \centering
        \includegraphics[width=0.8\linewidth]{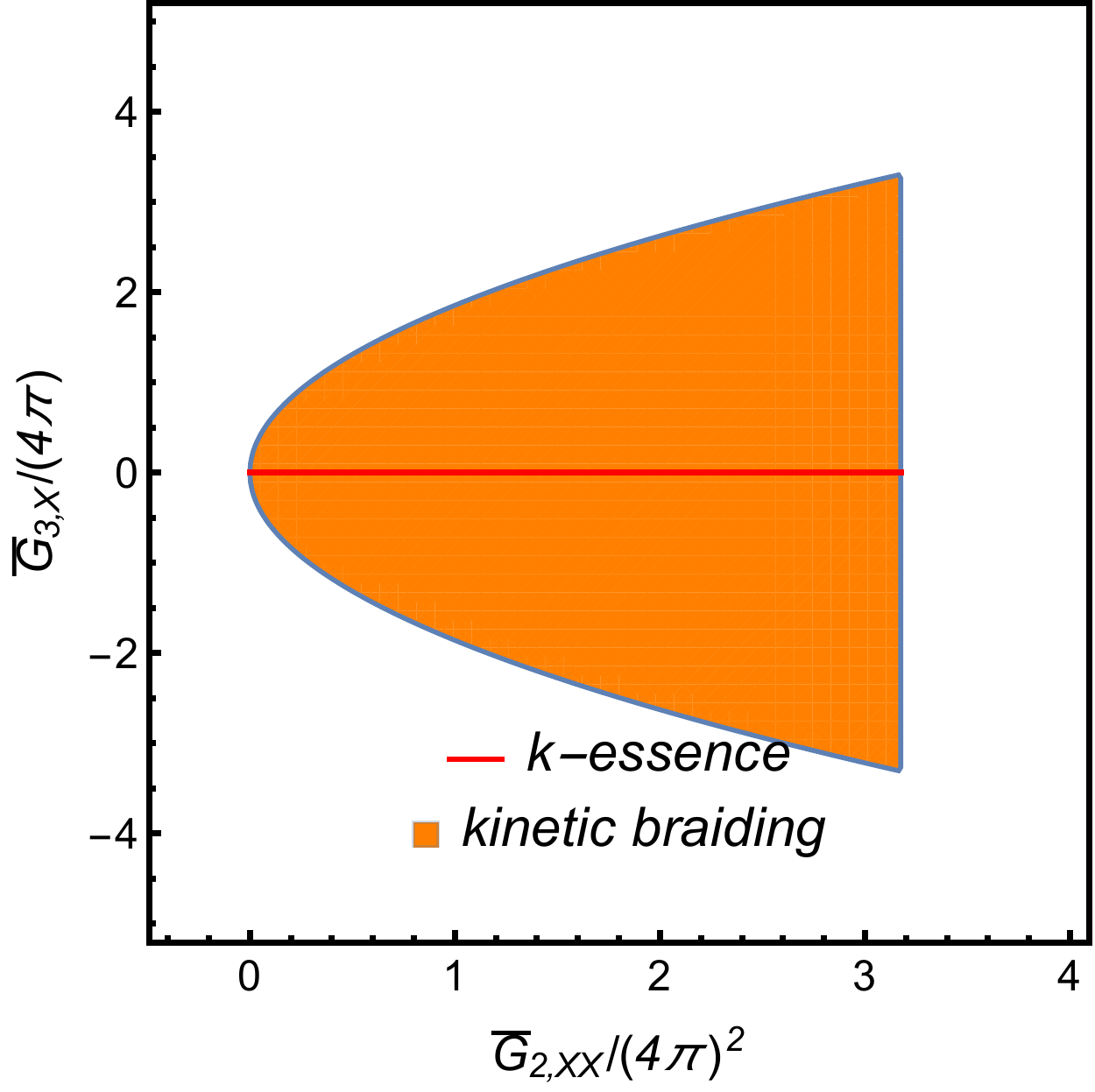}
    \end{subfigure}

     \centering
    \begin{subfigure}{0.48\linewidth}
    \centering
        \includegraphics[width=0.8\linewidth]{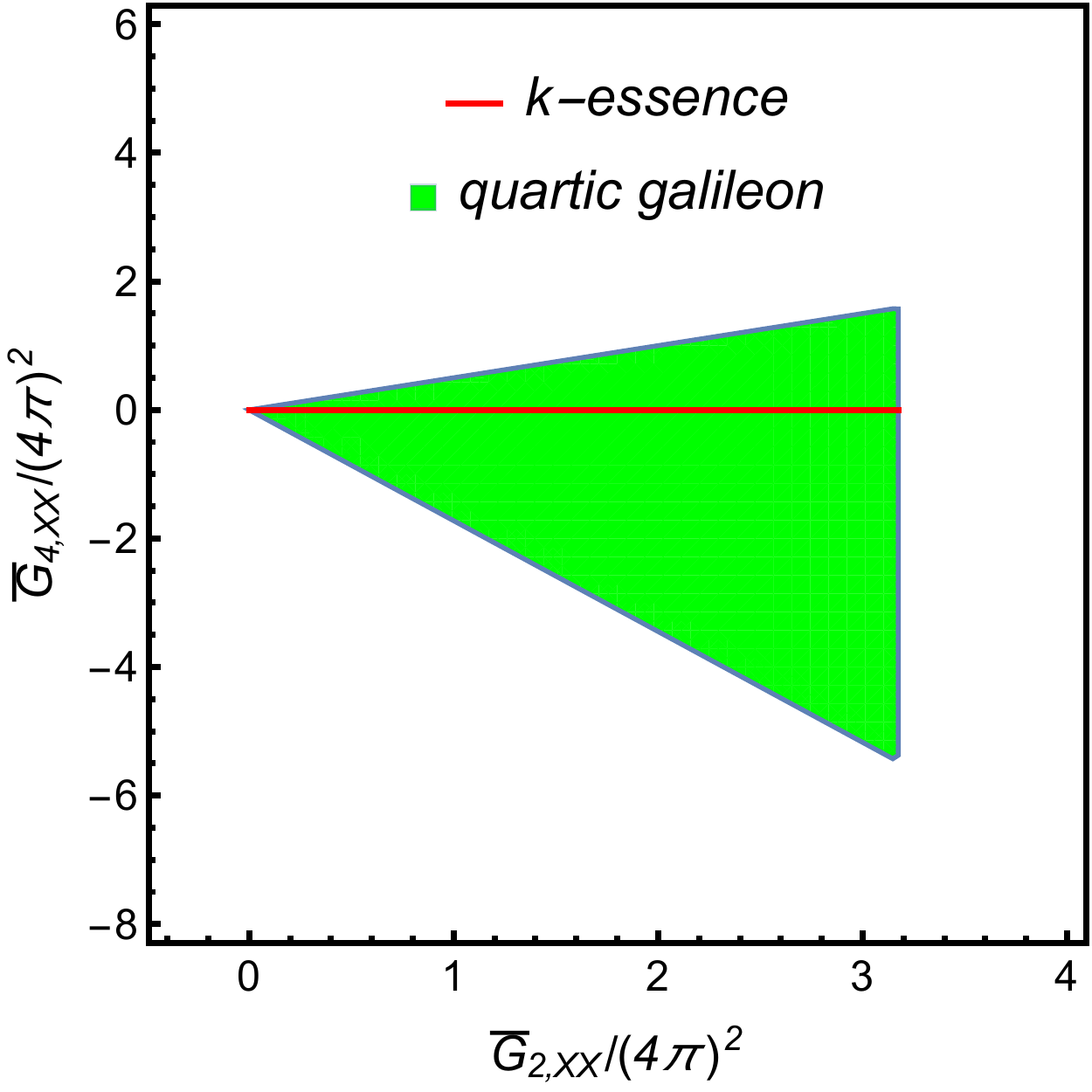}
    \end{subfigure}
     \centering
    \begin{subfigure}{0.48\linewidth}
    \centering
        \includegraphics[width=0.805\linewidth]{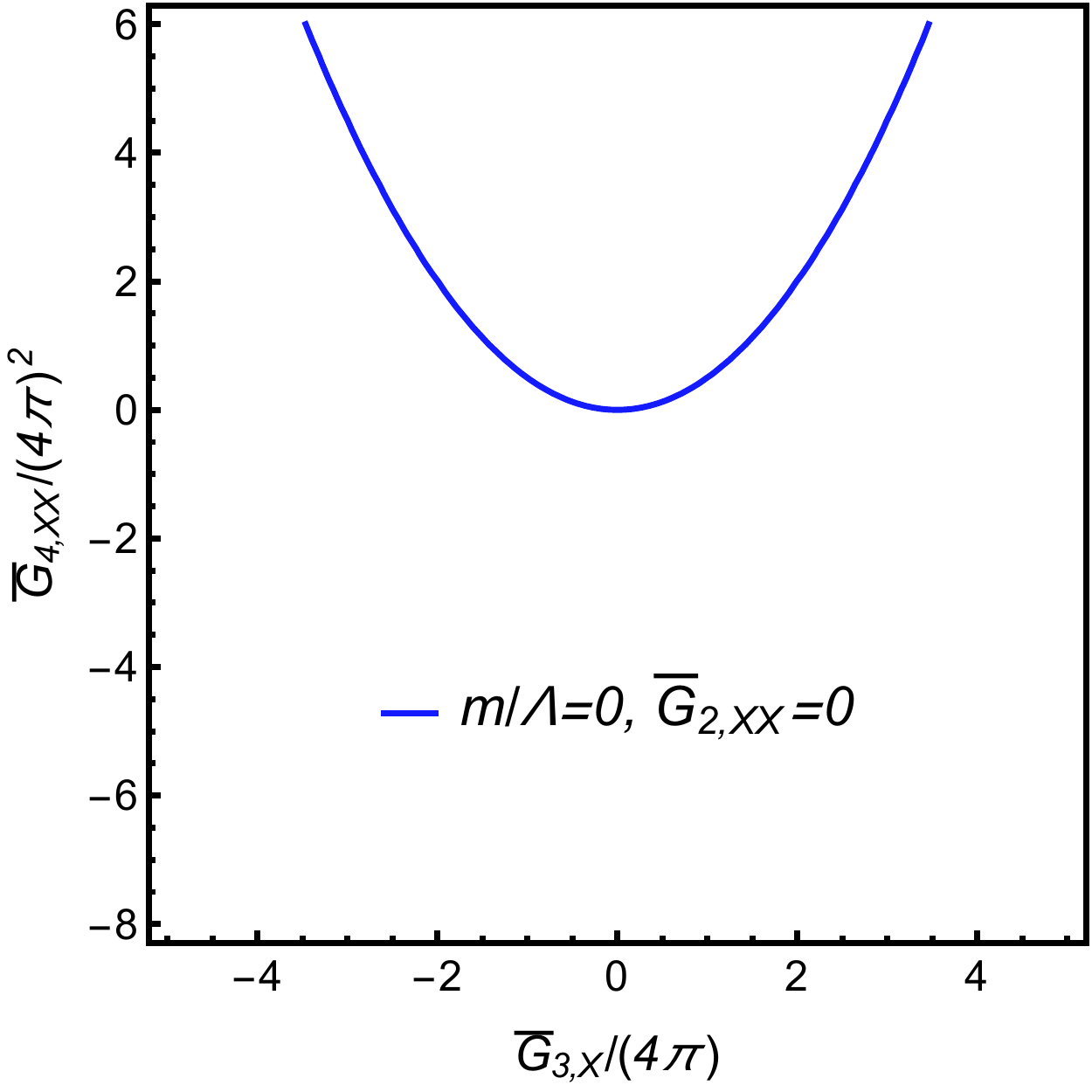}
    \end{subfigure}

    \caption{\label{fig:sshtm0} Fully crossing symmetric positivity bounds on shift symmetric Horndeski theory ($m=0$). The parameters $\bG_{2,XX}$, $\bG_{3,X}$ and ${\bG_{4,XX}}$ can be constrained by these bounds ({\it top left}), and a few cross-sections of the 3D positivity region, which correspond to some more specific models, are also shown ({\it other subfigures}). The 3D blue region is the positivity region of generic shift-symmetric Horndeski theory, the red straight line segment is that of k-essence, the orange plane is that of cubic galileon and kinetic braiding theory, and the green region is that of quartic galileon. }
\end{figure}

To obtain the dispersion relation, we need to compute the $\phi\phi\rightarrow\phi\phi$ scattering amplitude. For simplicity, we shall consider a canonical kinetic term\footnote{For a theory with a non-canonical kinetic term, we require $\bG'_{2,X}>0$ so that there is no ghost instability. Note that the no-ghost condition is essential for the S-matrix to be well-defined and thus for the positivity bounds to be derived. To canonically normalize the kinetic term, we can do a field redefinition $\phi\rightarrow(\bG'_{2,X})^{1/2}\phi\equiv\tilde{\phi}$ such that $\tilde{X}\equiv-(\partial\tilde{\phi})^2/2= X/\bG'_{2,X}$. For example, for the terms $\bG_{2,XX}X^2$ and $\bG_{3,X}X\square\phi$, we have the relations $ \bG_{2,XX}X^2=\bG_{2,XX}\tilde{X}^2/(\bG'_{2,X})^{2}$ and $\bG_{3,X}X\square\phi=\bG_{3,X}\tilde{X}\square\tilde{\phi}/(\bG'_{2,X})^{3/2}$. Thus, we can define $\tilde{\bG}_{2,\tilde{X}\tilde{X}}=\bG_{2,XX}/(\bG'_{2,X})^{2}$ and $\tilde{\bG}_{3,\tilde{X}}=\bG_{3,X}/(\bG'_{2,X})^{3/2}$ to translate our results to the case with a non-canonical kinetic term. That is, for a theory with a non-canonical kinetic term, we can replace ${\bG}_{2,{X}{X}}$ and ${\bG}_{3,{X}}$ in the positivity bounds in this paper with $\bG_{2,XX}/(\bG'_{2,X})^{2}$ and $\bG_{3,X}/(\bG'_{2,X})^{3/2}$ respectively, along with other similar replacements.}
\be
 \bG'_{2,X}=\bG_{2,X}+2\bG_{3,\phi}=1 .
\ee
(In many models, including k-essence and galileon, the existence of a cosmologically viable attractor solution actual requires a kinetic term with the wrong sign. The positivity bounds here do not apply to these models.)
By integration by parts, the relevant interaction vertices for $\phi\phi\rightarrow\phi\phi$ scattering can be extracted from the following Lagrangian terms
\bal\label{app3}
    &   \frac{\bG_{3,X}+3{\bG}_{4,\phi X}}{3\Lambda^3}\delta_{\alpha\beta}^{\mu\nu}\phi\phi_\mu^\alpha\phi_\nu^\beta, \quad \frac{\bG_{2,\phi X}+2\bG_{3,\phi\phi}}{4\Lambda}\phi^2\phi_\mu^\mu, \quad \frac{3\bG_{2,XX}+4\bG_{3,\phi X}}{24\Lambda^4}\phi^\mu\phi_\mu\phi^\nu\phi_\nu, \notag \\
    &\frac{\bG_{3,\phi X}+3{\bG}_{4,\phi\phi  X}}{6\Lambda^4}\delta_{\alpha\beta}^{\mu\nu}\phi^2\phi_\mu^\alpha\phi_\nu^\beta, \quad \frac{3\bG_{4,XX}-2{\bG}_{5,\phi X}}{12\Lambda^6}\delta_{\alpha\beta\gamma}^{\mu\nu\rho}\phi\phi_\mu^\alpha\phi_\nu^\beta\phi_\rho^\gamma.
\eal
where we have defined $\phi_\mu\equiv \partial_\mu\phi$, $\phi_\nu^\mu\equiv \partial^\mu\partial_\nu\phi$ and the generalized Kronecker delta is defined as, for example, $\delta_{\alpha\beta\gamma}^{\mu\nu\gamma}\equiv \delta_\alpha^{[\mu}\delta_\beta^\nu\delta_\gamma^{\rho]}$ with the anti-symmetrization defined without the $1/3!$ factor. The mass squared is given by\footnote{The mass squared $m^2=-\bG_{2,\phi\phi}/\bG'_{2,X}$ for a theory with a non-canonical kinetic term.}
\be
m^2\equiv -\bG_{2,\phi\phi} .
\ee
Note that free of the tachyonic instability is also essential for the positivity bounds to be derived. Then, it is straightforward to calculate the amplitude of $\phi\phi\rightarrow\phi\phi$ in the generic Horndeski theory, which is given by
\begin{equation}\label{app6}
  A^{\rm H}(s,t)=A^{\rm H}_{4}+A^{\rm H}_s + A^{\rm H}_t + A^{\rm H}_u ,
\end{equation}
where
\begin{align}\label{app7}
  A^{\rm H}_4&=\frac{3\bG_{2,XX}+4\bG_{3,\phi X}}{12\Lambda^4}(s^2+t^2+u^2-4m^4) \notag \\
  &+\frac{\bG_{3,\phi X}+3{\bG}_{4,\phi\phi X}}{3\Lambda^4}\left(16m^4-(s^2+t^2+u^2)\right)
  +\frac{3\bG_{4,XX}-2{\bG}_{5,\phi X}}{2\Lambda^6} s t u ,
\end{align}
and
\begin{equation}\label{app8}
  A^{\rm H}_z(s,t)=\frac{1}{m^2-z} \left(\frac{\bG_{2,\phi X}+2\bG_{3,\phi\phi}}{2\Lambda}(z+2m^2) -\frac{\bG_{3,X}+3{\bG}_{4,\phi X}}{2\Lambda^3}z(z-4m^2)  \right)^2 ,
\end{equation}
where $A^{\rm H}_z$ denotes the $z$ channel amplitude with double three-point vertex insertions connected by a propagator. With adjusted power-counting, this amplitude is consistent with the corresponding one in \cite{Melville:2019wyy}.

In the last section, we have computed the positivity bounds on the $c^{2,0}(m)$ and $c^{2,1}(m)$ coefficients. For generic Horndeski theory, these two coefficients are given by
\begin{align}\label{app10}
  { \Lambda^4} c^{2,0}(m)&=\frac{1}{2}(\bG_{2,XX}-4{\bG}_{4,\phi\phi X}+6(\bG_{2,\phi X}+2\bG_{3,\phi\phi}){\bG}_{4,\phi  X}+2\bG_{3,X}(\bG_{2,\phi X}+2\bG_{3,\phi\phi}))  \notag \\
  &\qquad+\frac{1}{6}\(\frac{m}{\Lambda}\)^2(9\bG_{3,X}^2  +81 {\bG}_{4,\phi X}^2 -12 {\bG}_{4,XX} +8{\bG}_{5,\phi X}+54\bG_{3,X} {\bG}_{4,\phi X}), \\ \label{app11}
  { \Lambda^6} c^{2,1}(m)&=\frac{1}{4}(3\bG_{3,X}^2+18\bG_{3,X}{\bG}_{4,\phi X}+27{\bG}_{4,\phi X}^2-6{\bG}_{4,XX}+4{\bG}_{5,\phi X}).
\end{align}
Therefore, for generic Horndeski theory, the two-sided positivity bounds that make use of the full crossing symmetry are given by
\begin{equation}\label{genb}
  \begin{cases}
    &\Lambda^4 c^{2,0}(m)\,\bar{c}^{2,1}_{\rm min}(m)\leq \Lambda^6 c^{2,1}(m)\leq \Lambda^4 c^{2,0}(m)\,\bar{c}^{2,1}_{\rm max}(m) \\
    &0\leq \Lambda^4 c^{2,0}(m)\leq \Lambda^4 c^{2,0}_{\rm max} (m).
  \end{cases}
\end{equation}
where $c^{2,0}_{\rm max} (m)$ is the upper bound that can be extracted from Figure \ref{fig:1} and $\bar{c}^{2,1}_{\rm min}(m)$ and $\bar{c}^{2,1}_{\rm max}(m)$ are respectively the lower and upper bounds that can be extracted from Figure \ref{fig:c21}.

\subsection{(Softly broken) shift symmetry}

To illustrate the constraining power of the two-sided positivity bounds, in this subsection, we apply them to a few popular Horndeski models that are shift-symmetric $\phi\to \phi+c$, as well as their counterparts where the shift symmetry is softly broken by a mass term. In these models, the functions $G_i(\phi,X)$ are reduced to $G_i(X)$, and they still constitute a large class of scalar-tensor theories widely used in the early or late time cosmology. Nevertheless, the parameter space that can be constrained by the positivity bounds now shrinks to the space spanned by $\bG_{2,XX}$, $\bG_{3,X}$ and ${\bG}_{4,XX}$ only, as we focus on the tree level scattering. The constraints Eq.(\ref{genb}) can now be written as
\begin{eqnarray}\label{sshtcon}
\left\{
\begin{aligned}
  &\frac{(3\bG_{3,X}^2-6{\bG}_{4,XX})}{4}-\left(\frac{\bG_{2,XX}}{2}+\frac{1}{2}\(\frac{m}{\Lambda}\)^2(3\bG_{3,X}^2-4 {\bG}_{4,XX})\right)\bar{c}^{2,1}_{\rm min}(m) \geq 0  \\
  &\frac{(3\bG_{3,X}^2-6{\bG}_{4,XX})}{4}-\left(\frac{\bG_{2,XX}}{2}+\frac{1}{2}\(\frac{m}{\Lambda}\)^2(3\bG_{3,X}^2-4 {\bG}_{4,XX})\right)\bar{c}^{2,1}_{\rm max}(m) \leq 0  \\
  & 0\leq\frac{\bG_{2,XX}}{2}+\frac{1}{2}\(\frac{m}{\Lambda}\)^2(3\bG_{3,X}^2-4 {\bG}_{4,XX}) \leq \Lambda^4 c^{2,0}_{\rm max}(m).
\end{aligned}
\right.
\end{eqnarray}
where again $c^{2,0}_{\rm max} (m)$ is the upper bound that can be extracted from Figure \ref{fig:1} and $\bar{c}^{2,1}_{\rm min}(m)$ and $\bar{c}^{2,1}_{\rm max}(m)$ are respectively the lower and upper bounds extractable from Figure \ref{fig:c21}.

\begin{figure}[ht]
       \centering
    \begin{subfigure}{0.48\linewidth}
    \centering
        \includegraphics[width=0.9\linewidth]{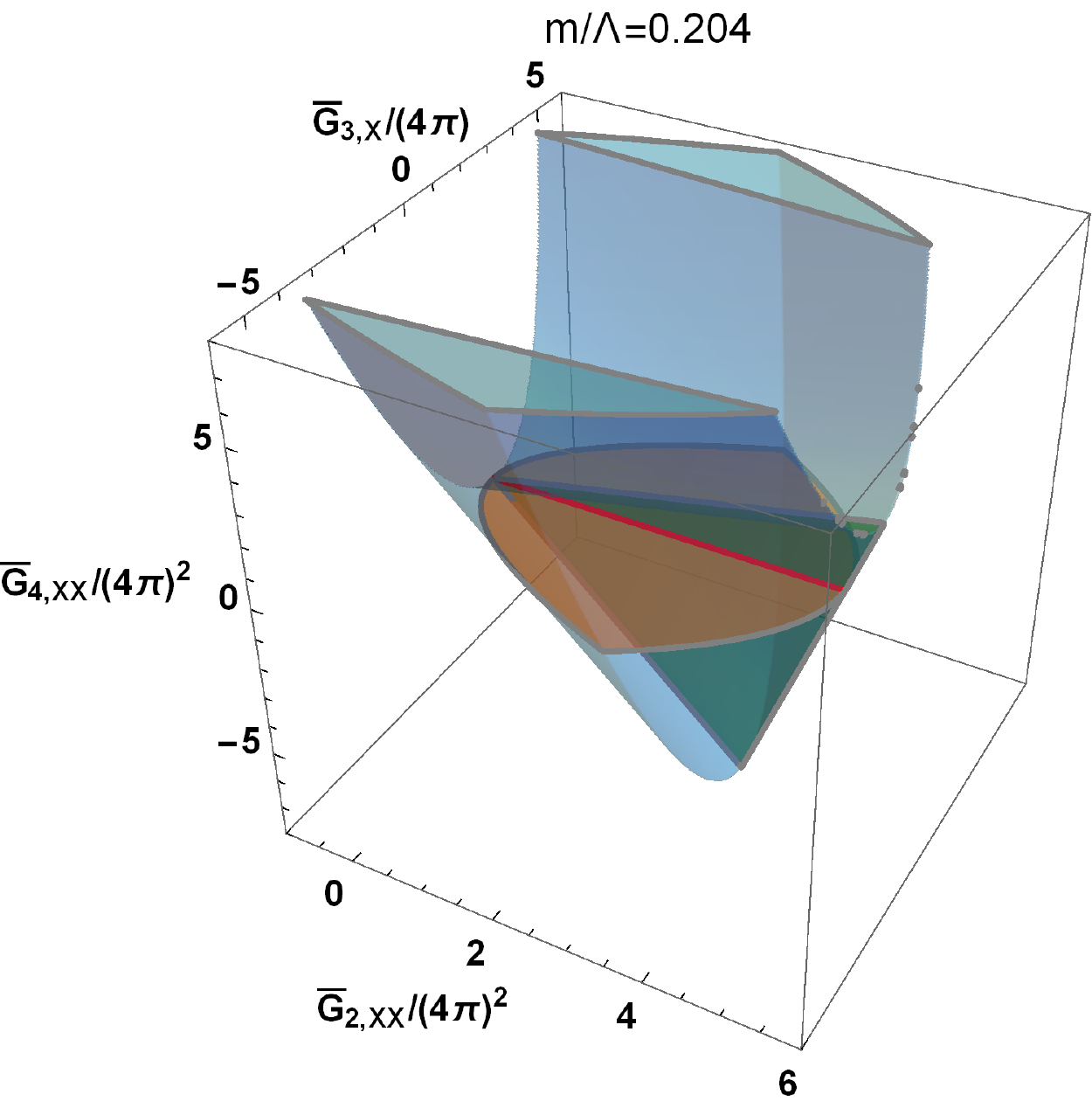}
    \end{subfigure}
     \centering
    ~\begin{subfigure}{0.485\linewidth}
    \centering
        \includegraphics[width=0.8\linewidth]{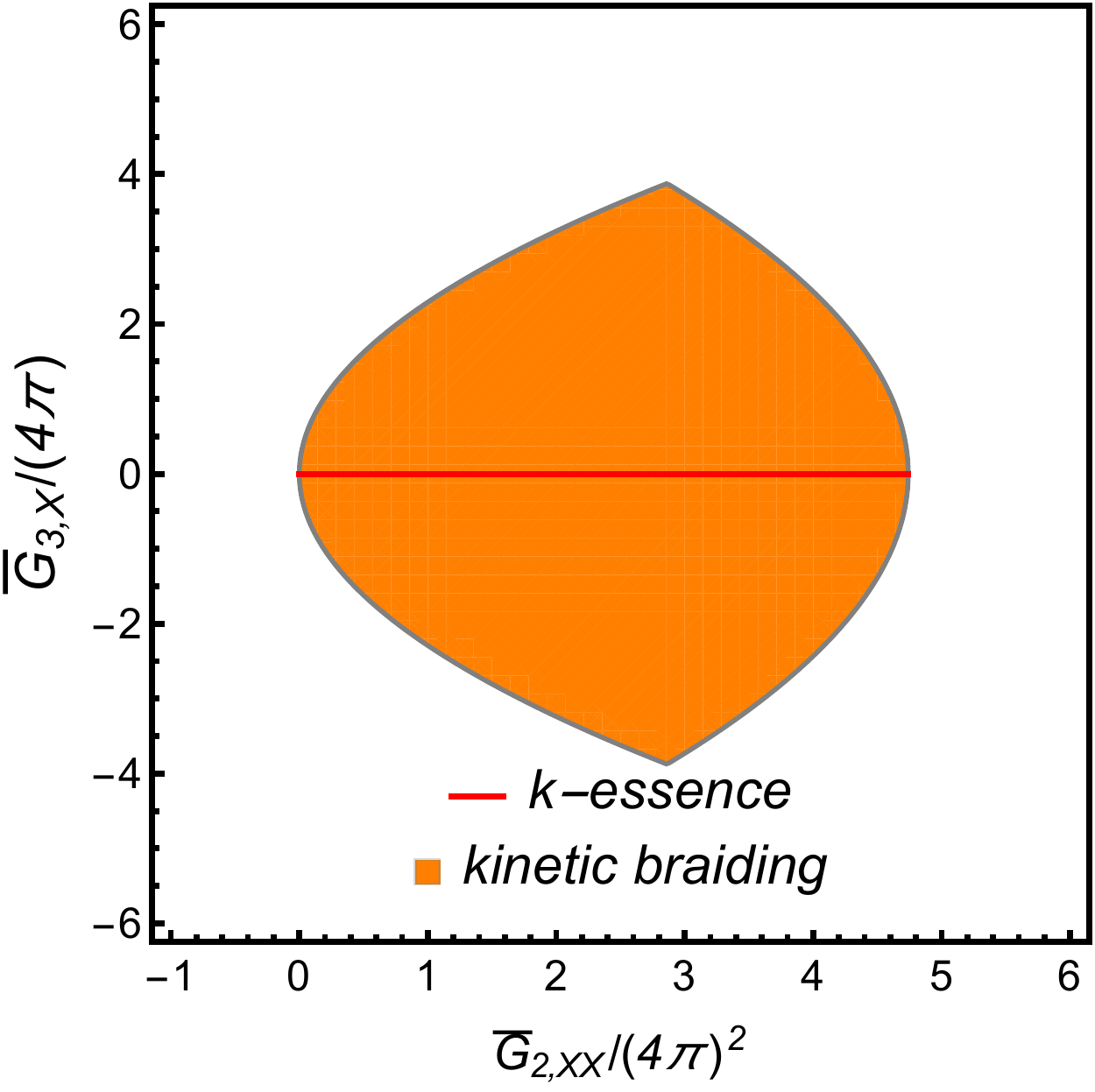}
    \end{subfigure}

     \centering
    \begin{subfigure}{0.48\linewidth}
    \centering
        \includegraphics[width=0.83\linewidth]{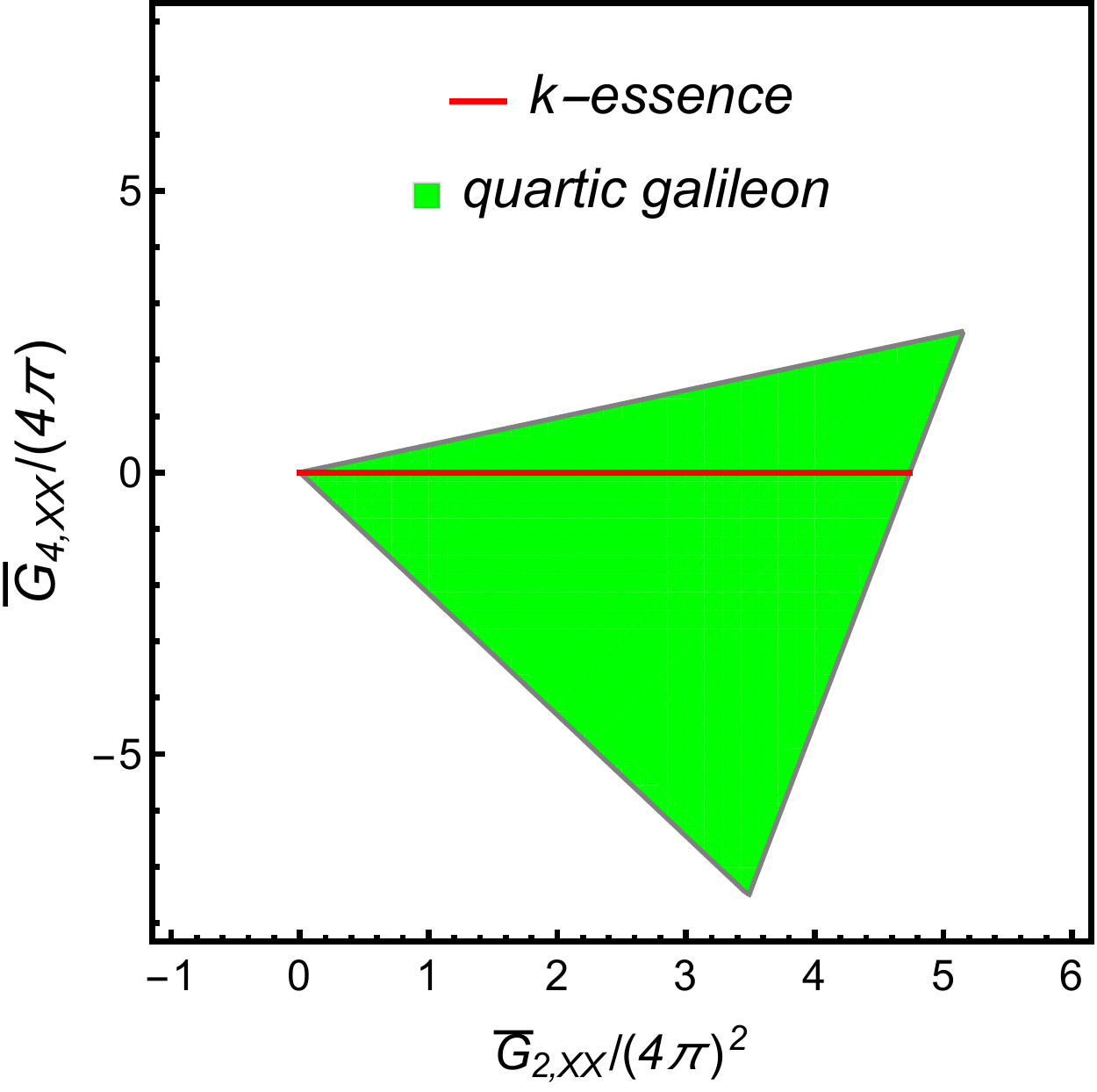}
    \end{subfigure}
     \centering
    \begin{subfigure}{0.48\linewidth}
    \centering
        \includegraphics[width=0.835\linewidth]{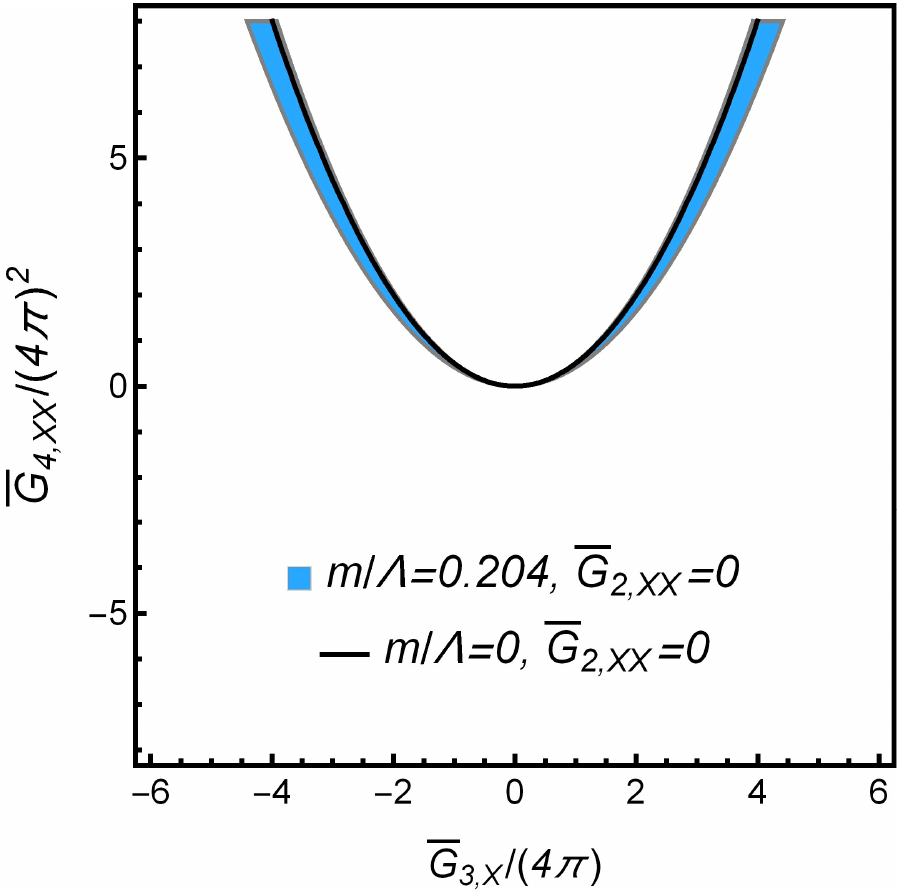}
    \end{subfigure}

    \caption{\label{fig:sshtm1} Fully crossing symmetric positivity bounds on softly broken shift symmetric Horndeski theory with mass $m/\Li=0.204$. The setup is the same as Figure \ref{fig:sshtm0} except for a nonzero mass $m/\Li=0.204$. In the bottom right subfigure, the black solid line corresponds to the massless case in Figure \ref{fig:sshtm0}. The massive case enlarges the positivity regions. }
\end{figure}

\begin{figure}[htbp]

       \centering
    \begin{subfigure}{0.48\linewidth}
    \centering
        \includegraphics[width=0.9\linewidth]{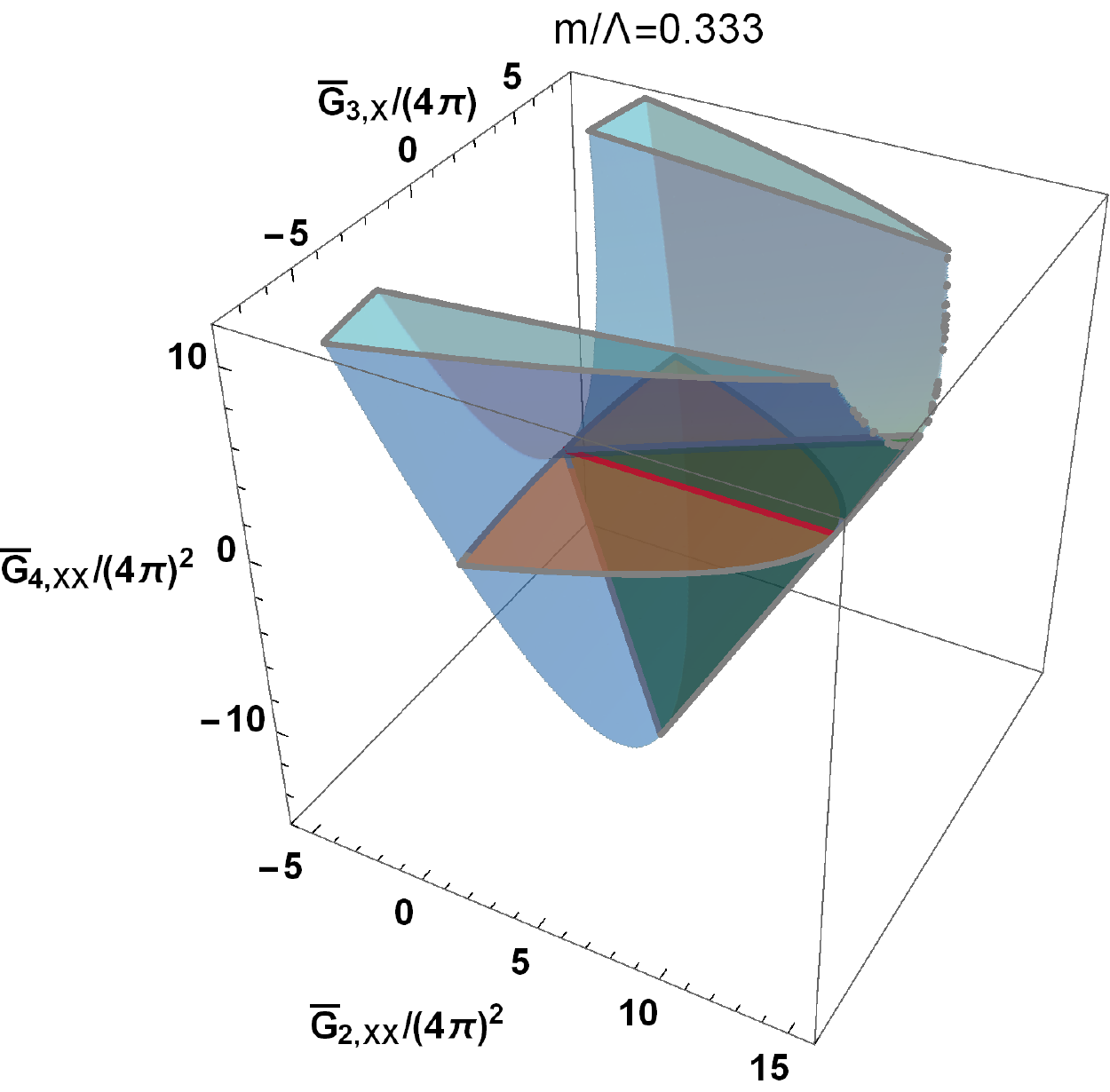}
    \end{subfigure}
     \centering
    ~\begin{subfigure}{0.485\linewidth}
    \centering
        \includegraphics[width=0.8\linewidth]{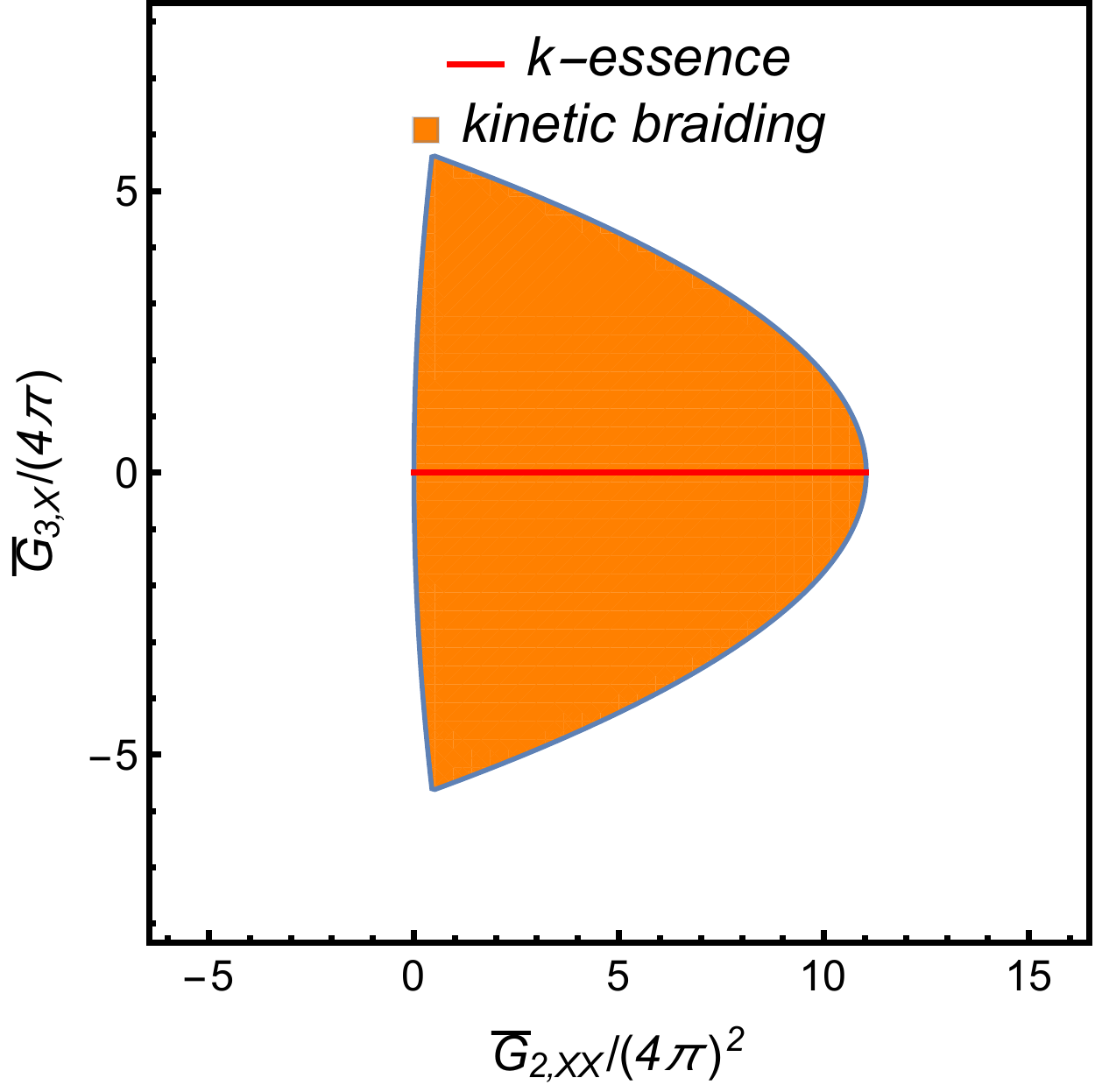}
    \end{subfigure}

     \centering
    \begin{subfigure}{0.48\linewidth}
    \centering
        \includegraphics[width=0.83\linewidth]{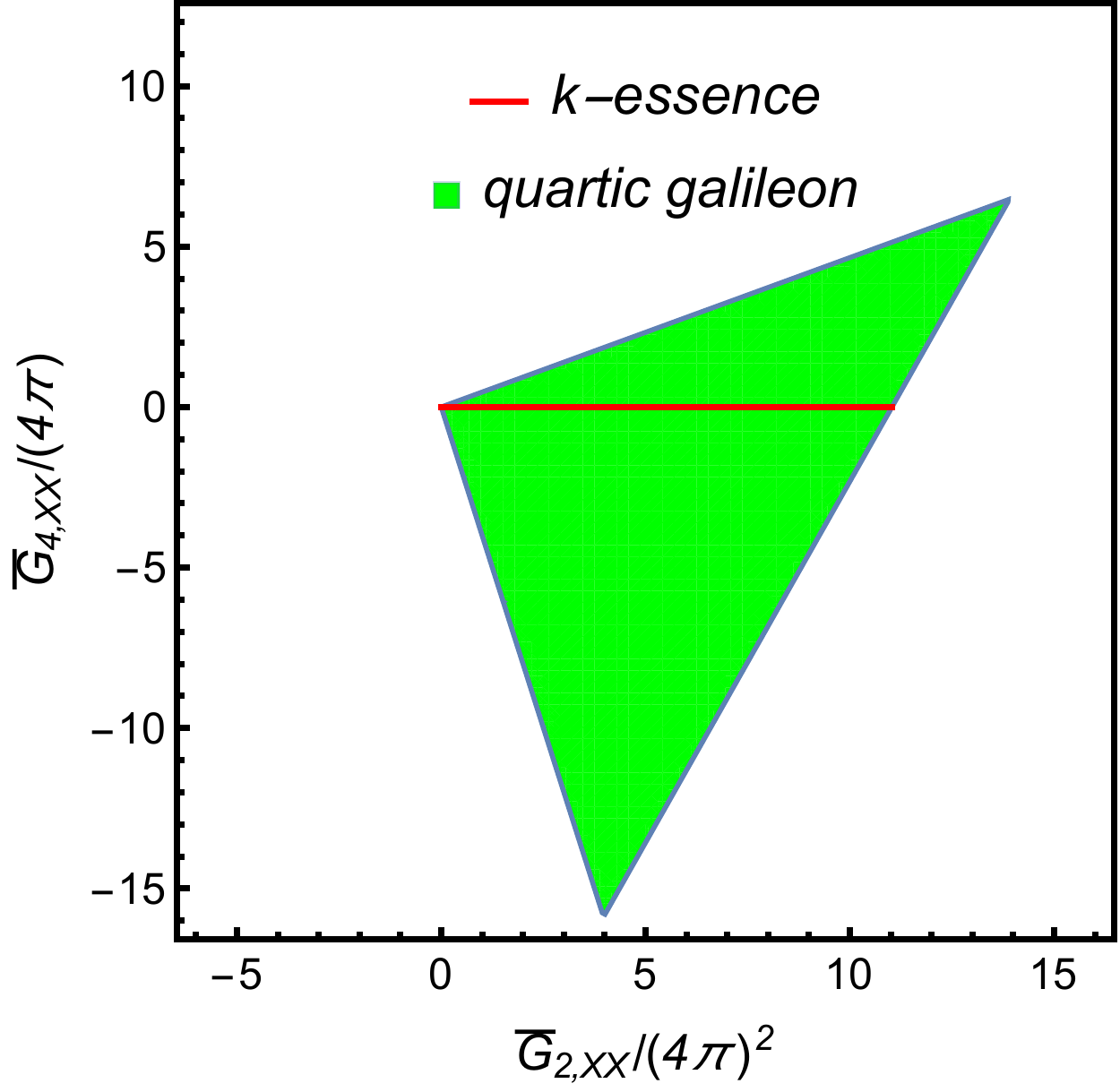}
    \end{subfigure}
     \centering
    \begin{subfigure}{0.48\linewidth}
    \centering
        \includegraphics[width=0.845\linewidth]{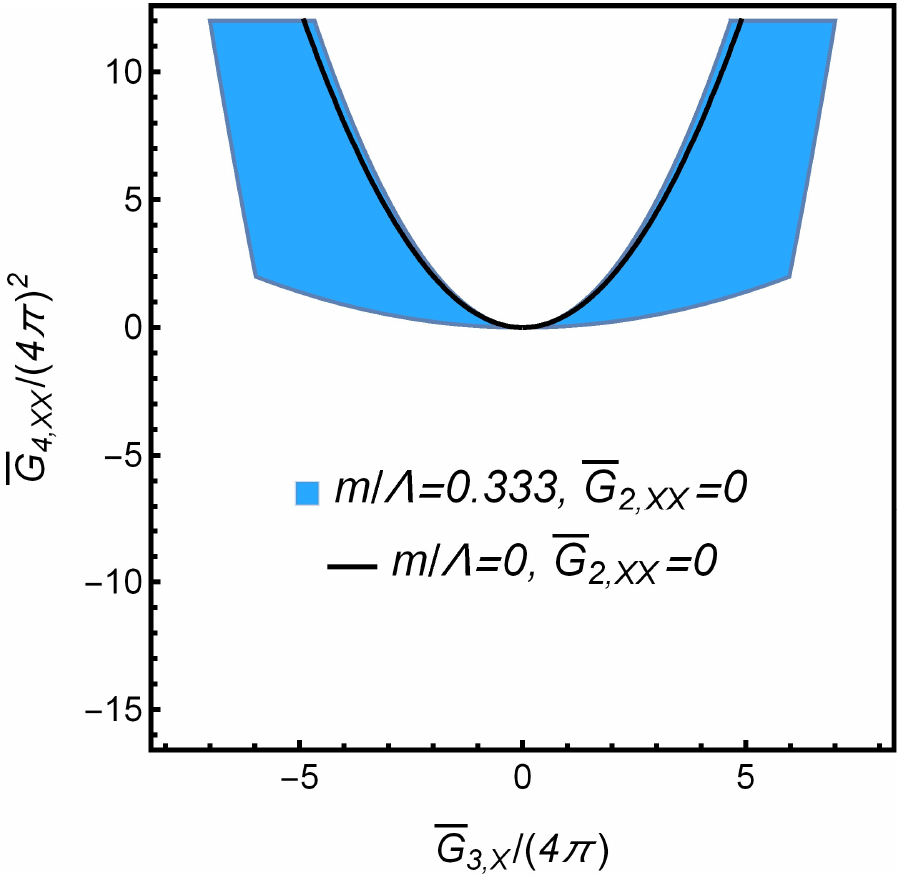}
    \end{subfigure}

    \caption{\label{fig:sshtm2}  Same as Figure \ref{fig:sshtm1} except for a different mass $m/\Li=0.333$. }
\end{figure}

In Figure \ref{fig:sshtm0}, we plot the positivity region for shift symmetric Horndeski theory. The region is not convex simply because we plot it using the $\bG_{3,X}$ parameter, while the constraints Eq.(\ref{sshtcon}) are linear in terms of $\bG_{3,X}^2$. This is also why the cross sections in the $\bG_{3,X}$ and ${\bG}_{4,XX}$ planes are bounded parabolic curves. In Figure \ref{fig:sshtm0}, we have also plotted a few cross-sections of the 3D positivity region, which correspond to a few specific shift symmetric models that will be discussed in the following subsections. It is interesting to see how these specific models are distributed in the generic positivity region. The presence of a mass term $-m^2 \phi^2/2$ softly breaks shift-symmetry, which enlarges the positivity region, as we see in Figure \ref{fig:sshtm1} and Figure \ref{fig:sshtm2}. (A mass term might be generated through a spontaneous symmetry breaking mechanism due to some non-trivial scalar background. The scale of the background dynamics is typically assumed to be smaller than the cutoff, so the generated mass might be well below the cutoff.)  According to the naive dimensional analysis (see around \eref{app1}), $\bG_{3,X}$ is typically around $4\pi$ and ${\bG}_{4,XX}$ is typically around $(4\pi)^2$, which is consistent with our numerical results in Figure \ref{fig:sshtm0}, Figure \ref{fig:sshtm1} and Figure \ref{fig:sshtm2}.

\begin{figure}
  \centering
    \begin{subfigure}{0.49\linewidth}
    \centering
        \includegraphics[width=1\linewidth]{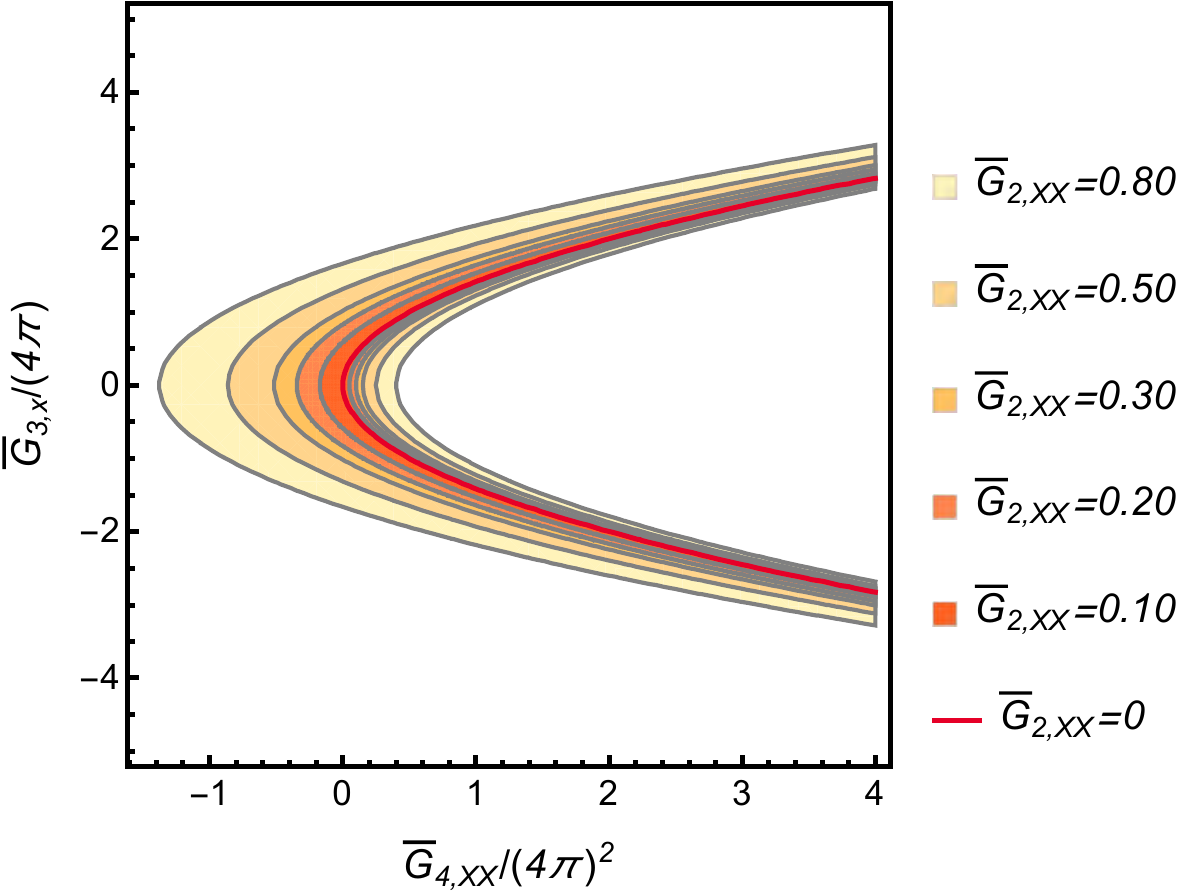}
    \end{subfigure}
    \centering
    \begin{subfigure}{0.49\linewidth}
    \centering
        \includegraphics[width=1\linewidth]{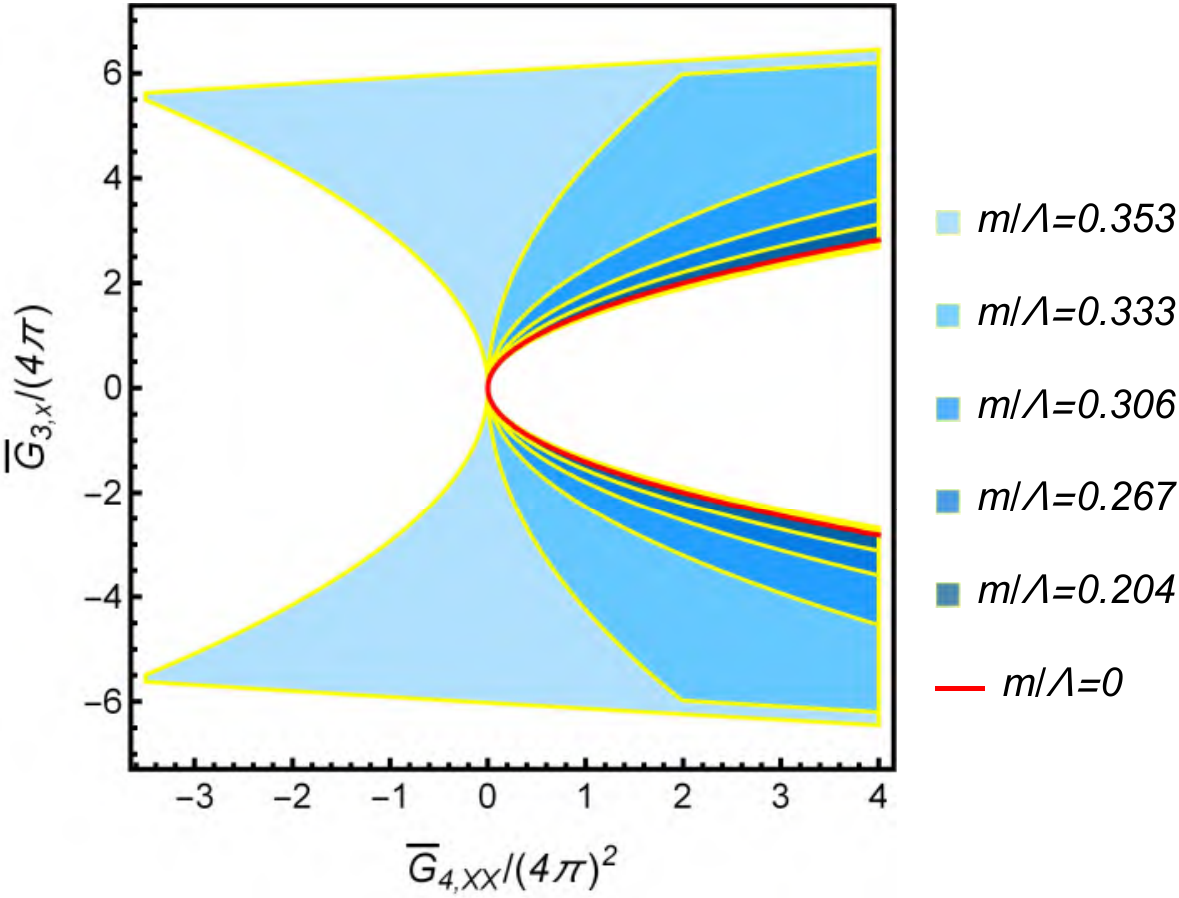}
    \end{subfigure}
    \centering
  \caption{ ({\it Left}) Two-sided positivity bounds in the $\bG_{3,X}$-${\bG}_{4,XX}$ plane for various $\bG_{2,XX}$, the mass being set to zero. ({\it Right})  Two-sided positivity bounds for various masses, $\bG_{2,XX}$ being set to zero. }
  \label{fig:sX}
\end{figure}

The lowest order forward positivity bound tells us that the coefficient in front of the $s^2$ term in the amplitude needs to be positive.  As we see in the bounds Eq.(\ref{sshtcon}), this can be achieved by either a nonzero mass ($m\neq 0$) or a $X^2$ term in the Lagrangian ($\bG_{2,XX}\neq 0$). The pure Galilleon theory does not have the $s^2$ term and so marginally violates positivity bounds. The effects of adding the mass term and adding the $X^2$ term are different in extending the positivity region of the model. For the case of a nonzero $\bG_{2,XX}$ (and $m=0$), as we see in Figure \ref{fig:sX}, it extends the positivity region along the horizontal direction, fattening the parabola, especially the bottom part of the parabola. To be more precise, for a nonzero $\bG_{2,XX}$, the positivity region is bounded by two non-intersecting parabolas, with the gap between them proportional to the value of $\bG_{2,XX}$.
%For the case of $\bG_{2,XX}=0$, it shrinks to a parabola, marginally violating the positivity bounds.
On the other hand, for the case of a nonzero mass (and $\bG_{2,XX}=0$), the allowed region is bounded by a pair of parabolas that intersect at the origin on the $\bG_{3,X}$-${\bG}_{4,XX}$ plane and two almost horizontal straight lines. Note that the mass parameter $m$ only appears as $(m/\Lambda)^2$ in the positivity bounds, which explains how the positivity region scales with $m$. As the mass increases, the intersecting angle between the two parabolas at the origin also increases. Unlike the non-vanishing $X^2$ case, the positivity region now is larger away from the origin but quite limited near it. For either small $\bG_{2,XX}$ or $m$, we have a very narrow band of region permitted by the positivity bounds, which is consistent with the qualitative result of Ref \cite{Tolley:2020gtv}.

It is worth noting that dimensionally the terms involving $\bG_{2,XX}$ are suppressed by $\mathcal{O}(1/\Lambda^4)$, while the terms involving $\bG_{3,X}$ and ${\bG}_{4,XX}$ are suppressed by $\mathcal{O}(1/\Lambda^3)$ and $\mathcal{O}(1/\Lambda^6)$ respectively. So, from \eref{sshtcon},  we see that adding a $\bG_{2,XX}$ term is much more effective to alter the results of the positivity bounds than letting the scalar be massive.

In the following subsections, we shall introduce a few specific models that have already been plotted in Figure \ref{fig:sshtm0},  Figure \ref{fig:sshtm1} and Figure \ref{fig:sshtm2}, and discuss the implications of the positivity bounds for these models.

\subsubsection{K-essence}
\label{sec:secks}

The k-essence models \cite{Armendariz-Picon:1999hyi, Armendariz-Picon:2000nqq, Armendariz-Picon:2000ulo} are a class of scalar-tensor theories with kinetic interactions, which can be used to model the accelerated universe in either the early or later cosmic epoch, thanks to the nonlinear kinetic structure of the models that allow for attractor solutions to generally exist.
The k-essence Lagrangian is given by
\begin{equation}\label{eq5-11}
  \mathcal{L}_{\text{k-essence}}=\frac{M_P^2R}{2}+M_P^2\Lambda^2G_2(\phi,X) ,
\end{equation}
where $G_2(\phi,X)$ is a general function of $\phi$ and $X=-\pd_\mu \phi \pd^\mu \phi/(2\Lambda^4)$. Here we only consider softly broken shift symmetric cases, i.e., $M_P^2\Lambda^2G_2(\phi,X)=-\f12 m^2\phi^2 + P(X)$, which are often referred to as $P(X)$ theory in the absence of the mass term. The DBI model \cite{Alishahiha:2004eh} is a special case of $P(X)$, which satisfies positivity bounds, while it is well-known that the anti-DBI model violates causality \cite{Tolley:2009fg} and thus positivity bounds. With $\bG_{3,X}$ and $\bG_{4,XX}$ vanishing, the positivity bounds Eq.(\ref{sshtcon}) reduce to
\begin{equation}\label{keb}
  0\leq\frac{\bG_{2,XX}}{2}\leq \Lambda^4 c^{2,0}_{\rm max}(m) .
\end{equation}
Therefore, in Figure \ref{fig:sshtm0}, the positivity region of this model is a red straight line segment for either the massless or massive case, and $\bG_{2,XX}$ is always positive. This is highly degenerate from the 3D point of view, but this line segment is well within the 3D positivity region. As the mass of the scalar increases, the allowed range for $\bG_{2,XX}$ also enlarges, as shown in Figure \ref{fig:sshtm1} and Figure \ref{fig:sshtm2}.

\subsubsection{Kinetic gravity braiding}

Kinetic gravity braiding models are a class of scalar-tensor theories that contain second derivatives of scalar in the Lagrangian without producing extra ghost degrees of freedom. This kind of terms introduce  kinetic mixings between the scalar and tensor modes, schematically taking the form $(\cdots)\partial g\partial \phi$ ($g$ being the metric). This kind of braiding in the Lagrangian will render the energy-momentum tensor to differ from that of a perfect fluid, giving rise to a range of novel features \cite{Deffayet:2010qz}.  Kinetic braiding theory is basically a subclass of Horndeski theory given by the following Lagrangian
\begin{equation}\label{eq5-14}
  \mathcal{L}_{\rm KB}=\frac{M_P^2}{2}R  +M_P^2\Lambda^2G_2(X) + \frac{M_P^2}{\Lambda} G_3(X)\square\phi  - \f12 m^2\phi^2 ,
\end{equation}
where we have assumed shift symmetry for $\phi$ except for the mass term. In this theory, there can exist attractor solutions which represent accelerated expansion with constant scalar kinetic energy \cite{Deffayet:2010qz} (see also \cite{Kobayashi:2010cm, Kobayashi:2011nu} for the inflationary scenario). For this theory, the $G_4$ term is absent, so the fully crossing symmetric positivity constraints are given by
\begin{eqnarray}\label{kbcons}
\left\{
\begin{aligned}
  &\frac{3\bG_{3,X}^2}{4}-\left(\frac{\bG_{2,XX}}{2}+\frac{3}{2}\(\frac{m}{\Lambda}\)^2\bG_{3,X}^2\right)\bar{c}^{2,1}_{\rm min}(m) \geq 0  \\
  &\frac{3\bG_{3,X}^2}{4}-\left(\frac{\bG_{2,XX}}{2}+\frac{3}{2}\(\frac{m}{\Lambda}\)^2\bG_{3,X}^2\right)\bar{c}^{2,1}_{\rm max}(m) \leq 0  \\
  & 0\leq\frac{\bG_{2,XX}}{2}+\frac{3}{2}\(\frac{m}{\Lambda}\)^2\bG_{3,X}^2 \leq \Lambda^4 c^{2,0}_{\rm max}(m).
\end{aligned}
\right.
\end{eqnarray}

For the massless case, the positivity bounds are a 2D region that is enclosed by a parabola and a straight line segment on the $G_{4}(X)=0$ plane, as shown in Figure \ref{fig:sshtm0}. As expected, the k-essence red line segment is embedded within this region. For the massive case, the right boundary grows into another parabola, which becomes increasingly curved as the mass increases. The left boundary, on the other hand, becomes flatter, as shown in Figure \ref{fig:sshtm1} and Figure \ref{fig:sshtm2}. All in all, as expected, the positivity region grows as the mass increases.

In Figure \ref{fig:sshtm0} and Figure \ref{fig:sshtm1} and Figure \ref{fig:sshtm2}, we see that the boundary of the positivity region has two kinks that are symmetric with respect to $\bG_{3,X}$. Kinks at the boundary can sometimes be identified with theories with special features. However, the kinks here arise simply because we are using two different methods to derive the bounds on $c^{2,0}$ and on the rest coefficients respectively. For the massless case, we have a sharp, vertical right boundary because in that case $\bG_{2,XX}$ has a sharp upper bound  $2\Lambda^4 c^{2,0}_{\rm max}(0)$, as one can see in the bounds (\ref{kbcons}). For the massive case, the right boundary becomes a parabola because the mass dependence introduces a $\bG_{3,X}^2$ term for the upper bound on $\bG_{2,XX}$.

\subsubsection{Galileon}

We then consider modified galileon models. Galileon theory finds its applications in many gravitational models. The cubic galileon was discovered in the decoupling limit of the DGP model \cite{Nicolis:2004qq}, and the galileon there can be identified as the brane bending mode \cite{Luty:2003vm}. Generic galileon theory, which is defined by a scalar with generalized symmetry $\phi\rightarrow\phi+v_\mu x^\mu+c$ ($v_\mu, c$ being constant), was then proposed as an effective IR modification of gravity that nonlinearly interpolates between negligible deviation in small scales and sizable modifications in large lengths \cite{Nicolis:2008in}. Generic galileon theory also inspired the construction of dRGT massive gravity \cite{deRham:2009rm}, which solves the long standing Boulware-Deser ghost problem. In the dRGT model, the galileon is essentially the helicity-0 mode of the massive graviton. On the other hand, based on galileon theory, generalized galileon was proposed as a generalization of scalar-tensor theory with at most second order field equations \cite{Deffayet:2009mn}, re-discovering Horndeski theory.

In terms of positivity bounds, the cubic galileon also inspired the proposal of the lowest order forward limit bound \cite{Adams:2006sv}. Because the strict galileon symmetry explicitly forbids the $X^2$ term, pure galileon theory lies on the boundary of the positivity bounds, that is, it is ruled out by positivity marginally. However, this can be amended by adding a mass term or a $X^2$ term. Indeed, the positivity bounds from Ref \cite{Bellazzini:2017fep} have shown that these terms must be not too small and at most one-loop suppressed compared to the galileon terms. Adding the mass term is particularly natural to do because away from the decoupling limit the galileon mode in massive gravity models becomes massive \cite{deRham:2016nuf}. Also, due to the Galileon non-renormalization theorem, adding the mass term does not give rise to extra terms that break the galileon symmetry under loop corrections \cite{deRham:2017imi}. Positivity bounds without the full crossing symmetry have been used to constrain massive galileon \cite{deRham:2017imi}, which generally has a viable positivity region. In the limit of a small mass term, it has been argued that fully crossing symmetric positivity bounds force the cutoff to be parametrically close to the mass scale \cite{Tolley:2020gtv}. Here, we shall explicitly chart the mass and $X^2$ dependence in these slightly modified galileon theories, and examine whether adding these terms can alleviate the tension between the positivity bounds and the observations, if the galileon models are used to describe the late time cosmic evolution.

Since we are constraining the coefficients in the decoupling limit, the following results are applicable generically for models with the galileon modes, whose avartars include dRGT massive gravity, covariant galileon and so on. With the $X^2$ and mass terms, the Lagrangian is given by
\bal\label{gallag}
\mc{L}^{\rm Gal} &= - \frac{1}{2} (\partial \phi)^2  + \frac{g_k}{4}(\partial\phi)^4 - \frac{m^2}{2}\phi^2 + \frac{g_3}{3!\Lambda^3}\phi\left((\square\phi)^2-(\partial_\mu\partial_\nu\phi)^2\right)
\nn
&~~~ + \frac{g_4}{4!\Lambda^6}\phi\left((\square\phi)^3-3\square\phi(\partial_\mu\partial_\nu\phi)^2+2(\partial_\mu\partial_\nu\phi)^3 \right) + \cdots
\eal
where $g_k, g_2, g_3,...$ are constants. In terms of the Horndeski $G_i$ functions, we have
$M_P^2\Lambda^2 G_2=X+g_k X^2, ~M_P^2 G_3/\Lambda=g_3X/2, ~ M_P^2{G}_4=g_4X^2/12$. The fully crossing symmetric positivity bounds now reduce to
\begin{eqnarray}\label{galcons}
\left\{
\begin{aligned}
  &\frac{3g_3^2-4g_4}{16}-\( g_k +\frac{1}{24}\( \frac{m}{\Lambda}\)^2(9g_3^2-8g_4)\)\bar{c}_{\rm min}^{2,1}(m) \geq 0 \\
  &\frac{3g_3^2-4g_4}{16}-\( g_k +\frac{1}{24}\( \frac{m}{\Lambda}\)^2(9g_3^2-8g_4)\)\bar{c}_{\rm max}^{2,1}(m) \leq 0  \\
  & 0\leq g_k+\frac{1}{24}\(\frac{m}{\Lambda}\)^2(9g_3^2-8g_4) \leq \Lambda^4 c^{2,0}_{\rm max}(m).
\end{aligned}
\right.
\end{eqnarray}
The positivity bounds on the modified galileon model can be found in Figure \ref{fig:sshtm0}, Figure \ref{fig:sshtm1}, Figure \ref{fig:sshtm2} and Figure \ref{fig:sX}, many of whose features have been discussed previously. We reiterate that the most important feature is that a small mass (relative to the cutoff) gives rise to a very narrow strip of positivity region, suggesting that some levels of fine-tuning are required. A relative large mass allows for a wider positivity region, but that reduces the applicability range of the theory as an EFT, since the hierarchy between the mass and the cutoff then becomes more restricted.

\section{Bounds on beyond Horndeski theories}
\label{sec:sec5}

Horndeski theory is the most general scalar-tensor theory with second-order equations of motion. Beyond Horndeski theories have equations of motion with more than second derivatives, and yet some of them are free of Ostragradski ghosts \cite{Zumalacarregui:2013pma, Gleyzes:2014dya, Gleyzes:2014qga}. These theories along with part of Horndeski theory can evade Ostragradski's theorem because they are degenerate systems. In particular, the kinetic matrix of Horndeski theory is degenerate within the scalar sector, while beyond Horndeski theories allow for degeneracies between the scalar and the metric in the kinetic matrix \cite{Langlois:2015cwa}. More general beyond Horndeski models have been uncovered along this line of reasoning, which are known as degenerate higher order scalar tensor (DHOST) theories \cite{Langlois:2015cwa}.  Nevertheless, these beyond Horndeski models can be related to Horndeski theory by redefining the metric with a combination of conformal and disformal transformations \cite{Gleyzes:2014qga, Crisostomi:2016czh}. In the presence of matter sources, the field redefinition will induce nontrivial couplings. Despite this, it is sometimes more convenient or beneficial to use the beyond Horndeski models directly, rather than employ their Horndeski counterparts. In this section, we shall apply our positivity bounds directly on the DHOST models.

A popular class of DHOST models are the one proposed in the original paper  \cite{Langlois:2015cwa} where only quadratic terms of the second-order scalar derivatives and a non-minimally coupled Ricci scalar are included:
\begin{equation}\label{dhost}
\mathcal{L}_{\rm d}=  \sqrt{-g}\(  G(\phi, X) R -\f{1}2 (\nd \phi)^2 -\frac{1}{2}m^2\phi^2 + \sum_{i=1}^5 \mathcal{L}_i \)
\end{equation}
with
\bal
\mathcal{L}^{\rm d}_1 &= \frac{1}{\Lambda^2} A_1(\phi, X) \nabla_\mu\nabla_\nu\phi \nabla^\mu\nabla^\nu\phi, \\
\mathcal{L}^{\rm d}_2 &=\frac{1}{\Lambda^2} A_2(\phi, X)(\square \phi)^2, \\
\mathcal{L}^{\rm d}_3 &=\frac{1}{\Lambda^6} A_3(\phi, X)(\square \phi) \nabla_\mu\phi \nabla^\mu\nabla^\nu\phi \nabla_\nu\phi, \\
\mathcal{L}^{\rm d}_4 &=\frac{1}{\Lambda^6} A_4(\phi, X) \nabla_\mu\phi \nabla^\mu\nabla^\rho\phi \nabla_\rho\nabla_\nu\phi \nabla^\nu\phi, \\
\mathcal{L}^{\rm d}_5 &=\frac{1}{\Lambda^{10}} A_5(\phi, X)\left(\nabla^\mu\phi \nabla_\mu\nabla_\nu\phi \nabla^\nu\phi\right)^2 ,
\eal
where $G$ and $A_i$ are functions of $\phi$ and $X\equiv -(\nabla \phi)^2/(2\Lambda^4)$ that need to satisfy some degeneracy conditions for a few different classes \cite{Langlois:2015cwa}. Note that here we have chosen $G$ to be of mass dimension 2, rather than dimensionless as we have done for other free functions, to simplify the equations later. We shall explicitly list these degeneracy conditions for some classes shortly. We have added a kinetic term for the scalar, to avoid strong coupling problems, and also its mass term\footnote{We assume that $\bar A_1=\bar A_2=0$ so that there are no quadratic scalar terms of form $(\Box\phi)^2$ coming from $\mathcal{L}^{\rm d}_1$ and $\mathcal{L}^{\rm d}_2$. This is without loss of generality because one can always redefine $\phi\to \phi +(...)\Box\phi$ to eliminate these terms, which adjusts the other coefficients. We can assume that it has already been done for our Lagrangian.}, which clearly does not affect the DHOST cancelling between higher derivative terms. Our strategy is to first taylor-expand $G$ and $A_i$ around the background values $\phi_0=0$ and $X_0=0$, denoting their derivatives as  $\bar G_{,\phi X}=\pd^2 G(\phi_0,X_0)/\pd(\phi /\Lambda) \pd X$, $\bar A_i {}_{,X} =\pd A_i(\phi_0,X_0)/\pd X$, etc., similar to \eref{eq5-gi}, and then to directly derive the bounds on these derivatives in the decoupling limit. The decoupling limit can again be effected by sending $M_P\to \infty$ and keeping $\Lambda$ fixed. After that, we impose the degeneracy conditions of the DHOST models. Again, we see that the vertices involving the graviton are suppressed. Specifically, since we have set $\bar A_1=\bar A_2=0$, the relevant terms for the $\phi\phi\to\phi\phi$ scattering are schematically $\bG_{,\phi\phi}\phi\phi\partial^2h/(M_P \Lambda^2)$ and $\bG_{,X}\phi\partial^2\phi\partial^2h/(M_P\Lambda^4)$, which vanish in the decoupling limit.

To compute the $\phi\phi\to \phi\phi$ amplitude, we first derive the vertices for the Feynman diagrams. The three-point vertices include two operators:
\begin{equation}\label{threeV2}
  \frac{1}{\Lambda^3}\bar{A}_{1,\phi}\phi\phi_{\mu\nu}\phi^{\mu\nu}, \qquad  \frac{1}{\Lambda^3}\bar{A}_{2,\phi}\phi(\square \phi)^2.
\end{equation}
and the relevant four-point vertices are
\begin{align}
  &\frac{ \bA_{1,\phi\phi}}{2\Lambda^4} \phi^2\phi_{\mu\nu}\phi^{\mu\nu},  \qquad&-\frac{\bA_{1,X}}{2\Lambda^6} \phi_\rho\phi^\rho\phi_{\mu\nu}\phi^{\mu\nu}, &\qquad\qquad \frac{\bA_{2,\phi\phi}}{2\Lambda^4} \phi^2(\square\phi)^2 \notag, \\
  -&\frac{\bA_{2,X}}{2\Lambda^6}\phi_\rho\phi^\rho(\square\phi)^2, & \frac{\bA_{3}}{\Lambda^6}\square\phi\phi^\mu\phi_{\mu\nu}\phi^\nu, & \qquad\qquad  \frac{\bA_{4}}{\Lambda^6}\phi^\mu\phi_{\mu\nu}\phi^{\nu\rho}\phi_\rho.
\end{align}
where, as previously mentioned, $\bar A_i{}_{,\phi X}$ and so on are evaluated on the background $\phi_0=0$.

After reading out the Feynman rules, a straightforward computation gives the $\phi\phi\to \phi\phi$ amplitude
\begin{align}\label{Adamp}
   A^{\rm d}(s,t) &= \f{12\bA_{2,\phi}m^4}{\Lambda^4}+\frac{4\bA_{2,X}m^6}{\Lambda^6}     +  \f{\bA_{1,\phi\phi}}{\Lambda^4}(s^2+t^2+u^2-4m^2)
   \nn
   &~~~ + \frac{2\bA_3}{\Lambda^6} m^2 (st+su+ut-4m^2)  -\frac{\bA_{1,X}}{2\Lambda^6}\left( (s-2m^2)^3+(t-2m^2)^3+(u-2m^2)^3\right)
    \notag \\
   &~~~   -\frac{\bA_4}{2\Lambda^6}\big{(} s(s-2m^2)^2 +t(t-2m^2)^2 +u(u-2m^2)^2\big{)}  \notag \\
    &~~~   +A^{\rm d}_s+A^{\rm d}_t+A^{\rm d}_u ,
\end{align}
with
\begin{equation}\label{eq5-7}
  A^{\rm d}_z\equiv \frac{1}{\Lambda^6(m^2-z)}\left( 2\bA_{1,\phi}\(\frac{3z^2}{4}-m^2 z+m^4\)+2\bA_{2,\phi}m^2(m^2+2z)\right)^2 ,
\end{equation}
where $A^{\rm d}_z$ denotes the $z$ channel amplitude with double three-point vertex insertions connected by a propagator. For the generic DHOST theory given by Eq.(\ref{dhost}), the fully crossing symmetric positivity bounds can be obtained through Eq.(\ref{genb}),  in which the coefficients $c^{2,0}$ and $c^{2,1}$ are given by

\begin{equation}\label{gendh}
  \begin{aligned}
    \Lambda^4 c^{2,0} & = 2\bar{A}_{1,\phi\phi}+\frac{m^2}{\Lambda^2}\(-\f{21}2\bA_{1,\phi}^2+2\bA_{1,X}-24\bA_{1,\phi}\bA_{2,\phi}-2\bA_3 \), \\
    \Lambda^6 c^{2,1} & = \frac{1}{4}\(27\bar{A}_{1,\phi}^2+6\bar{A}_{1,X}+ 6\bar{A}_4\).
  \end{aligned}
\end{equation}

In the following of this section, we will discuss a few specific models as classified by the degeneracy conditions and explore their parameter spaces allowed by the triple symmetric positivity bounds.

\subsection{N-I class DHOST}

We now consider the N-I class (adopting the classification of \cite{Crisostomi:2016czh}) of DHOST theory defined by Lagrangian (\ref{dhost}) subject to the constraints
 \begin{equation}\label{NI}
 \begin{aligned}
 A_{2}&=-A_{1}\neq -G/X,  \\
 A_4&=\frac{1}{8\left(G-A_1 X\right)^2} {\left[4 G\left(3\left(A_1-2 G_X\right)^2-2 A_3 G\right)-A_3 X^2\left(16 A_1 G_X+A_3 G\right)\right.} \\
&\left.+4 X\left(3 A_1 A_3 G+16 A_1^2 G_X-16 A_1 G_X^2-4 A_1^3+2 A_3 G G_X\right)\right], \\
 A_5&=\frac{1}{8\left(G-A_1 X\right)^2}\left(2 A_1-A_3 X-4 G_X\right)\left[A_1\left(2 A_1+3 A_3 X-4 G_X\right)-4 A_3 G\right] .
\end{aligned}
 \end{equation}
 In the following, we shall disregard the last condition with $A_5$ above, as $A_5$ does not center the positivity bounds. In this class, we are especially interested in the subset of models that are subject to an additional constraint
\be\label{NIGW}
A_1(\phi,X)=0.
\ee
This constraint is motivated by the fact that the speed of gravitational waves $c_{GW}^2=G/(G-\Li^2 A_1X)$ at late times has been constrained to be very close to unity \cite{Copeland:2018yuh, Sakstein:2017xjx, Creminelli:2017sry, Ezquiaga:2017ekz}. However, it has been cautioned that these observational constraints rely on validity of the EFT at scales very close to the cutoff \cite{deRham:2018red}. Even assuming that these constraints are robust, there is still a possibility, albeit ad hoc and fine-tuning, that the gravitational waves can deviate from unity in the earlier universe. Therefore, we shall also consider the cases of $A_1\neq 0$ in the following subsections.

\begin{figure}[h]
    \centering
    \includegraphics[width=0.6\linewidth]{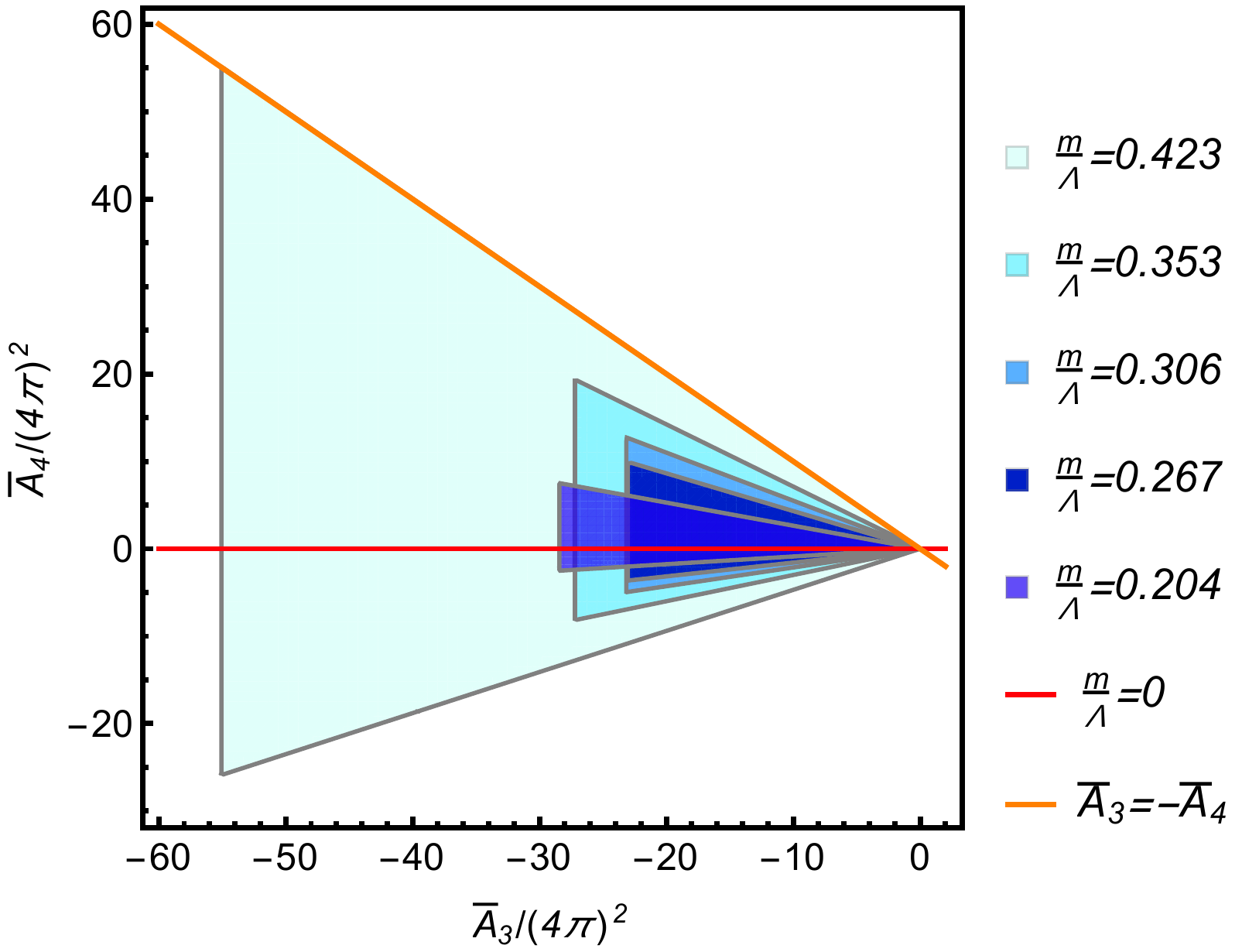}
    \caption{\label{fig:NI} Positivity allowed regions in the N-I class of DHOST theory with $A_1=0$ (the unity condition of the speed of gravitational waves) for various scalar masses. The solid red line indicates the massless case, and the orange line denotes the case of a small $G_{,X}$. The small $G_{,X}$ case (Newton's coupling being nearly constant across the spacetime)  is in tension with the positivity bounds, as it requires $m/\Li\simeq 0.423$, unless $\bar A_4=-\bar A_3=0$.}
\end{figure}

For this well-motivated subclass with $A_1=0$, the two-sided positivity bounds are given by
\begin{eqnarray}\label{dGWcons}
\left\{
\begin{aligned}
  \frac{3}{2} \bA_4 +2(m/\Lambda)^2 \bA_3 \, \bar{c}^{2,1}_{\rm min}(m) \geq 0 \\
     \frac{3}{2} \bA_4 +2(m/\Lambda)^2 \bA_3 \, \bar{c}^{2,1}_{\rm max} (m)\leq 0 \\
    0\leq(m/\Lambda)^2(-2\bA_3) \leq \Lambda^4 c^{2,0}_{\rm max}(m).
\end{aligned}
\right.
\end{eqnarray}
where $\Lambda^4 c^{2,0}_{\rm max} (m)$ is the upper bound that can be extracted from Figure \ref{fig:1} and $\bar{c}^{2,1}_{\rm min}(m)$ and $\bar{c}^{2,1}_{\rm max}(m)$ are respectively the lower and upper bounds that can be extracted from Figure \ref{fig:c21}. Generally, the positivity bounds do not contain $A_2$ because the $A_2$ terms do not contribute to the twice subtracted dispersion relation. In this subclass, of course, $A_2$ happens to vanish by the degeneracy condition $A_2=-A_1=0$. Consequently, the parameter space that can be constrained by the positivity bounds is two dimensional, furnished by the parameters $\bA_3$ and $\bA_4$. The blue regions in Figure \ref{fig:NI} illustrate the allowed regions by the fully crossing symmetric positivity constraints for different masses, before imposing the degeneracy conditions. We see that the preliminary positivity regions are triangular areas with one of their vertices sitting at the origin.

Let us look at the remaining non-trivial degeneracy condition
\begin{equation}\label{Adegen1}
8 (A_3 + A_4) G-48 \Li^{-2} G_{,X}^2- 8  A_3 G_{,X} X+\Li^2 A_3^2  X^2 =0 ,
\end{equation}
more carefully. This degeneracy condition is satisfied by the whole $A_3$, $A_4$ and $G$ functions, so the background functions $\bar A_3$, $\bar A_4$ and $\bar G$ also satisfy the same condition
\begin{equation}\label{Adegen2}
8 (\bar A_3 + \bar A_4) \bar G-48 \Li^{-2} \bar G_{,X}^2- 8  \bar A_3 \bar G_{,X} X+\Li^2 \bar A_3^2  X^2 =0 .
\end{equation}
(One can taylor-expand it around the background and every order of the expansion must vanish individually.) In a generic case, for arbitary $\bar A_3$ and $\bar A_4$, a suitable $\bar G_{,X}$, which we can not constrain in our formalism, can make \eref{Adegen2} satisfied, and the positivity regions in Figure \ref{fig:NI} are the bounds one can extract. The same conclusion can also be reached if both $\bar G$ and $\bar G_{,X}$ scale as $M_P^2$ and we take the decoupling limit $M_P\to \infty$ of \eref{Adegen1}, in which case \eref{Adegen2} becomes $-48 \Li^{-2} \bar G_{,X}^2=0$, which does not impose any restrictions on $\bar A_3$ and $\bar A_4$.  In Figure \ref{fig:NI}, we again observe that, similar to what we saw in Horndeski theory, letting the scalar have a mass can enlarge the theory space of DHOST theory that is consistent with positivity.

On the other hand, if $\bar G$ scales as $M_P^2$ and $\bar G_{,X}$ scales as $\Li^2$ such that one has $G\sim M_P^2+\Li^2 \bar G_{,X} X+...$, the decoupling limit of \eref{Adegen2} becomes $8 (A_3 + A_4) G=0$, which leads to
\be
\label{A2A4A3deg}
\bar A_4=-\bar A_3 .
\ee
Then the degeneracy condition corresponds to the orange line in Figure \ref{fig:NI}. This is a well-motivated scenario where the gravitational coupling function $G$ is mostly constant throughout the spacetime. We see from Figure \ref{fig:NI} that, unless $\bar A_4=-\bar A_3=0$, this kind of DHOST theory can only be consistent with positivity bounds if the mass of the scalar is almost comparable with the cutoff: $m/\Li \simeq 0.423$, which means that such a theory has a perilously narrow range of applicability. The same conclusion also applies if $G$ is only a function of $\phi$, for which case $\bar G_{,X}=0$ and again leads to \eref{A2A4A3deg} and thus the tension between the positivity bounds and the applicabililty of DHOST theory as an EFT.

In the massless case, as we see in Figure \ref{fig:NI}, the positivity bounds require $\bA_4=0$, and there is a range of $\bar A_3$ satisfying the positivity bounds. When $\bar G_{,X}$ vanishes or scales as $\Li^2$, however, the degeneracy conditions require $\bA_3=-\bA_4=0$. Therefore, the theory in this case can not have nonzero background values of $A_3$ and $A_4$.

In summary, if the speed of gravity is required to be close to unity and Newton's coupling is required to be slowly varying across spacetime, the leading terms of DHOST theory are severely constrained by the fully crossing symmetric positivity bounds.

\subsection{M-I class DHOST}

\begin{figure}[ht]
    \center
    \includegraphics[width=0.6\linewidth]{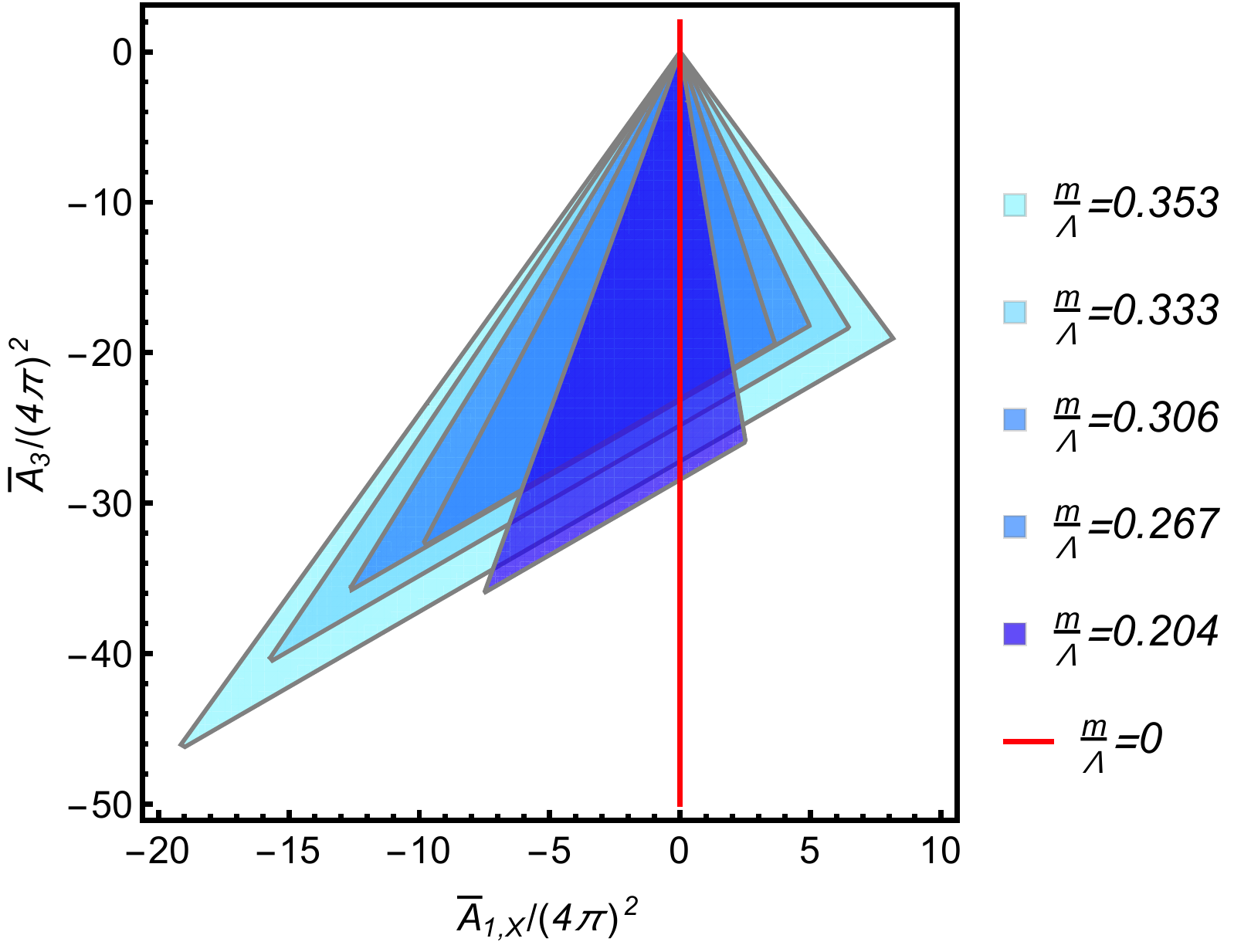}
    \caption{\label{fig:MI} Positivity allowed regions in the M-I class of DHOST theory for various masses.  The solid red line indicates the massless case.}
\end{figure}

We now consider the M-I class of DHOST theory, which is given by Lagrangian (\ref{dhost}) along with the following degeneracy condition
\begin{equation}\label{MIde}
  A_4=-2A_1/X .
\end{equation}
 Again we have dropped a condition involving $A_5$, as it does not affect out results here. Locality requires $A_4$ not to blow up as $X$ goes to zero, so $A_{1}$ has to take the form of $A_1=XF(\phi,X)$, where $F(\phi,X)$ is regular as $X$ goes to zero. Evaluating at the background $\phi_0=X_0=0$, the fully crossing symmetric positivity constraints can be written as
 \begin{eqnarray}\label{MIcons}
\left\{
\begin{aligned}
  -3\bar{A}_{1,X}/2 - \(\frac{m}{\Lambda}\)^2 (2\bar{A}_{1,X}-2\bar{A}_3)  \bar{c}^{2,1}_{\rm min}(m) \geq 0\\
  -3\bar{A}_{1,X}/2 - \(\frac{m}{\Lambda}\)^2 (2\bar{A}_{1,X}-2\bar{A}_3)  \bar{c}^{2,1}_{\rm max}(m) \leq 0 \\
  0\leq(m/\Lambda)^2 (2\bar{A}_{1,X}-2\bar{A}_3) \leq \Lambda^4 c^{2,0}_{\rm max}(m).
\end{aligned}
\right.
\end{eqnarray}

In Figure \ref{fig:MI}, we plot the corresponding positivity allowed regions for various scalar masses. We see that the two-dimensional positivity regions are again closed triangles. In the beyond Horndeski model of \cite{Gleyzes:2014dya}, the degeneracy conditions become
\begin{equation}\label{bHde}
  A_2=-A_1,\qquad A_3=-A_4=2A_1/X .
\end{equation}
In this case, the triple crossing bounds are reduced to the following simple constraints
\begin{equation}\label{MIbH}
  m/\Lambda\geq 0.423,\qquad  -54.78 \leq \bar A_{1,X}\leq 0 ,
\end{equation}
which impose a lower bound on the mass $m$, which is about the same order as the cutoff $\Lambda$. (The lower for $\bar{A}_{1,X}$ can be obtained when $m/\Lambda=0.423$.)

Therefore, the beyond Horndeski model of \cite{Gleyzes:2014dya} is in tension with fully crossing positivity bounds, as in this model a healthy hierarchy between $m$ and $\Li$ is ruled out by these pure theoretical bounds, even if the speed of gravity constraints do not rule them out.

\subsection{M-II class DHOST}

\begin{figure}[ht]
    \center
    \includegraphics[width=0.6\linewidth]{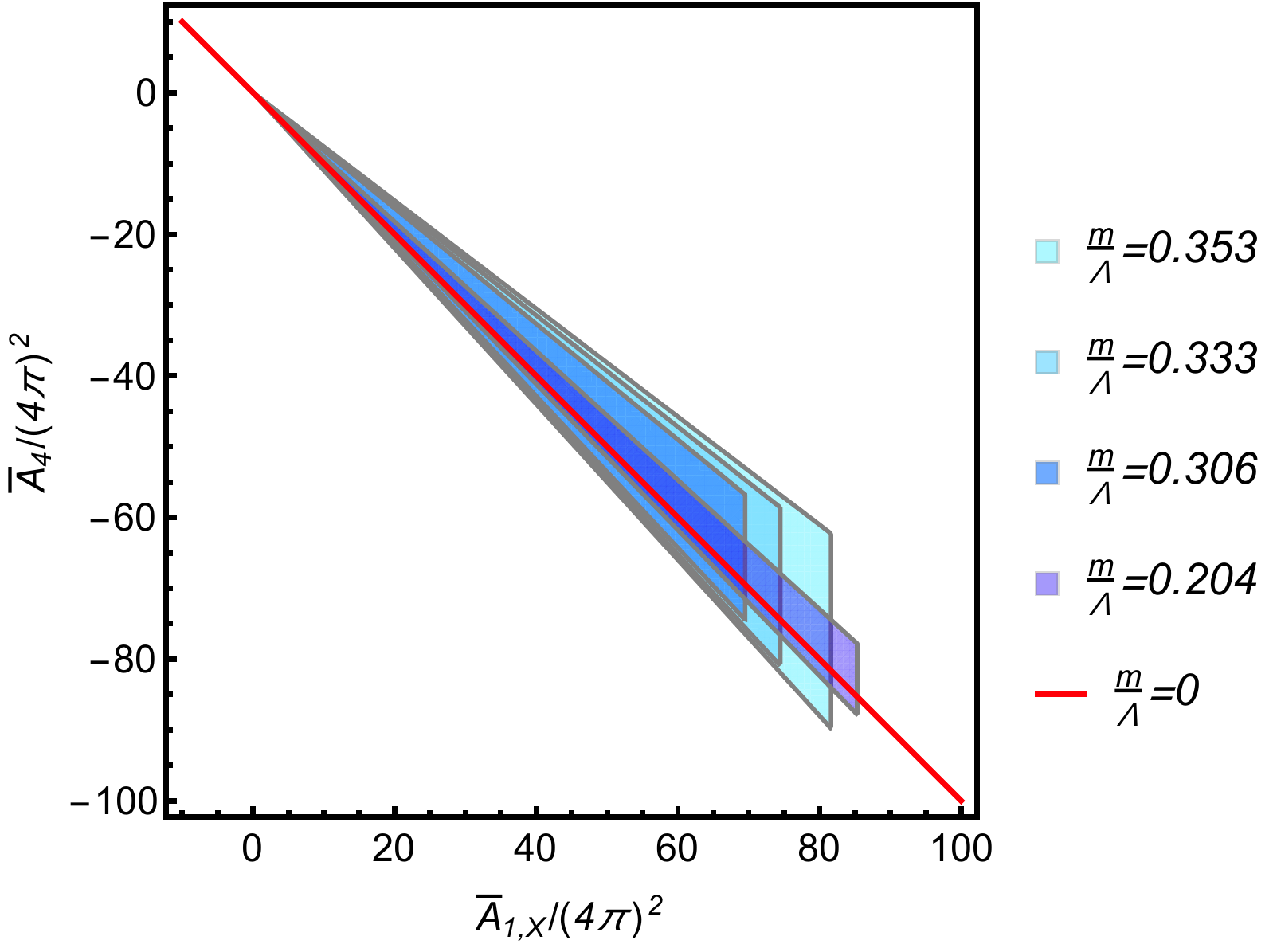}
    \caption{Positivity allowed regions in the M-II class of DHOST theory for various masses.  The solid red line indicates the massless case. }
    \label{fig:MII}
\end{figure}

Similar to the M-I class, the M-II class can be obtained by imposing the following degeneracy conditions
\begin{equation}\label{MII}
  A_2=-\frac{A_1}{3},\qquad A_3=\frac{2A_1}{3X}
\end{equation}
Again, locality demands $A_1=XF(\phi,X)$ so that $A_3$ is regular when $X$ goes to zero. Then, we can obtain the following triple crossing symmetric positivity bounds
\begin{eqnarray}\label{MIIcons}
\left\{
\begin{aligned}
  \frac{3}{2}(\bar{A}_{1,X}+\bar{A}_4) - \frac{2}{3}\(\frac{m}{\Lambda}\)^2 \bar{A}_{1,X}  \bar{c}^{2,1}_{\rm min}(m) \geq 0\\
  \frac{3}{2}(\bar{A}_{1,X}+\bar{A}_4) - \frac{2}{3}\(\frac{m}{\Lambda}\)^2 \bar{A}_{1,X}  \bar{c}^{2,1}_{\rm max}(m) \leq 0 \\
  0\leq \frac{2}{3}\(\frac{m}{\Lambda}\)^2 \bar{A}_{1,X} \leq \Lambda^4 c^{2,0}_{\rm max}(m).
\end{aligned}
\right.
\end{eqnarray}
In Figure \ref{fig:MII}, we plot the positivity allowed regions on the $\bar{A}_{1,X}$-$\bar{A}_4$ plane for various masses. The positivity regions are again triangular areas, which become thin diagonal strips if we assume a healthy hierarchy between $m$ and $\Li$.

\subsection{Combining Horndeski and DHOST}

\begin{figure}[htbp]

       \centering
    \begin{subfigure}{0.46\linewidth}
    \centering
        \includegraphics[width=1\linewidth]{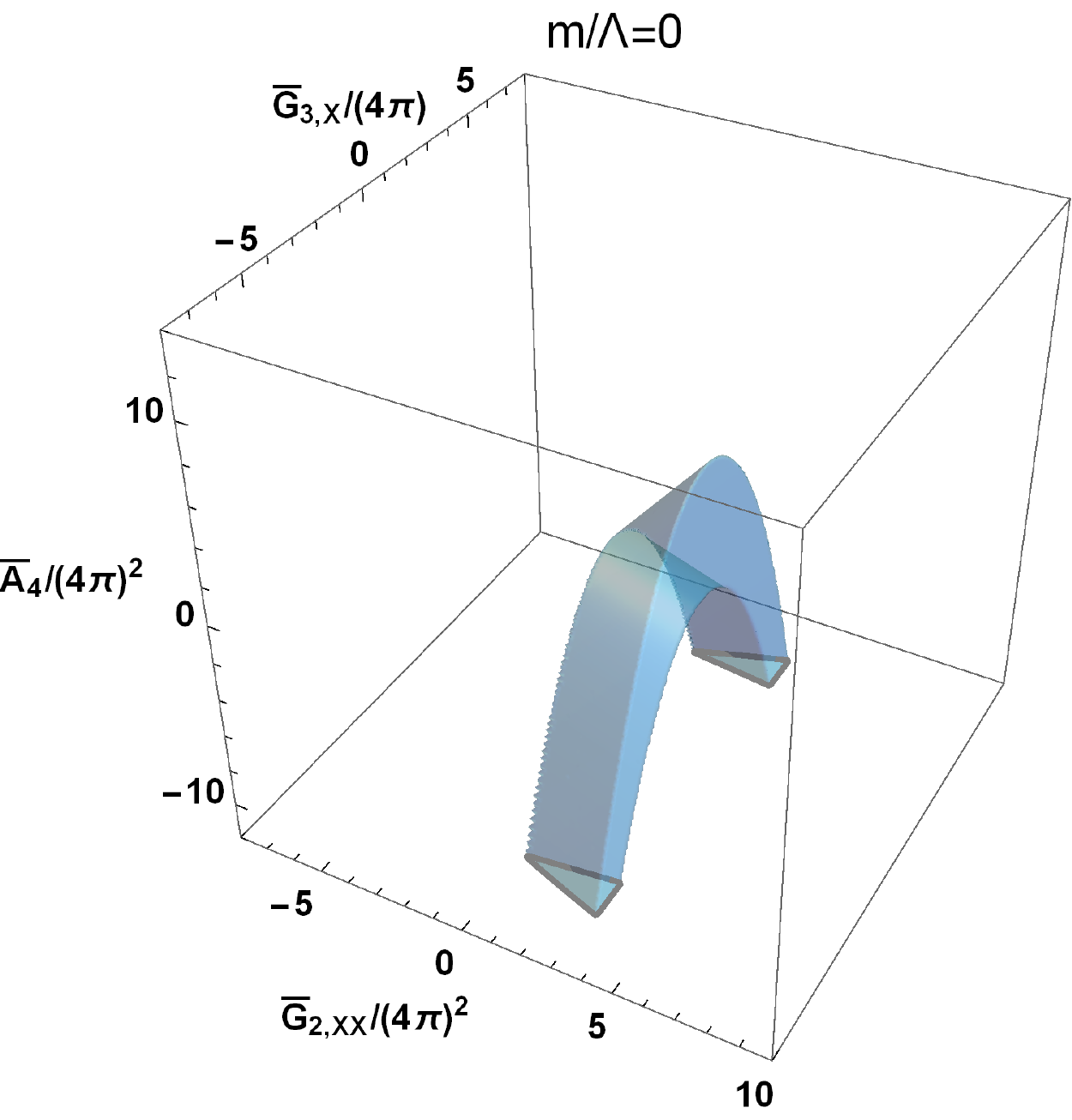}
    \end{subfigure}
     \centering
    \begin{subfigure}{0.46\linewidth}
    \centering
        \includegraphics[width=1\linewidth]{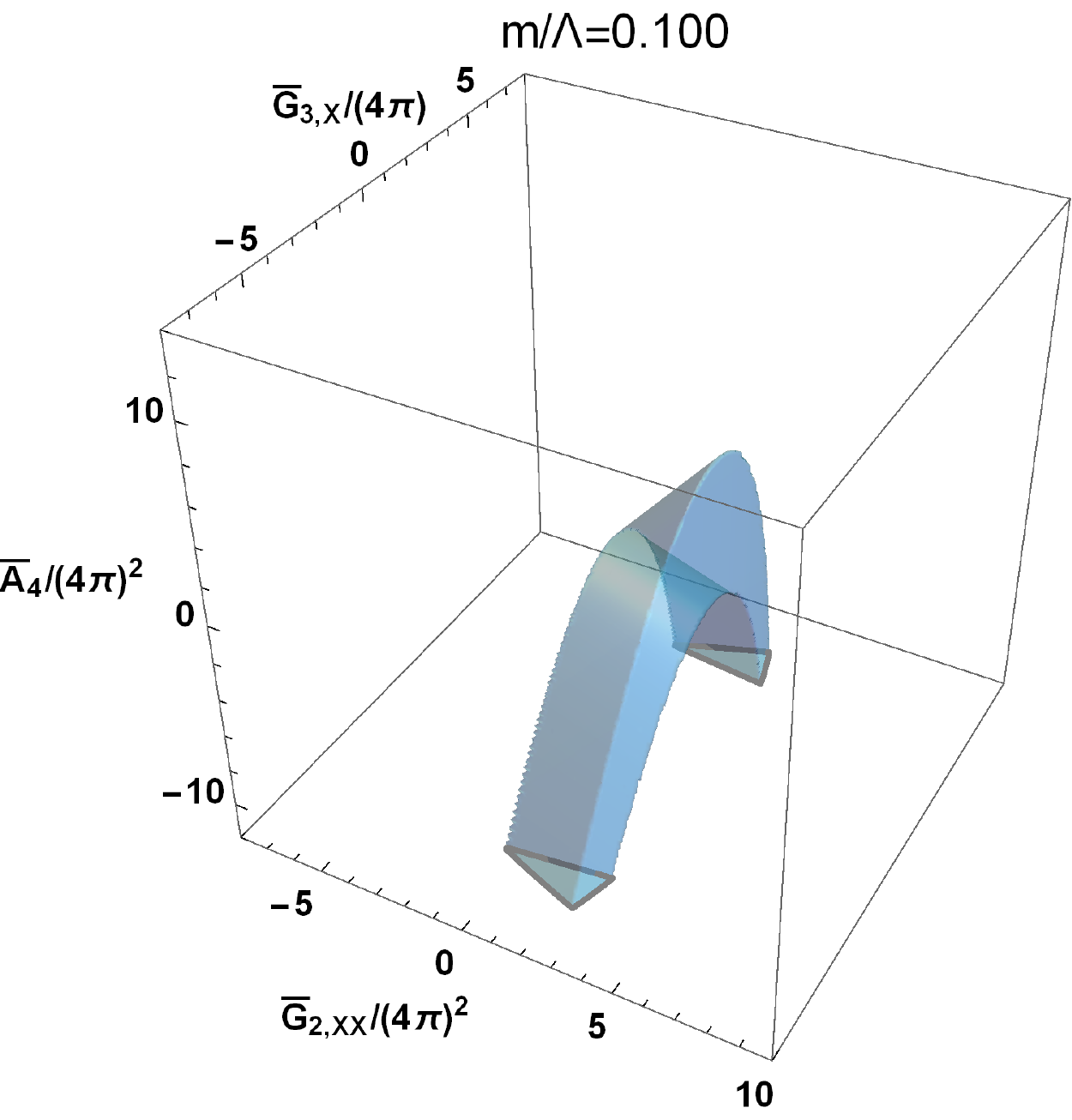}
    \end{subfigure}
\vskip 12pt
     \centering
    \begin{subfigure}{0.46\linewidth}
    \centering
        \includegraphics[width=1\linewidth]{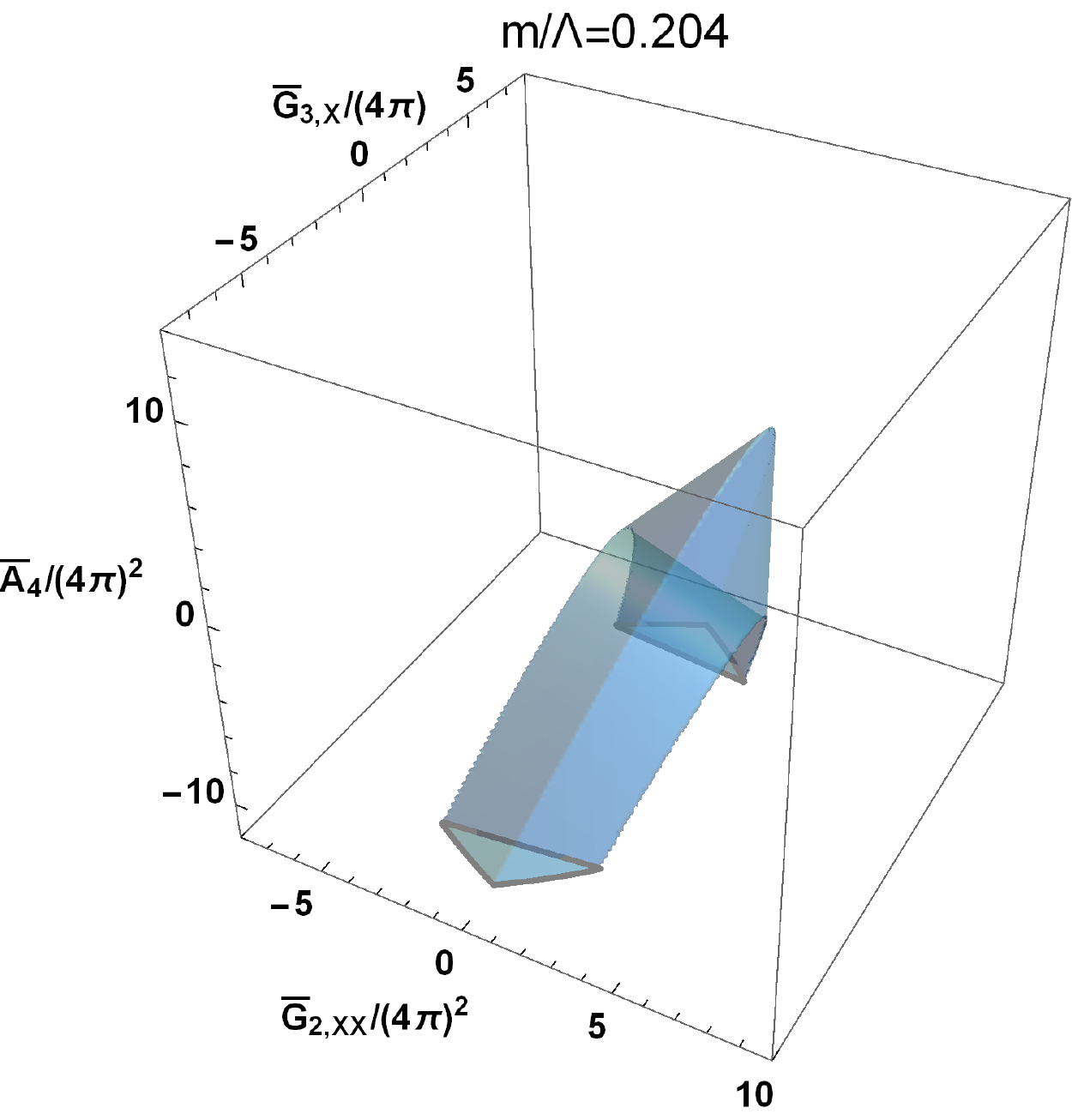}
    \end{subfigure}
     \centering
    \begin{subfigure}{0.46\linewidth}
    \centering
        \includegraphics[width=1\linewidth]{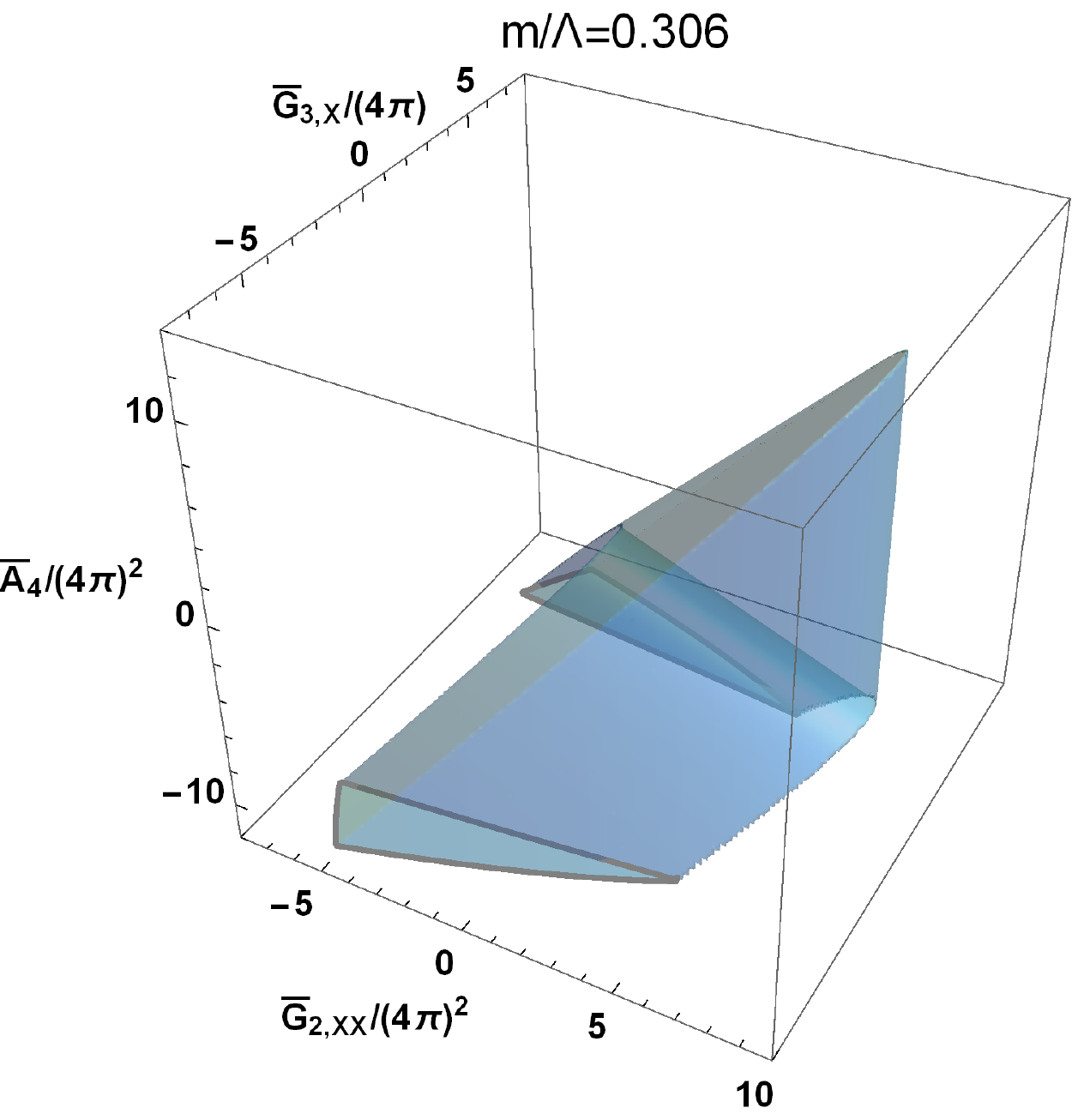}
    \end{subfigure}

    \caption{\label{fig:cDGW} Fully crossing symmetric positivity bounds on a class of models combining Horndeski and DHOST terms (see \eref{cD}) subject to the speed of gravity constraints. }
\end{figure}

In this subsection, we shall consider a class of models combining certain Horndeski and DHOST terms, which is referred to as ${}^2$N-I$+{}^3$N-I \cite{BenAchour:2016fzp} and
given by the following Lagrangian
\begin{equation}\label{cD}
\begin{aligned}
  \mathcal{L}^{\rm cD}=&\mathcal{L}_2^{\rm H}+\mathcal{L}_3^{\rm H}+G(\phi,X) R -\frac{A_1(\phi,X)}{\Lambda^2} \( (\square\phi)^2-(\nabla_\mu\nabla_\nu\phi)^2\)+\mathcal{L}_3^{\rm d}
  +\mathcal{L}_4^{\rm d} \\
  &+\frac{1}{\Lambda^3} C(\phi,X) G^{\mu\nu}\nabla_\mu\nabla_\nu \phi+\frac{B(\phi,X)}{\Lambda^5} \left((\square\phi)^3-3\square\phi(\nabla_\mu\nabla_\nu\phi)^2+2(\nabla_\mu\nabla_\nu\phi)^3 \right).
\end{aligned}
\end{equation}
These models are free of instabilities in desirable cosmological backgrounds \cite{deRham:2016wji, Langlois:2017mxy}.  Again, by computing the $\phi\phi\rightarrow\phi\phi$ amplitude in the decoupling limit, we can derive the fully crossing symmetric positivity bounds:
\begin{equation}\label{genbComb}
  \begin{cases}
    &\Lambda^4 c^{2,0}(m)\,\bar{c}^{2,1}_{\rm min}(m)\leq \Lambda^6 c^{2,1}(m)\leq \Lambda^4 c^{2,0}(m)\,\bar{c}^{2,1}_{\rm max}(m) \\
    &0\leq \Lambda^4 c^{2,0}(m)\leq \Lambda^4 c^{2,0}_{\rm max} (m).
  \end{cases}
\end{equation}
where
\begin{align}\label{cDamp}
  \Lambda^4 c^{2,0}(m) &= \frac{1}{2}\bG_{2,XX}+2\bar{A}_{1,\phi\phi}-3\bar{A}_{1,\phi}(\bar{G}_{2,\phi X}+2\bG_{3,\phi\phi})+\bar{G}_{3,X}(\bG_{2,\phi X}+2\bG_{3,\phi\phi}) \notag \\
  & ~~~+\frac{1}{6}\(\frac{m}{\Lambda} \)^2 \( 9\bar{G}_{3,X}^2+81\bar{A}_{1,\phi}^2+12\bar{A}_{1,X}-48\bar{B}_{,\phi}-54\bar{G}_{3,X}\bar{A}_{1,\phi} -12\bar{A}_3\),
   \\
  \Lambda^6c^{2,1}(m)&=\frac{3\bG_{3,X}^2}{4}-\frac{9\bG_{3,X}\bar{A}_{1,\phi}}{2}+\frac{ 27\bar{A}_{1,\phi}^2}{4} +\frac{3}{2}\bar{A}_{1,X}-6\bar{B}_{,\phi }+\frac{3}{2}\bar{A}_4 .
\end{align}

If we take into account the observational constraints on the speed of gravitational waves \cite{Langlois:2017dyl} and the decay of gravitational waves into dark energy \cite{Creminelli:2018xsv}, this class of models is reduced to the following Lagrangian
\be
\label{cDGW}
\mathcal{L}^{\rm cDGW}=\Lambda^4 G_{2}(\phi,X)+\Lambda G_3(\phi,X) +G(\phi,X) R
  +\frac{1}{\Lambda^6} A_4(\phi, X) \nabla_\mu\phi \nabla^\mu\nabla^\rho\phi \nabla_\rho\nabla_\nu\phi \nabla^\nu\phi ,
\ee
subject to the degeneracy condition $A_4=3G_{,X}^2/(2\Lambda^2 G)$. In this case, the triple crossing symmetric bounds of this subclass take a much simpler form when $G_i\,$s are shift-symmetric:
\begin{eqnarray}
\label{cDGWcons}
\left\{
\begin{aligned}
  \frac{3}{4}(\bG_{3,X}^2+2\bar{A}_4) - \( \bG_{2,XX}/2+ \frac{3}{2}\(\frac{m}{\Lambda}\)^2\bG_{3,X}^2 \) \bar{c}^{2,1}_{\rm min}(m) \geq 0\\
     \frac{3}{4}(\bG_{3,X}^2+2\bar{A}_4) - \( \bG_{2,XX}/2+ \frac{3}{2}\(\frac{m}{\Lambda}\)^2\bG_{3,X}^2 \) \bar{c}^{2,1}_{\rm max} (m)\leq 0 \\
    0\leq \bG_{2,XX}/2+ \frac{3}{2}\(\frac{m}{\Lambda}\)^2\bG_{3,X}^2 \leq \Lambda^4 c^{2,0}_{\rm max}(m).
\end{aligned}
\right.
\end{eqnarray}
There are three theory parameters in these bounds: $\bG_{2,XX}$, $\bG_{3,X}$ and $\bar{A}_4$. In Figure \ref{fig:cDGW}, we plot the bounds on these three parameters for a few masses. We see that the mass only significanlty affect the bounds when it is very close to the cutoff.

\acknowledgments

We would like to thank Dong-Yu Hong, Shi-Lin Wan, Zhuo-Hui Wang and Guo-Dong Zhang for helpful discussions. SYZ acknowledges support from the Fundamental Research Funds for the Central Universities under grant No.~WK2030000036, from the National Natural Science Foundation of China under grant No.~12075233 and 12247103, and from the National Key R\&D Program of China under grant No. 2022YFC220010.\\

\appendix

\section{Numerical setup}
\label{app:appA}

In this appendix, we provide more details about how to implement the linear programs in Section \ref{sec:optbounds} using the SDPB package, and also show some convergence tests for our numerical results.

\subsection{Implementation}

In Section \ref{sec:optbounds}, the problem of obtaining optimal positivity bounds has been formulated as solving a series of linear programs. To solve these linear programs, we need to impose linear inequalities such as that of \eref{c21} for all $\mu\geq \tilde\Lambda^2$ and $l$. $l$ takes discrete values from 0 to $\infty$. Numerically, we can truncate it at a finite value, $l=0,2,...,l_{\rm max}$. The UV scale $\mu$, on the other hand, is a continuous decision variable. One way to deal with $\mu$ would be discretize it. However, for the current problem, the constraints for all $\mu\geq \tilde\Lambda^2$ can actually be imposed exactly. The SDPB package is able to solve a linear program where the linear inequality constraints contain polynomials of a positive continuous variable, by converting it a standard semi-definite program \cite{Simmons-Duffin:2015qma}. The $C^{2i,j}_l(\mu)$ and $N^{p,q}_l(\mu)$ quantities are not quite polynomials of a positive continuous variable, but we can easily massage them into the right forms. To this end, we first define
\be
\mu\equiv\tilde{\Lambda}^2(1+x),
\ee
where now $x$ is a positive continuous variable. After that, we only need to multiply the linear inequalities such as that of \eref{c21} with appropriate powers of $(1+x+2m^2/\tilde\Lambda^2)$ and $(1+x)$, and then the resulting linear inequalities are admissible by the SDPB package.

Let us illustrate how it works in practice via the example of obtaining the bound on the $\bar{c}^{2,2}\text{-}\bar{c}^{2,1}$ plane with the lowest order null constraint. For notational convenience, we define the vector $\xi$ of the decision variables
\begin{equation}\label{4-1}
  \xi \equiv (\alpha_{2,0},\alpha_{2,1} , 1 , \lambda_{1,3} ).
\end{equation}
where we have normalized the variable $\alpha_{2,2}$ to 1, thanks to the fact that all the constraints are linear inequalities in terms of the decision variables.  We also define the vector $V_l(x)$:
\begin{equation}\label{4-2}
  V_l(x)\equiv (1+x+2m^2/\Li^2)^8(1+x)^6 (C_l^{2,0},C_l^{2,1},C_l^{2,2},N_l^{1,3}) ,
\end{equation}
whose components are polynomial functions of $x$ if truncated to a finite order of $m^2$. Then, the linear inequalities become $\xi \cdot V_l(x)\geq 0$, and acting $\langle ~\rangle$ on this inequalities gives $\alpha_{2,0} {c}^{2,0} + \alpha_{2,1} {c}^{2,1} +  {c}^{2,2}\geq 0$. For a given $\bar{c}^{2,1} = \bar{c}^{2,1}_{0}$, we can search all $\xi$ vectors for the maximum of $-\alpha_{2,0}-\alpha_{2,1} \bar{c}^{2,1}_{0}$, which will give the lower bound on $\bar{c}^{2,2}$. That is, the linear program we can put into the SDPB package to solve is
\begin{equation}\label{lpmax}
  \begin{cases}
    \text{maximize: } O_{\rm max}=\xi\cdot(-1,-\bar{c}^{2,1}_{0},0,0) \\
    \text{subject to: } \xi \cdot V_l(x) \geq 0 , \quad \text{for any } x>0 \text{ and } l=0,2,4 \dots, l_{\rm max}.
  \end{cases}
\end{equation}
The lower bound on $\bar{c}^{2,2}$ for a given $\bar{c}^{2,1} = \bar{c}^{2,1}_{0}$ is $\bar{c}^{2,2}\geq O_{\rm max}$.

\begin{figure}
  \centering
    \begin{subfigure}{0.465\linewidth}
    \centering
        \includegraphics[width=0.9\linewidth]{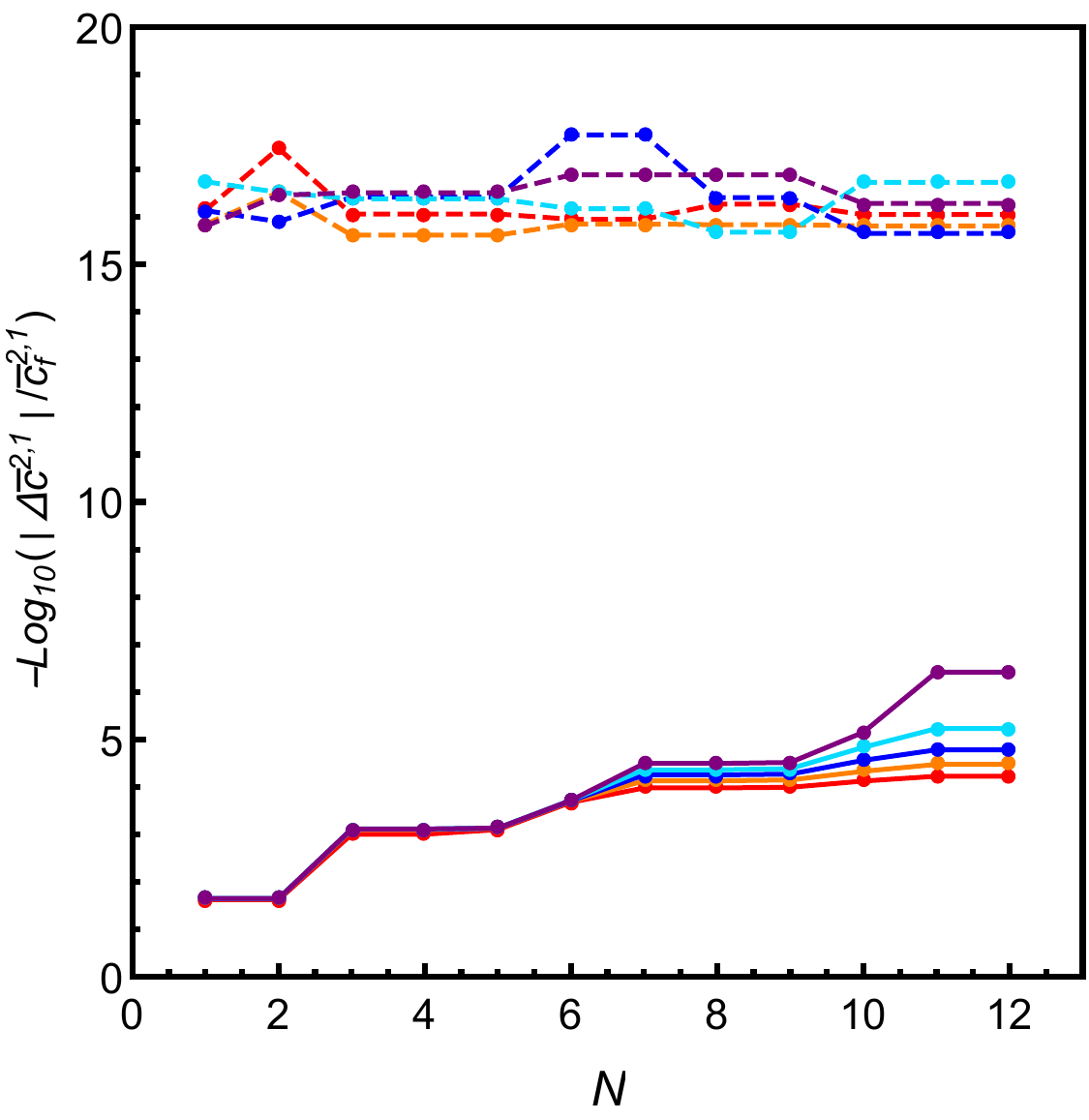}
    \end{subfigure}
    \centering
  \begin{subfigure}{0.47\linewidth}
    \centering
        \includegraphics[width=0.9\linewidth]{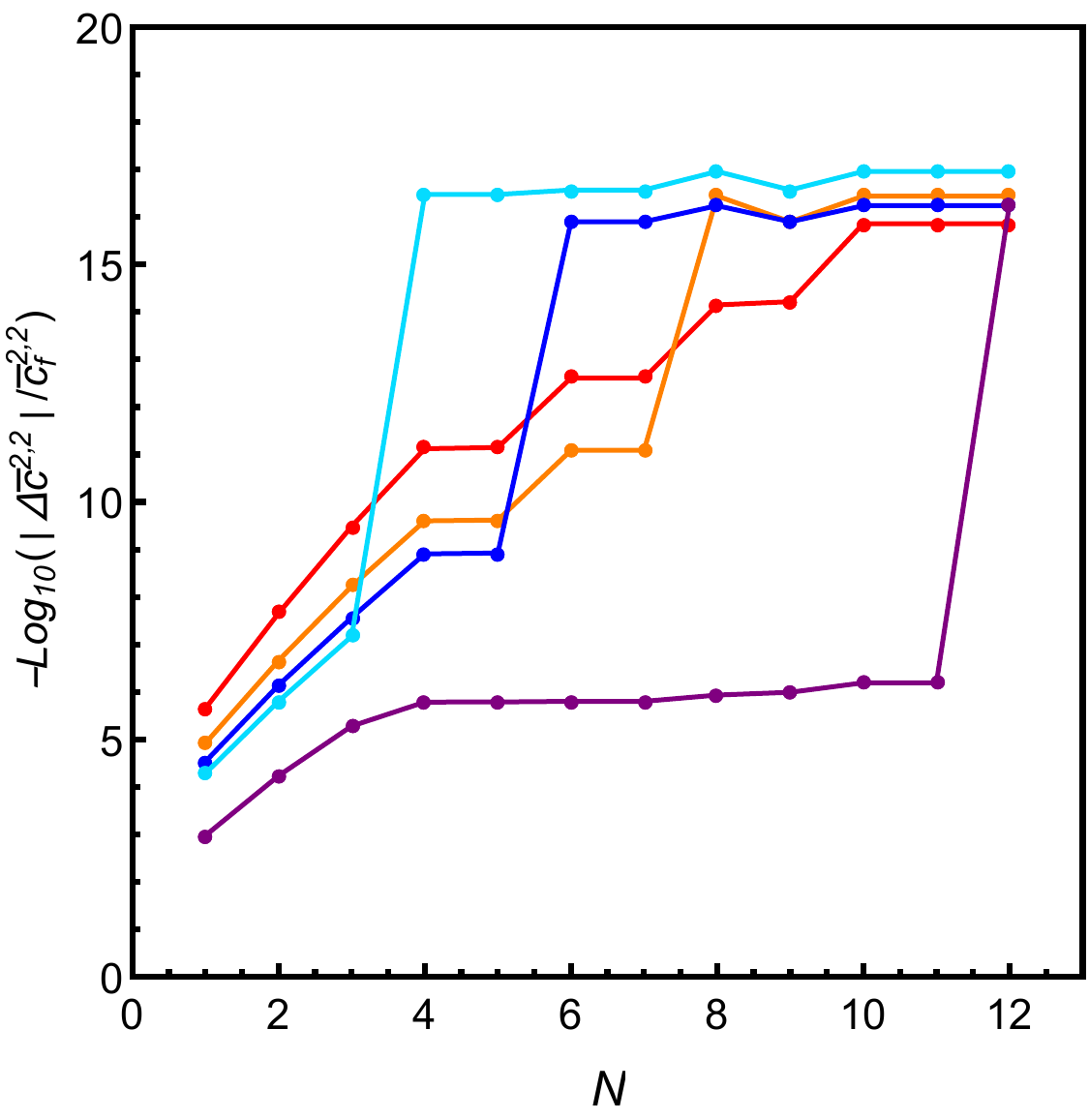}
    \end{subfigure}
    \centering
    \begin{subfigure}{0.47\linewidth}
    \centering
        \includegraphics[width=0.9\linewidth]{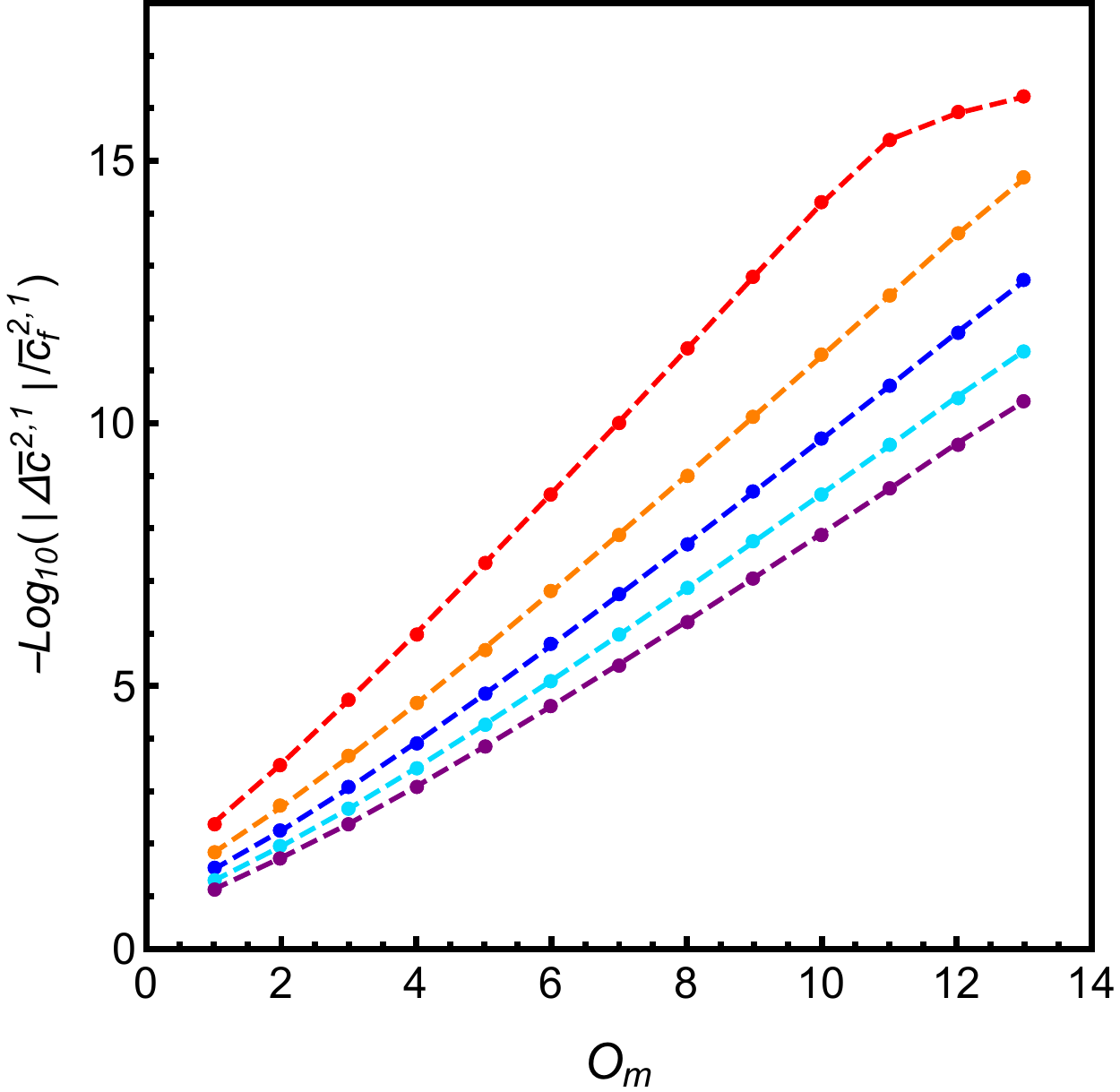}
    \end{subfigure}
     \centering
    \begin{subfigure}{0.47\linewidth}
    \centering
       \includegraphics[width=0.9\linewidth]{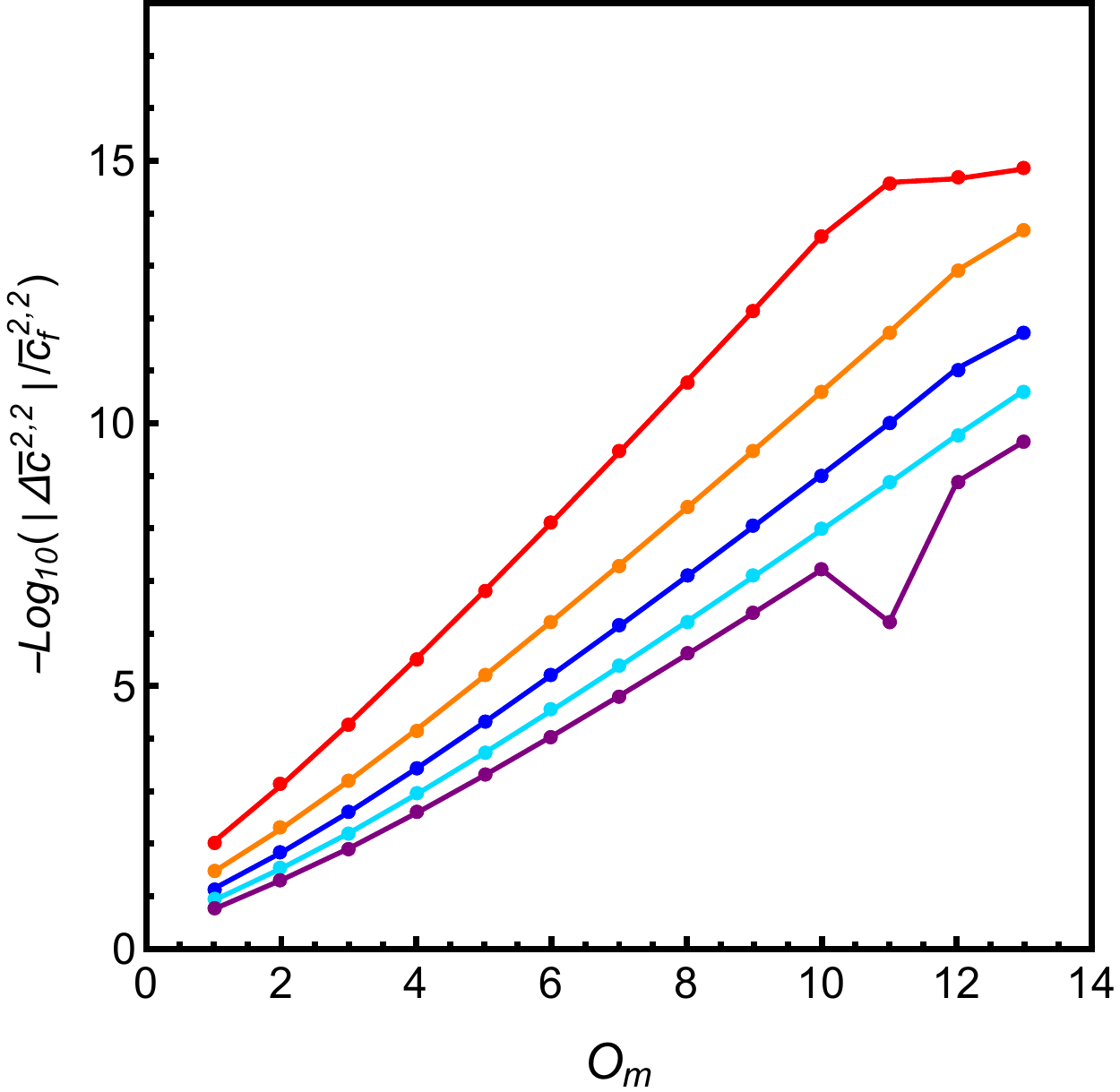}
    \end{subfigure}
    \centering
    \begin{subfigure}{0.47\linewidth}
    \centering
        \includegraphics[width=0.9\linewidth]{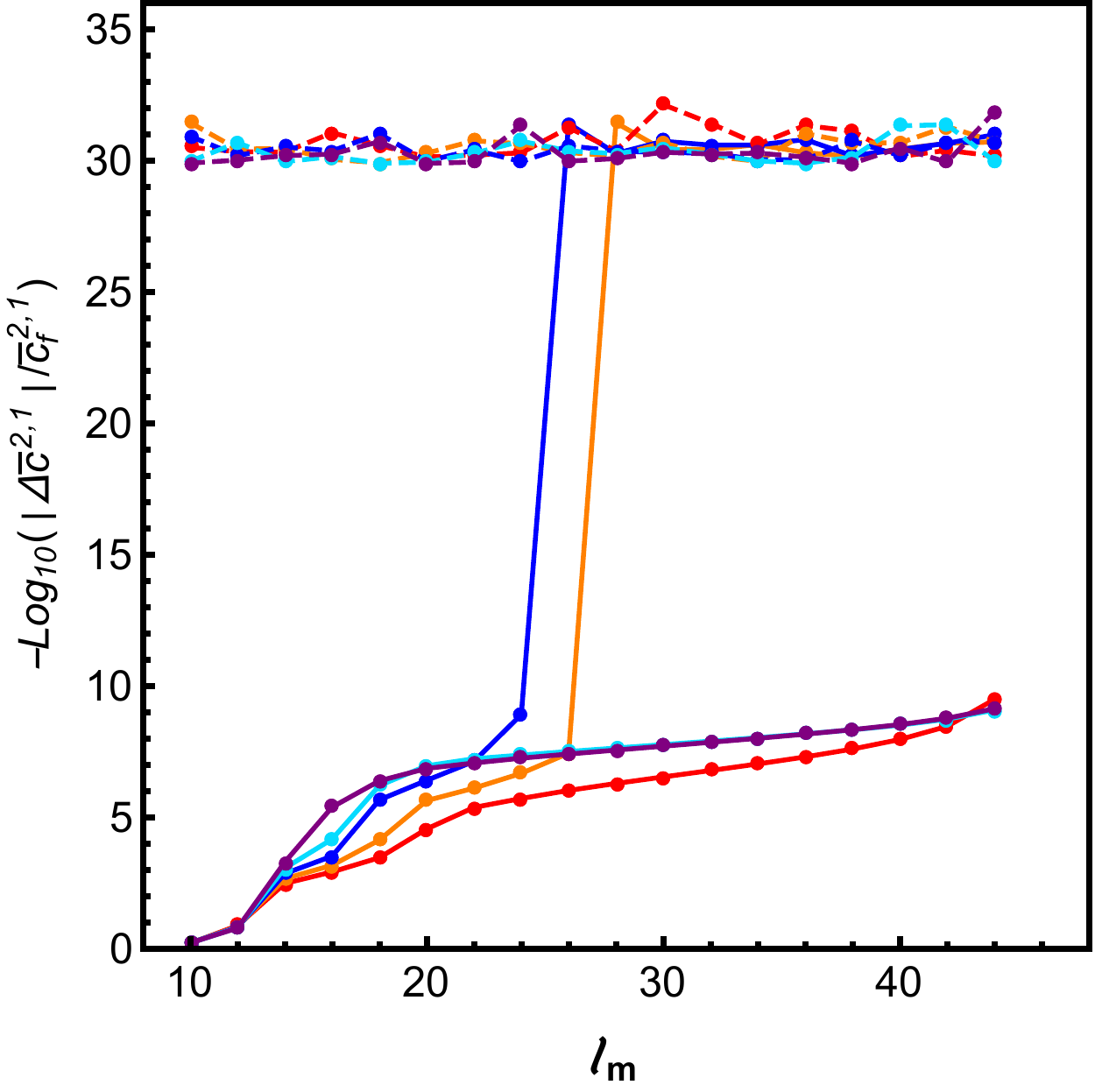}
    \end{subfigure}
    \centering
    \begin{subfigure}{0.47\linewidth}
    \centering
        \includegraphics[width=0.9\linewidth]{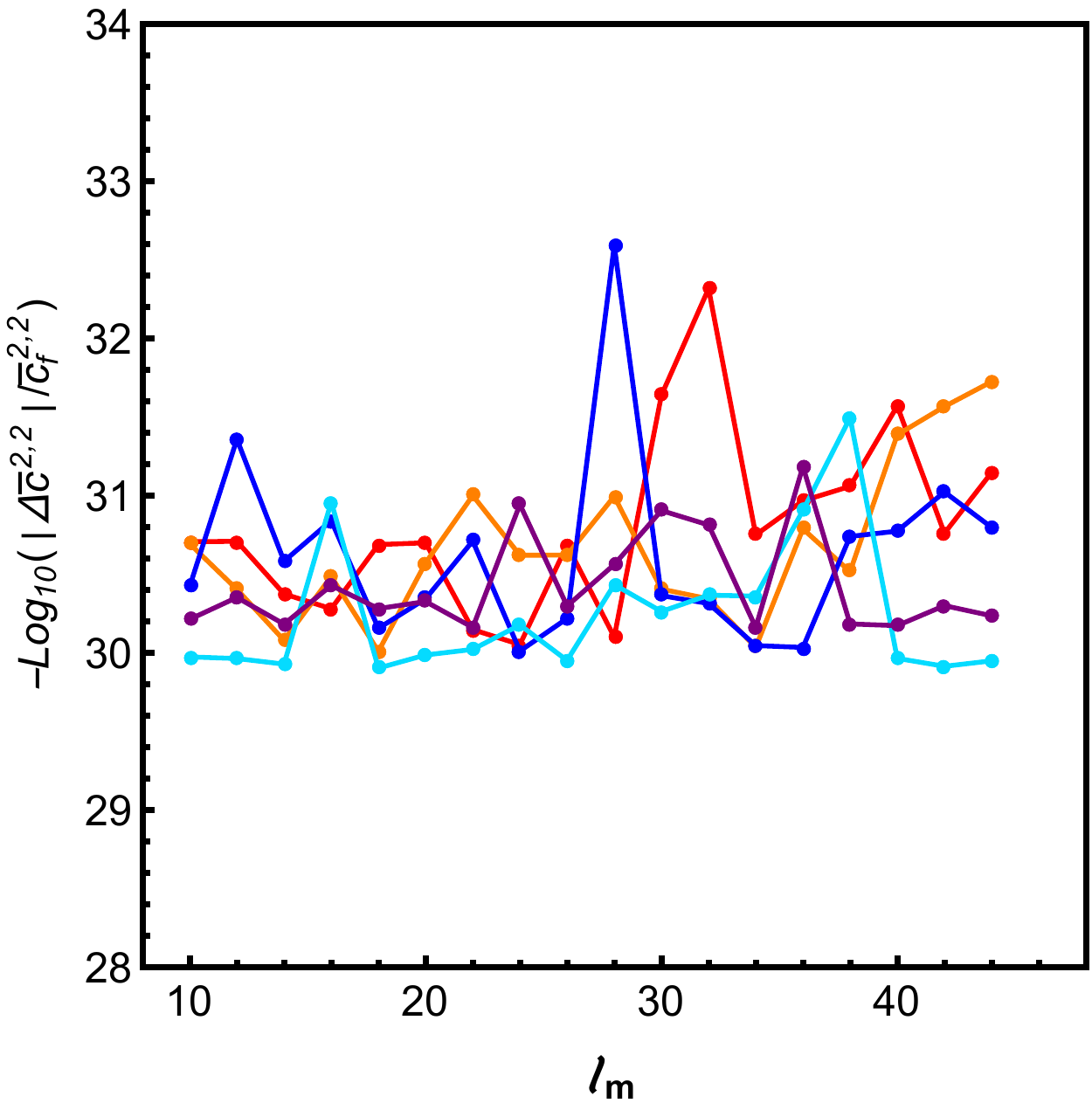}
    \end{subfigure}

    \centering
    \begin{subfigure}{0.9\linewidth}
    \centering
        \includegraphics[width=0.9\linewidth]{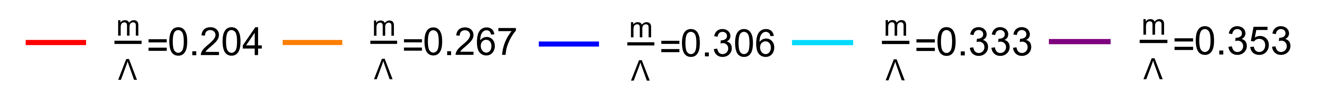}
    \end{subfigure}
  \caption{Convergence tests for the positivity bounds on $\bar{c}^{2,1}$ and $\bar{c}^{2,2}$. We have defined $\Delta \bar{c}^{2i,j}\equiv \bar{c}^{2i,j} -\bar{c}^{2i,j}_f$ where $\bar{c}^{2i,j}_f$ is the value of $\bar{c}^{2i,j}$ with the maximal $N$, $O_\mathrm{m}$ or $l_{\rm max}$. $N$ is the number of null constraints, $O_\mathrm{m}$ is the mass truncation order, and $l_{\rm max}$ is the maximal spin cutoff. The solid (dashed) lines are the upper (lower) bounds. The lower bounds on $\bar{c}^{2,2}$ are strictly zero.}
  \label{fig:afig1}
\end{figure}

On the other hand, the upper bound $\bar{c}^{2,2}$ can be obtained by changing the vector of decision variables to be $\eta \equiv (\alpha_{2,0},\alpha_{2,1} , -1 , \lambda_{1,3} )$, and solving the following linear program
\begin{equation}\label{lpmin}
  \begin{cases}
    \text{minimize: } O_{\rm min}=\eta \cdot(1,\bar{c}^{2,1}_{0},0,0) \\
    \text{subject to: } \eta \cdot V_l(x) \geq 0 , \quad \text{for any } x>0 \text{ and } l=0,2,4 \dots, l_{\rm max}.
  \end{cases}
\end{equation}
The upper bound on $\bar{c}^{2,2}$ for a given $\bar{c}^{2,1} = \bar{c}^{2,1}_{0}$ is then $\bar{c}^{2,2}\leq O_{\rm min}$. Including more null constraints and scanning sufficiently many $\bar{c}^{2,1}$, we will get Figure \ref{fig:c21}.

In outputting the figures in Section \ref{sec:sec3},  the number of null constraints used is typically $N=11$, the truncation order of the $m^2$ expansion is up to $O_{\mathrm{m}}=11$, and a partial wave truncation of $l_{\rm max}=40$ is usually sufficient. We set the precision parameter in SDPB to be 1024 and the value of the maximal iterations to be 1000.

\subsection{Convegence tests}

\begin{figure}
  \centering
  \begin{subfigure}{0.48\linewidth}
    \centering
        \includegraphics[width=0.955\linewidth]{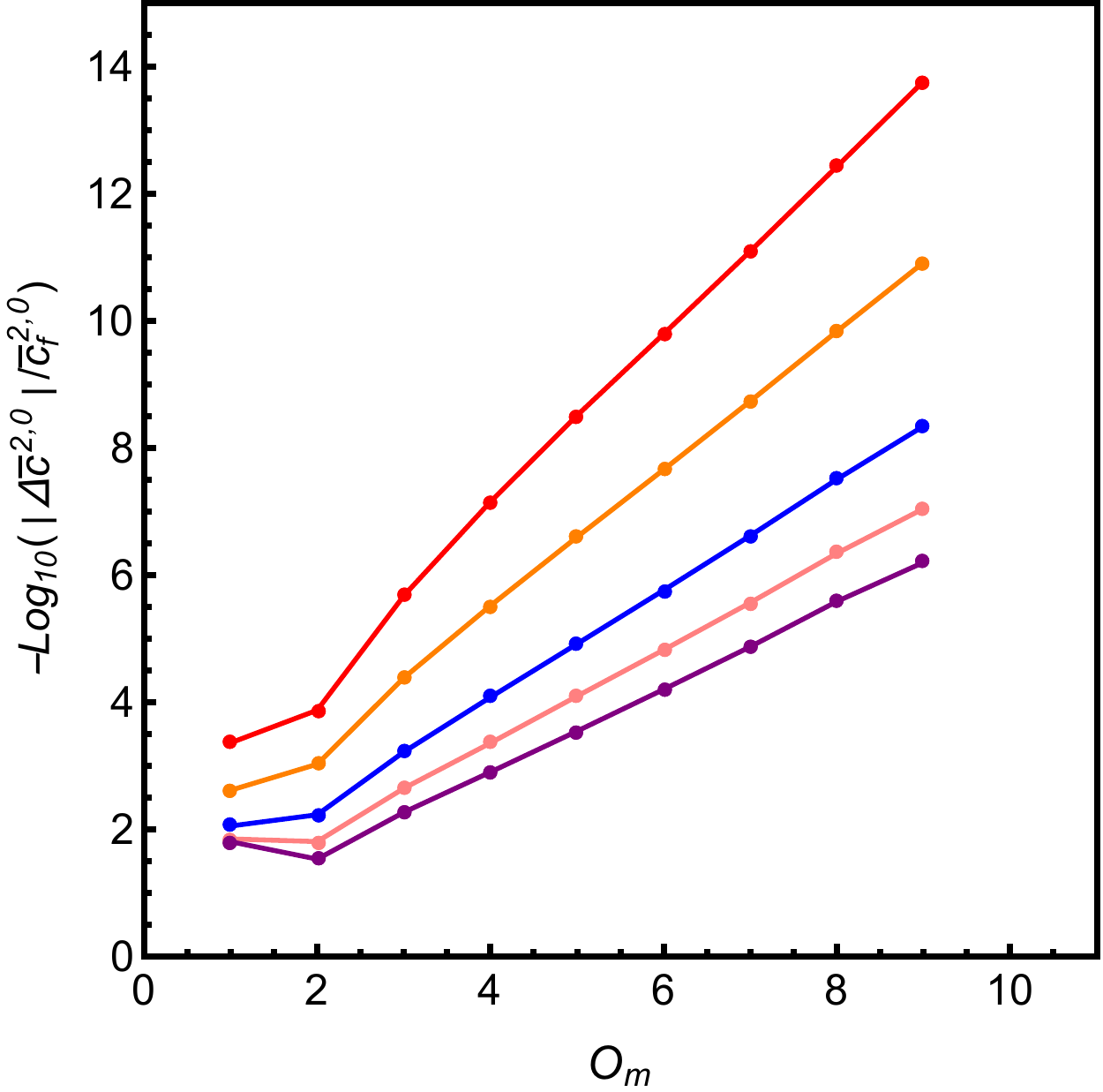}
    \end{subfigure}
    \centering~~
    \begin{subfigure}{0.48\linewidth}
    \centering
        \includegraphics[width=1\linewidth]{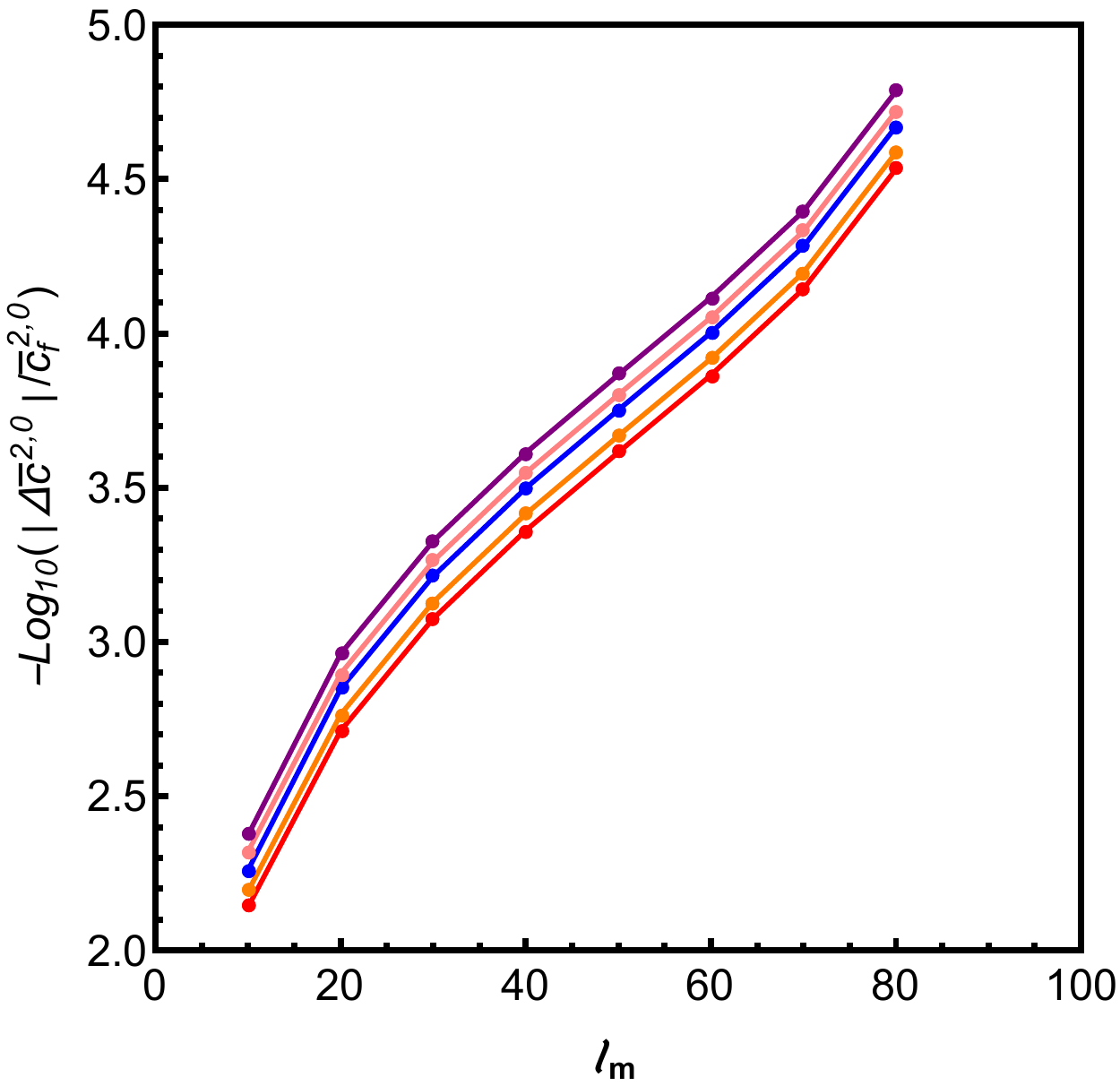}
    \end{subfigure}

    \centering
    \begin{subfigure}{0.9\linewidth}
    \centering
        \includegraphics[width=0.9\linewidth]{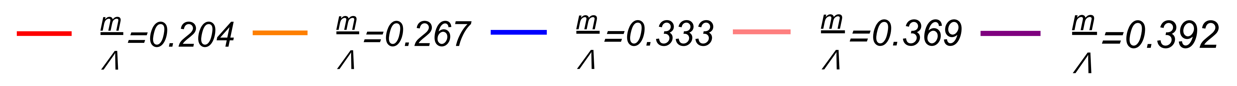}
    \end{subfigure}
  \caption{Convergence tests for the positivity bounds on $\bar{c}^{2,0}$. We have defined $\Delta \bar{c}^{2,0}\equiv \bar{c}^{2,0} -\bar{c}^{2,0}_f$ where $\bar{c}^{2,0}_f$ is the value of $\bar{c}^{2,0}$ with the maximal $l_{\rm max}$ or $O_\mathrm{m}$. $O_\mathrm{m}$ is the mass truncation order and $l_{\rm max}$ is the maximal spin cutoff. The solid (dashed) lines are the upper (lower) bounds. The lower bounds on $\bar{c}^{2,0}$ are strictly zero.}
  \label{fig:afig3}
\end{figure}

To test the robustness of our numerical results, we here will show some convergence tests for the triple crossing positivity bounds on $\bar{c}^{2,1}$ and $\bar{c}^{2,2}$, as well as the upper bound on $c^{2,0}$, which uses a different method than the one constraining $\bar{c}^{2i,j}$. In Figure \ref{fig:afig1}, we plot how the triple crossing positivity bounds vary with the number of null constraints $N$, the mass truncation order $O_\mathrm{m}$ and the maximal spin cutoff $l_{\rm max}$, which are used to compute the bounds on $\bar{c}^{2,1}$ and $\bar{c}^{2,2}$. We find that the lower bounds are usually numerically much more accurate than the upper bounds. In Figure \ref{fig:afig3}, we plot how the upper bound on $c^{2,0}$ varies with $O_\mathrm{m}$ and $l_{\rm max}$. We see that all the physical truncation errors are well under control, even for a relatively large mass $m$. While the precision of the bounds increases steadily with $O_{\rm m}$, it is often easy to achieve good accuracy when the number of null constraints used reaches a threshold, after which the accuracy often can not be raised significantly. For the maximal spin cutoff $l_{\rm max}$, the upper bound $c^{2,0}$ improves steadily with a larger $l_{\rm max}$, while the bounds on $\bar{c}^{2i,j}$ are often stable after $l_{\rm max}$ reaches a threshold.

\bibliographystyle{JHEP}  %use JCAP style
\bibliography{refs}

\end{document}